# Machine Learning-Based Classification, Interpretation, and Prediction of High-Entropy-Alloy Intermetallic Phases

**Jie Qi[1], Diego Ibarra Hoyos[1], and S. Joseph Poon[1,2]**

[1] Department of Physics, University of Virginia, Charlottesville, VA 22904
[2] Department of Materials Science and Engineering, University of Virginia, Charlottesville, VA 22904

**Abstract**: The design of high-entropy alloys (HEA) with desired properties is challenging due to their large compositional space. While various machine learning (ML) models can predict specific HEA solid-solution phases (SS), predicting high-entropy intermetallic phases (IM) is underdeveloped due to limited datasets and inadequate ML features. This paper introduces feature engineering-assisted ML models that achieve detailed phase classification and high accuracy. By combining phase-diagram-based and physics-based features, it is found that the ML models trained on the Random Forest (RF) and Support Vector Machine (SVM) regressors, are able to classify individual SS and common IM (Sigma, Laves, Heusler, and refractory B2 phases) with accuracies ranging from 80 - 94%. The machine-learned features also enable the interpretation of IM formation. Furthermore, the efficacies of the RF, SVM, and neural network (NN) models are critically evaluated. The phase classification accuracies are found to decrease upon utilizing the NN model to train the datasets. The accuracy of the model prediction is validated by synthesizing 86 new alloys. This approach provides a practical and robust framework for guiding HEA phase design, particularly for technologically significant IM phases.

Corresponding author:
S. Joseph Poon
sjp9x@virginia.edu, +1 434-924-6792
382 McCormick Rd, Charlottesville, VA, 22904
Jie Qi
jq4xa@virginia.edu, +1 434-327-0269
382 McCormick Rd, Charlottesville, VA, 22904

## 1. Introduction

The high-entropy alloy (HEA) design concept founded on the vast chemical degrees of freedom and nearly inexhaustible compositions has brought about a new paradigm in alloy discovery. The design space of HEAs is inevitably enormous. For example, a pool of 30 elements can constitute 142,506 different five-component HEA systems. Further inclusion of atomic percentages can easily lead to millions and even trillions of possible compositions. Thus, the new alloy design paradigm poses the fundamental challenge of selecting alloy phases and compositions endowed with desirable structural and functional properties. Predictions of alloy phases have been actively under development since the birth of HEA, including those involving *ab-initio* simulation, density functional theory (DFT), Calculation of Phase Diagrams (CALPHAD), empirical parameters, Machine Learning (ML), and artificial intelligence (AI)[1–3]. Compared to first-principles calculations and CALPHAD, ML is computationally much less intensive and yet the method has shown high accuracies, typically 70 – 90 %, in classifying HEA phases[3–23]. Thus, it is not surprising that ML remains a primary method for identifying HEA phases on demand. Beyond phase classification, microstructures and properties are often optimized with the help of CALPHAD and DFT.

Machine learning models normally categorize the phases of HEA as face-centered cubic (FCC), body-centered cubic (BCC), FCC+BCC, hexagonal closed packed structure (HCP), solid solution phases (SS), amorphous phase (AM), non-specific intermetallic phase (IM), and single/multi-phase. However, two issues exist in the current ML phase classification models, namely, a low number of phase categories, and in some cases, a low level of detail within a classified category; that is, the categories are general instead of specific. As discussed in more detail below, many models only classify HEA phases into no more than three categories, because as additional categories are included, there is an increase in the complexity, and the challenge of attaining high accuracy increases. Only FCC, BCC, FCC+BCC, and HCP categories in these models represent specific phases. More general categories, such as SS, AM, IM, and single/multi-phase, correspond to unspecified structural phase groups. The low level of categorization detail gives limited guidance for HEA design, i.e., when a HEA is categorized as IM, it can be either B2 (ordered BCC), Laves, Sigma, or Mu phase. This report is dedicated to solving the two challenges mentioned above by addressing the questions: (1) Can we achieve a more specific/detailed IM classification? (2) Can we predict more phase categories simultaneously and accurately to guide HEA design?

Detailed IM classification and prediction are certainly important in advancing the ML design of HEA beyond the common phases. Current knowledge indicates that Laves, Sigma, B2, and Heusler ($L2_1$) phases are four of the most common IM in HEA[24]. Heusler[25] and B2 phases can improve the HEA structural and functional properties[26–30]. Heusler phases are known for their wide range of multifunctional properties, including magneto-optical, magnetocaloric, and spintronic properties[25]. In addition, the Heusler phase is reported to have a superior creep resistance[31,32], and its presence in a HEA SS host can improve the mechanical properties[28,33,34]. A B2 phase generally consists of two types in HEAs: AlNi type[29] and Al-X-Y type (X and Y are specific groups of refractory elements)[35]. The AlNi type is widely used as a strengthening precipitate in HEAs[29], while the Al-X-Y type can improve high-temperature mechanical properties, lower physical density and cost, and enhance oxidization-resistance over the traditional disordered BCC refractory HEAs[35]. On the other hand, some IM, such as Sigma and Laves phases, are well-known for their embrittling effects[36,37]. The need to achieve the predictive formation of beneficial IM while avoiding unfavorable IM has led us to develop a more accurate and interpretable phase prediction method.

Phase predictions can be made efficient using appropriate features in training ML models, especially in the case of a smaller dataset. Previously, we introduced a set of ML features derived from binary phase diagrams, called phenomenological features that were able to classify ~ 850 HEAs into six categories[3,21]. Recently, feature engineering (FE), which has been underused in data science-driven materials research, has been successfully adapted to formulating superconducting critical temperature equations[38] and designing HEA[7]. Corresponding algorithms such as the Genetic Programming and SISSO[39] can effectively improve the prediction accuracy, especially for the regression ML problems. HEA phase prediction is a complex problem that may not be efficiently executed by using individual features alone. Rather, features should interact with each other to expand the feature pool and transform the feature space through which the classification error is reduced. Apart from FE, certain ML algorithms (such as neural networks) inherently allow interaction among the features. This inherent capability may mitigate the need for employing engineered or transformed features as inputs to the ML model. However, the inherent use of transformed features at the expense of individual features often decreases the accuracy since now the model cannot isolate the dependencies on individual features. Conversely, the FE methodology adopted in the present study engineers the individual features before the application of ML algorithms. This process permits the utilization of both individual and transformed features in ML classification tasks. Furthermore, FE allows more flexible mathematical operations amongst individual features (as detailed in Section 3), creating a broader set of engineered features. This method thus enables the model to better capture complex patterns and relationships in the data, potentially improving its predictive power and interpretability.

In this article, we will describe a feature engineering strategy that synergistically blends phase diagram-based (PD) features, thermodynamic (Thermo) features, and Hume-Rothery rule (HR) features to interpret and predict the formation of different HEA phases. The unblended features are described in a recent review[3]. The feature engineering-based method enhances the accuracy of the ML models to near 90 % for nine HEA phases categories: BCC, FCC, HCP, FCC+BCC, AlNi type B2+, Sigma+, Laves+, Heusler+, and Al-X-Y type B2+. The symbol "+" denotes possible coexisting phases. As such, the present method predicts more phase categories with a higher level of specificity and accuracy than other reported methods to date. In addition, we provide a feature importance analysis on the Thermo and HR factors to interpret the driving forces for specific phase formation. The ML-trained features have given deeper insights into the stability of IM in the complex landscape of phase competition. To validate the model's predictive capability, alloy synthesis and characterization were conducted on 86 new compositions. The accurate and interpretable ML models presented herein can be integrated with other HEA property prediction models, based on which HEA compositions with targeted phases and properties can be designed with high reliability.

## 2. Overview of Methodology and Model Accuracy

The HEA phase classification methodology utilizes a two-layer method, as illustrated in Fig.1. The first layer corresponds with the multi-phase prediction model for SS (FCC, BCC, HCP, FCC+BCC) and common IM (AlNi type B2+, Laves+, and Sigma+). Categories FCC, BCC, and HCP are indicative of single phase HEAs. FCC+BCC corresponds to coexistence of FCCs, BCCs, or FCCs and BCCs. The category AlNi type B2+ applies to HEAs that exclusively form B2 as the IM phase, potentially alongside other SS phases. Finally, categories Laves+ or Sigma+ represent HEAs that form Laves or Sigma phases in combination with other phases. The model has an overall accuracy of 84 % in classifying specific phases in 835 HEAs, a 4% improvement over that reported by the authors earlier[21]. In particular, the accuracy for AlNi type B2+ is high at 90%, while the accuracies for Laves+ and Sigma+ are lower by ~10%. Accordingly, the second layer consists of four models that are grouped into two pairs for IM prediction. If a HEA is predicted as one of the commonly occurring Laves+ or Sigma+ in the first layer, then the verification from two models in the second layer, as shown on the left in Fig. 1, will result in accuracies above 90 % for both phases. On the other hand, if the multi-phase prediction model predicts no Laves+ or Sigma+ formation, two other models will evaluate whether the HEA can form IM Heusler or Al-X-Y type B2 phases, with accuracies of 92 % and 80 %, respectively. In other words, the two-layer method herein can predict single-phase HEAs as well as HEA composites comprising specific phases with high accuracy. Readers should be aware that Fig.1 is a general alloy design path in our research where we eliminate the formation tendency of the typically undesired Laves and Sigma phases before more computation is conducted to predict the formation of functionally important phases such as Heusler and Al-X-Y type B2 phase. The sub-models in second layer are not mutually exclusive. If a user so chooses, these models can be adjusted to operate concurrently on the same HEAs.

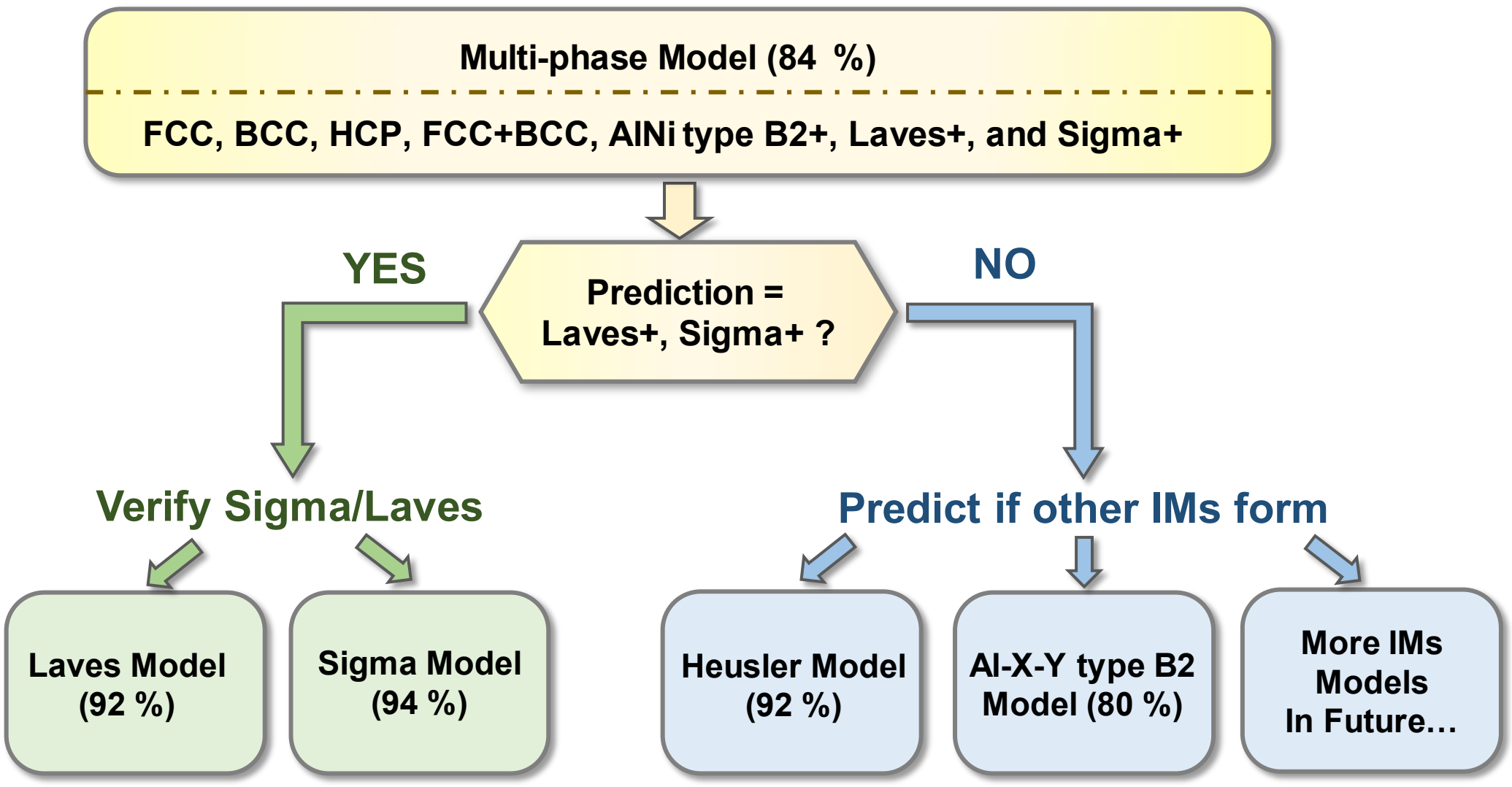

Figure 1: Two-layer method for predicting HEA phases. Pathways of modeling, with resulting classification accuracy (parentheses) for each model, are shown.

## 3. Method

### I. Machine Learning (ML)

In general, ML models are trained based on training datasets and then give predictions of new data points. ML is normally applied to regression and classification problems. The regression models predict continuous values, such as the hardness of HEAs. HEA phase prediction is a classification problem where HEAs are classified into different phase categories.

In this work, each HEA datum contains the phase category it belongs to and the features, which are the selected physical parameters for classifying and predicting the phase. ML classification algorithms, such as the Random Forest, Support Vector Machine (SVM), and Neural Network, serve to identify the relationship between features and phases in the feature space. Different classification algorithms are tested and compared, and the one with the highest accuracy is picked for each model. All computational processes related to this work, encompassing feature computation and ML, were carried out using the MATLAB programming language. Further specifics about these computations, and the comparison of different algorithms' performance can be found in Section 3 of the supplementary material. In the ML HEA phase prediction problem, feature construction and selection are the most crucial parts. This process is described below.

### II. Raw Features

Prior studies have utilized thermodynamic (Thermo), Hume-Rothery rule (HR), and phase diagram-based (PD) features[21] in HEA phase formation[3]. Thermo features represent the various thermodynamic driving forces for forming SS and IM. HR features influence phase formation from the atomic size mismatch and electron configuration aspects. In this work, Thermo features $\Delta S_{\mathrm{mix}}$[40], $\Delta \mathrm{H}_{\mathrm{mix}}$[41], $\Omega$[42], $\Phi$[43], $\eta$[44], and $\mathrm{k}_1^{\mathrm{cr}}$[45], and HR features $\delta$[41], $\frac{\mathrm{E}_2}{\mathrm{E}_0}$[46], $\Delta\chi$[47], and VEC[48–50] are used with their definitions given in Table 1. Following our previous study, alloy melting temperatures $\mathrm{T}_\mathrm{m}$ involved in the feature calculation are calculated by tracing the liquidus trends in the phase diagrams to capture the effect of alloying[21]. The PD features, introduced in the prior study[21], are extracted from binary phase diagrams. These PD features, $\mathrm{PFP}_{\mathrm{A1}}$, $\mathrm{PFP}_{\mathrm{A2}}$, $\mathrm{PFP}_{\mathrm{A3}}$, $\mathrm{PFP}_{\mathrm{B2}}$, $\mathrm{PFP}_{\mathrm{Laves}}$, $\mathrm{PFP}_{\mathrm{Sigma}}$, and PSP, were defined as representing the probability of forming FCC_A1, BCC_A2, HCP_A3, AlNi type B2, Laves, and Sigma phases, as well as the phase separation[21]. In contrast to the Thermo and HR features, which typically overlook alloy preparation methods, PD features inherently incorporate thermal processing associated with alloy preparation. For example, for as-cast HEAs, PD features are computed at a ML optimized phase formation temperature, approximately $0.8\mathrm{T}_\mathrm{m}$; for the as-annealed HEAs, PD features are determined at the respective annealing temperatures. A total of 17 Thermo, HR, and PD features are used as the raw features in this work.

Table 1: Definition of thermodynamic and Hume-Rothery rule features used in this work

| **Formula** | **Comments** |
|---|---|
| Mixing Entropy: $\Delta S_{mix} = -R\sum_{i=1}^{N} c_i \ln(c_i)$ | R: The gas constant. $c_i$: The atomic percentage of the i-th element for a N-component system. (Definitions of N and $c_i$ are the same elsewhere.) |
| Mixing Enthalpy: $\Delta H_{mix} = \sum_{i=1,i\neq j}^{N} 4\,\Delta H_{i,j}^{mix}\, c_i c_j$ | $\Delta H_{i,j}^{mix}$: The binary mixing enthalpy obtained from Miedema's model of i-j element pair. |
| $\Omega = \frac{T_m \Delta S_{mix}}{\lvert\Delta H_{mix}\rvert}$ | $T_m$: Alloy melting temperature. |
| $\Phi = \frac{\Delta G_{SS}}{-\lvert G_{max}\rvert}$ | $\Delta G_{SS}$: The Gibbs free energy change for forming a fully disordered SS phase. $\Delta G_{max}$: The larger absolute Gibbs free energy change of forming the strongest binary compound, or having phase segregation. |
| $\eta = \frac{-T_{ann}\Delta S_{mix}}{\lvert\Delta H_f\rvert}$ | $T_{ann}$: Annealing temperature. If $T_{ann}$ is not known, use $T_{ann} = 0.8\,T_m$. $T_m = \frac{\sum_{i\neq j} T_{i-j}\times c_i \times c_j}{\sum_{i\neq j} c_i \times c_j}$, where $T_{i-j}$ is the melting temperature of the i-j elements for the relative ratio of the two elemental concentrations $c_i$ and $c_j$ of the HEA composition. $\Delta H_f$: The most negative binary mixing enthalpy for forming IM[44]. |
| $k_1^{cr} = \frac{\left(1 - \frac{0.4T_m\,\Delta S_{mix}}{\Delta H_{mix}}\right)}{\frac{\Delta H_{IM}}{\Delta H_{mix}}}$ | $\Delta H_{IM}$: Mixing enthalpy of forming IM. When $k_1^{cr} < 1$, IM tends to form. Otherwise, SS tends to form. |
| Radius Mismatch: $\delta = \sqrt{\sum_{i=1}^{N} c_i \left[1 - \frac{r_i}{\sum_{j=1}^{N} c_j\, r_j}\right]^2}$ | $r_i$: The atomic radius of the i-th element. This definition is the same throughout the document. |
| $\frac{E_2}{E_0} \propto (\Delta d)^2 = \sum_{j\geq i}^{N} \frac{c_i c_j \lvert r_i + r_j - 2\bar{r}\rvert^2}{(2\bar{r})^2}$ | $\bar{r} = \sum_{i=1}^{N} c_i\, r_i$: Average atomic radius. $\Delta d$: The strain due to atomic radius difference. |
| Electronegativity Mismatch: | $\chi_i$: Electronegativity of i-th element. |
| $\Delta\chi = \sqrt{\sum_{i=1}^{N} c_i \left[\chi_i - \sum_{j=1}^{N} c_j \chi_j\right]^2}$ | |
| Mean Valence Electron Concentration: $\mathrm{VEC} = \sum_{i=1}^{N} c_i \, \mathrm{VEC}_i$ | $\mathrm{VEC}_i$: Valence electrons count of the i-th element. |

## III. Feature Engineering and Feature Selection

Feature engineering is a technique for developing and identifying the best math variations of raw features. The process includes feature construction, transformation, reduction, and selection.

The feature construction process collects the individual raw physical features that may influence phase formation. All relevant raw features are included regardless of the degree of importance they possess in determining the phase. Unimportant features are filtered out in the later steps.

The feature transformation process (Fig. 2A) transforms the raw features by first constructing mathematical variations $x^2, x^{-1}, \sqrt{x}, \ln(x)$, and $e^x$ for each feature X. The different expressions can mathematically change how features influence the phase prediction in ML algorithms. For example, $\ln(x)$ or $e^x$ may reduce or inflate the effect of the outliers compared to using feature X. Then, the feature pool is further expanded by grouping any two math variations, A and B, using operations A+B, A-B, A/B, and AB. This step creates some synergetic effects from multiple features. For example, the comparison effects (A-B, A/B) or joint effects (A+B, AB) may bring new insights into phase prediction. At this point, the feature transformation constructs a huge feature pool, which potentially includes engineered features more qualified for phase prediction than the raw features. The current work expands 17 raw features to ~ 25,000 engineered features. Then to select the best features from the pool, a systematic method including feature reduction and selection is provided below. The feature reduction and selection methods contain filtering, intrinsic, and wrapper methods:

1. Filtering method:

The Pearson Correlation Coefficient (PCC) between two features indicates their linear correlation strength. As shown in Fig. 2B, PCC values approaching +1, -1, or 0 indicates a strong positive, strong negative, or no linear correlation. Strongly correlated features are considered to be inter-substitutable in ML. Therefore, only one feature is kept from any pair with |PCC| > 0.9 in this work.

2. Intrinsic method:

Direct feature selection from the filtered-out features is computationally expensive and unnecessary as many features are irrelevant to phase formation. Therefore, a rapid ML method, logistic regression (LR) with L1 (or Lasso) regularization, is used to remove the irrelevant features (Fig. 2C). This algorithm will minimize the total prediction cost as follows:

$$J(\overrightarrow{W}) = \frac{1}{m}\sum_{j=1}^{m} \mathrm{Cost}\left[h_{\overrightarrow{w}}\left(\overrightarrow{F}^{(j)}\right), y^{(j)}\right] + \gamma \sum_{i=1}^{n} |w_i| \quad \text{(Eqn.1)}$$

Herein, $J(\overrightarrow{W})$ is the prediction cost with feature weight vector $\overrightarrow{W} = [w_1, w_2, \ldots, w_n]$. The first term is LR prediction cost $\mathrm{Cost}[h_{\overrightarrow{w}}(\overrightarrow{F}^{(j)}), y^{(j)}]$ calculated by the log-loss function[51], which is directly related to the classification error, wherein the cost function of predicting the j-th sample as $h_{\overrightarrow{w}}(\overrightarrow{F}^{(j)})$ while the correct category is $y^{(j)}$. $h_{\overrightarrow{w}}(\overrightarrow{F}^{(j)})$ is obtained based on feature weights $\overrightarrow{W}$ and feature values $\overrightarrow{F}^{(j)}$. m is the total sample count in the dataset. The second term is the regularization cost. n is the number of features. $\gamma$ is the regularization strength. $w_i$ is the i-th feature's weight in $\overrightarrow{W}$. To reduce $J(\overrightarrow{W})$, the first term tends to use more features to reduce the prediction error, while the second term tends to invalidate more features by zeroing their weight $w_i$. The trade-off between the two terms will activate the minimum number of essential features in ML. Tuning $\gamma$ changes the regularization strength and regulates the number of selected/activated features. After this step, about 100 features are retained.

3. Wrapper method:

Sequential learning (SL), shown in Fig. 2D, selects the best features iteratively from ~100 features. ML models built with different combinations of features are evaluated by the average error from thirty rounds of 5-fold cross-validations with different random seeds. The error is calculated by 1- *f1* score (The same definition on error is used throughout this article). SL starts with an empty feature set in the first round, tests each feature in ML algorithm independently, and picks the feature with the lowest ML classification error. In the subsequent rounds, each unselected feature is tested combinatorically with the previously picked ones. Finally, the best feature combination to minimize the classification error is constructed.

It is worth noting that all ML processes, including the LR algorithm in the intrinsic method section and the ML algorithms in the wrapper method, have been subjected to feature value normalization.

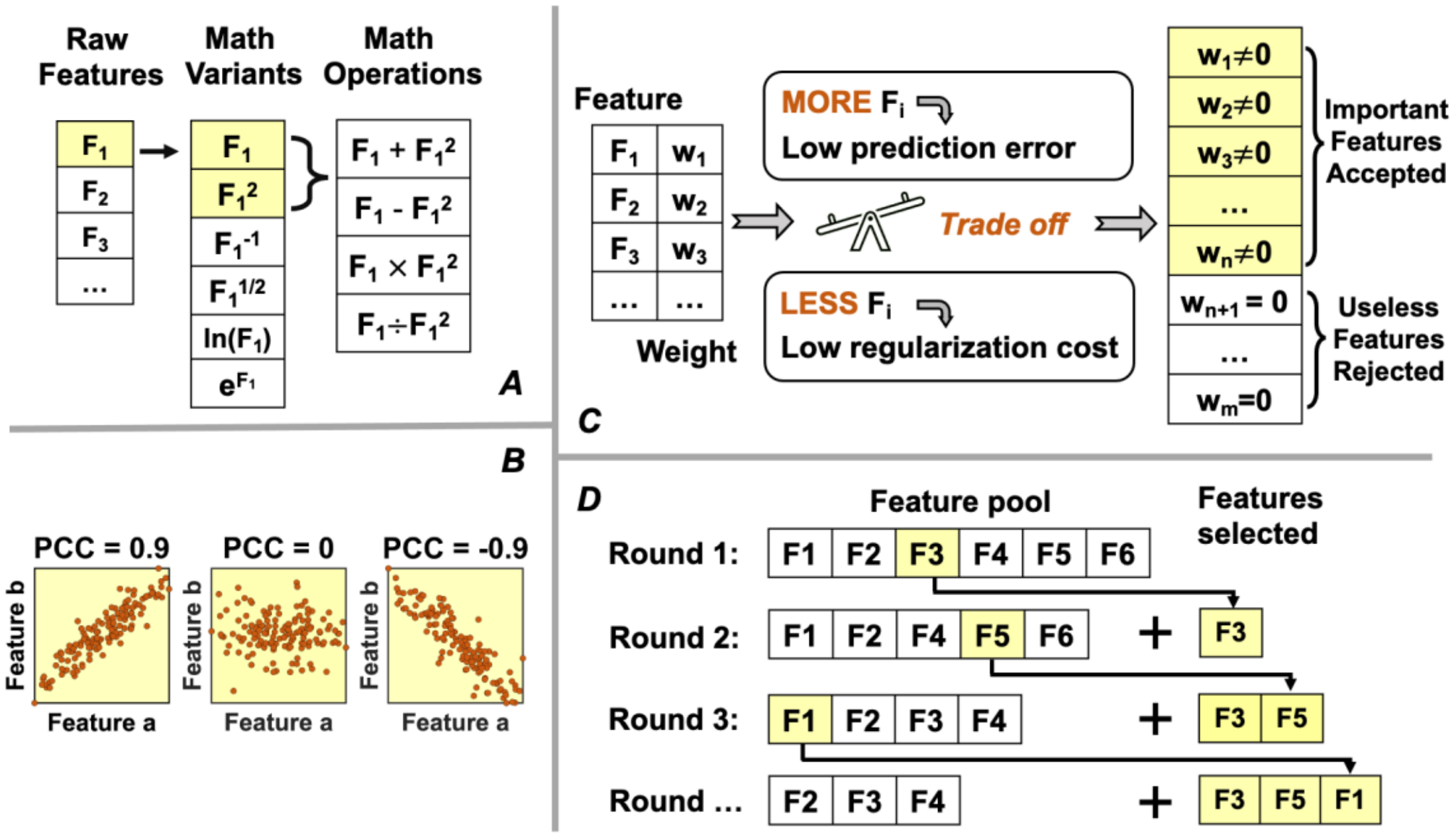

Figure 2. Process of feature engineering: (A) Feature expansion method; (B) PCC values reflect the linear correlations between two features; (C) Intrinsic method: LR with L1 regularization to eliminate features irrelevant to phase formation; (D) Wrapper method: SL selecting several best features for ML.

## IV. Experimental Method

Alloys for validation were synthesized using arc melting. Raw materials with a minimum purity of 99.97 wt.% were placed into a water-cooled copper crucible. Raw materials were melted five times under a high-purity argon atmosphere. Each melt was conducted for a minimum of a minute. The sample was flipped over between melts to ensure homogeneity. All HEAs were characterized in the as-cast state, consistent with most data used in training the presented ML models. The ML models are set in the high-temperature ranges most suitable for as-cast alloys or alloys annealed at high temperatures, e.g., ~ 0.8 of the melting temperatures[21]. Finally, alloys were polished using grinding papers with grit sizes 180, 320, 600, and 1200. X-Ray Diffraction (XRD) measurements were conducted on a PANalytical Empyrean diffractometer with Cu Kα radiation and a scanning rate of ~0.15 degree/s.

## 4. Results and Discussion

### Part 1: Multi-phase Prediction Model

As described in the methodology section, the multi-phase prediction model in the first layer (Fig. 1) has seven categories: FCC, BCC, HCP, FCC+BCC, AlNi type B2+, Laves+, and Sigma+. Different classification algorithms are tested, and the Random Forest (RF) classification algorithm is used to perform sequential learning (SL). In comparison, the Neural Network (NN) has relatively low accuracy, likely as a result of the large amount of training data required. The performance comparison across different ML algorithms can be found in Fig. S1 of the supplementary material. Thirty rounds of SL were conducted. Fig. 3A shows the overall classification errors and the error bars (standard deviation) plotted against the number of top-ranked raw features (labeled as “No FE”) and engineered features (labeled as “FE”), respectively. We only keep the first six engineered features to train the ML prediction model because adding more features only increases the risk of over-fitting disproportionately to the diminishing gains in accuracy. A list of these features is presented in Table 2. The FE classification error with six features is 0.161, 10 % smaller than the error without FE. Fig. 3B shows the classification errors of the individual phase category plotted against the number of top-ranked engineered features. HCP, AlNi type B2+, FCC, and BCC predictions have lower errors, while FCC+BCC, Sigma+, and Laves+ predictions are relatively less accurate. Therefore, the IM formation needs to be verified by the models discussed in the next section. Fig. 3C gives the database category size. The available HEA experimental data is continuously expanding. As shown in previous work[21], expanding the database may only improve marginally a well-trained model’s accuracy. Thus, we consider the up-to-date 835 data collected from literature are sufficient in training the ML.

We comprehensively evaluate the model performance based on the following criteria: ML accuracy, the level of detail on phase categories, the number of phase categories, and the number of features. This model’s prediction error is among the lowest compared to other similar models[6–12,17,20,22,23,52]. With the use of only six features, this model is able to classify up to seven phase categories with a high level of category detail, i.e., detailed phase content such as Heusler and Sigma can be identified instead of classifying them into a general category labeled as “IM”. In addition, we address the functionally important IM phase AlNi type B2+, Laves+, and Sigma+, which have rarely been explored by ML methods. Overall, our FE-assisted ML model shows high capability in classifying HEA phases.

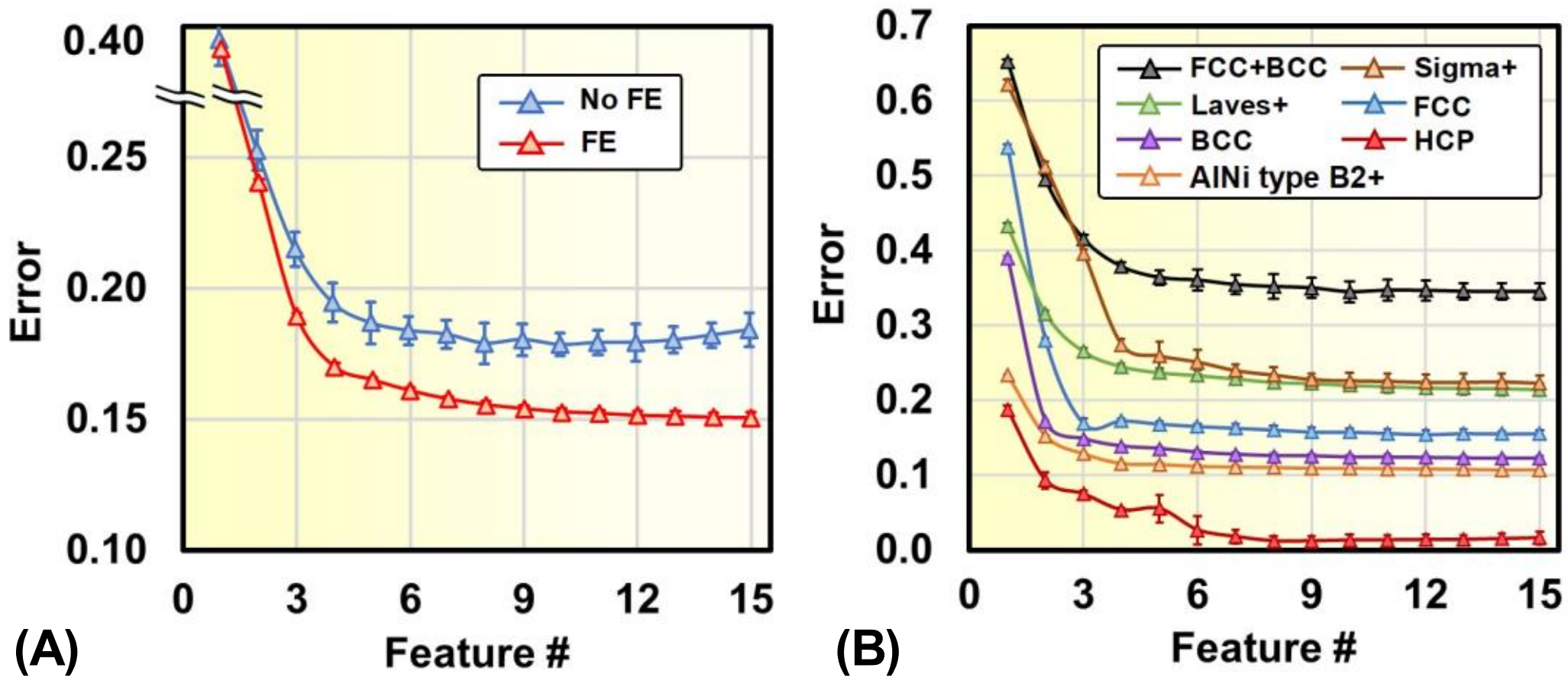

(C)

| Total HEA# | AlNi type B2+ | BCC | FCC | HCP | LAVES + | FCC+BCC | SIGMA + |
|---|---|---|---|---|---|---|---|
| 835 | 291 | 178 | 132 | 14 | 96 | 72 | 52 |

Figure 3. (A) Overall classification error of multi-phase prediction model (first layer) versus the number of top-ranked features is plotted with error bars (standard deviation). Results with and without FE are shown. (B) Classification errors for individual phase categories versus the number of engineered features. (C) The number of HEA data in the database and each phase category.

Table 2: Engineered Features selected for phase prediction.

| Prediction Model | Features |
|---|---|
| Multi-phase | $\eta$, $PFP_{A1} - e^{PFP_{A3}}$, $\frac{E_2}{E_0} \cdot \Delta H_{mix}$, $\Delta\chi^2 \cdot \sqrt{PFP_{Laves}}$, $PFP_{Sigma} \cdot \Phi$, $PFP_{A3}/e^{\delta}$ |
| Laves+ | $k_1^{cr}/\ln(PFP_{Laves})$, $\Delta H_{mix} \cdot \sqrt{\Omega}$, $PFP_{Laves} \cdot PFP_{A1}$, $\Phi \cdot \sqrt{PFP_{Laves}}$ |
| Sigma + | $\Delta\chi^2 \cdot \ln(PFP_{Sigma})$, $\Delta\chi \cdot VEC^2$, $PFP_{A1} \cdot \sqrt{PFP_{A3}}$, ${PFP_{B2}}^2/\ln(PSP)$ |
| Heusler + | $\delta/\Phi$, $PFP_{Sigma} \cdot {\Delta H_{mix}}^2$, $PFP_{B2}/{PFP_{A2}}^2$, $PFP_{B2} \cdot PFP_{A3}$ |
| Al-X-Y type B2+ | $\eta + \Delta\chi$, $\Delta S_{mix} \cdot VEC^2$, $PSP \cdot PFP_{A3}$ |

**Part 2: Laves+, Sigma+, Heusler+, and Al-X-Y Type B2+ Prediction Models**

The four models in the second layer (Fig. 1) use Support Vector Machine as the classification algorithm, given its reduced prediction error compared to other algorithms, as demonstrated in Fig. S1 of the supplementary material. For the Sigma+ and Laves+ prediction models, the appreciable imbalanced data distributions require special handling. For example, the Sigma+ prediction model database consists of 52 Sigma-containing HEAs ($HEA_{Sigma}$) and 783 HEAs without the Sigma phase ($HEA_{No\text{-}Sigma}$). The imbalance makes the ML model biased to the dominant category $HEA_{No\text{-}Sigma}$ and adversely affects the predictions for $HEA_{Sigma}$. Conventional methods of handling imbalanced databases include under-sampling and over-sampling methods such as the Random Over-sampling, Adaptive Synthetic Sampling Approach for Imbalanced Learning (ADASYN)[53] and Synthetic Minority Over-sampling Technique (SMOTE) [54]. The under-sampling method is used here since it is more accurate and does not artificially generate virtual data to balance the two categories as some over-sampling methods do. More comparisons and technical details about the under/over-sampling methods can be found in Section 4 of Supplementary Materials. The under-sampling method will randomly pick 52 samples from the $HEA_{No\text{-}Sigma}$ to constitute a ML database with the 52 $HEA_{Sigma}$. Thirty rounds of random samplings followed by sequential learning (SL) are conducted, and the average performance is presented.

Similarly, the Laves+ prediction model database consists of 96 Laves-containing HEAs ($HEA_{Laves}$) and 739 HEAs without the Laves phase ($HEA_{No\text{-}Laves}$). $HEA_{No\text{-}Laves}$ are under-sampled to 96 to constitute a ML database with the 96 $HEA_{Laves}$ in each of the thirty random sampling rounds.

Fig 4 A and B show how errors decrease with more features and compare the results with and without FE for Sigma+ and Laves+ predictions. In both models, only the first four engineered features will be kept for ML prediction, and we obtain low errors of 0.06 and 0.08 for Sigma+ and Laves+ predictions, respectively. FE suppresses the error by around 0.05 from No-FE results. The four features giving the lowest error among the thirty rounds are presented in Table 2.

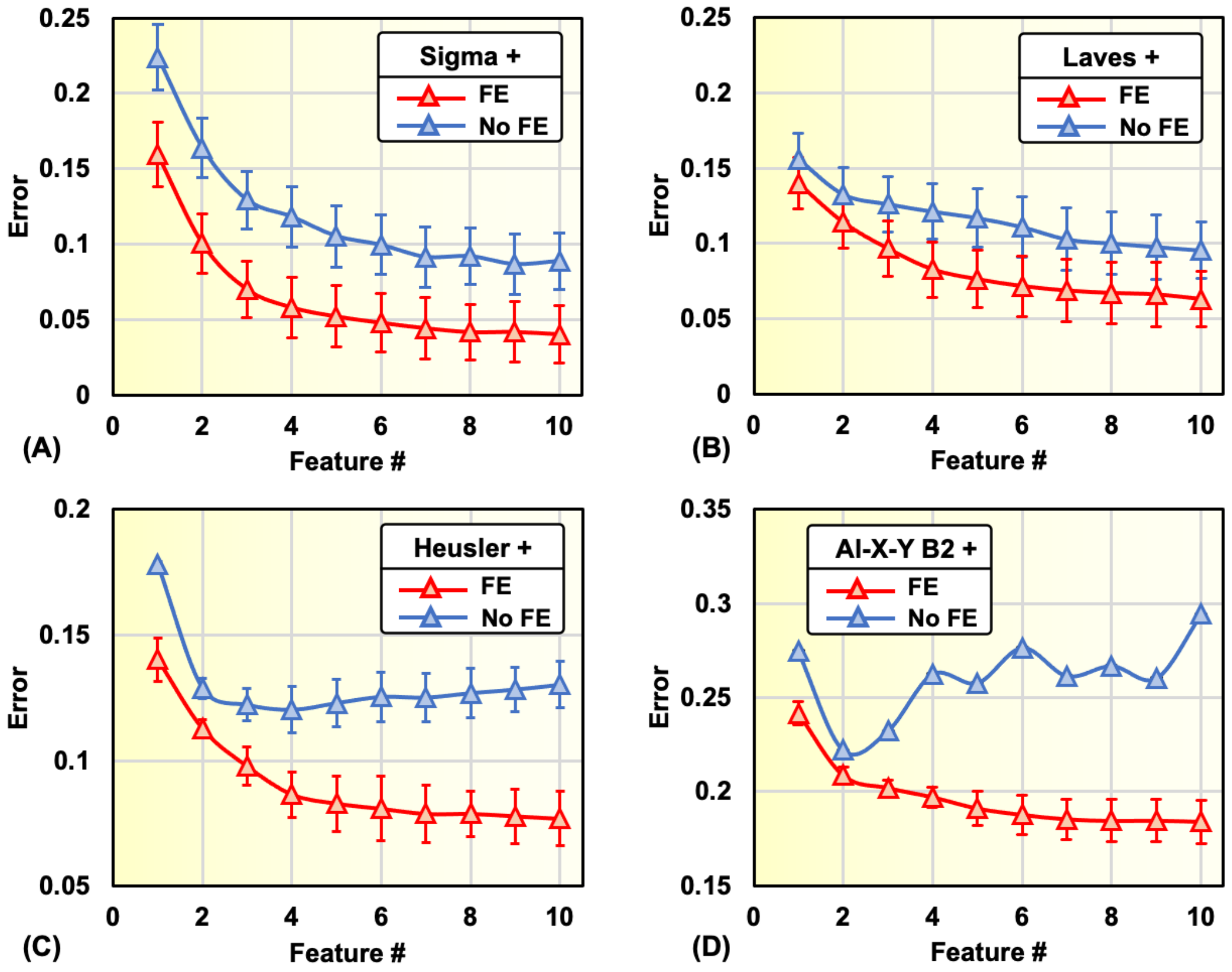

Figure 4. ML classification error decreases as the number of engineered features increases. The comparisons of the results between using and not using FE are presented for: (A) Sigma +; (B) Laves +; (C) Heusler +; and (D) Al-X-Y B2 + prediction models. Error bars (standard deviation) are presented in all plots. Small error bars may be invisible in figure (D).

A Heusler phase has a general composition $X_2YZ$, where X, Y, and Z symbolize specific groups of elements in the periodic table[55]. The database constitutes 77 HEAs containing the Heusler phase ($HEA_{L21}$), and 109 HEAs without the Heusler phase ($HEA_{Non\text{-}L21}$). $HEA_{Non\text{-}L21}$ are selected based on the criteria: (1) they include appropriate X, Y, and Z elements for forming the Heusler phase; and (2) they are annealed to ascertain the non-emergence of the Heusler phase. Thirty rounds of SL are conducted. The average classification errors for using FE and not using FE are presented in Fig. 4C. As more features are included, FE error becomes saturated, and No-FE error increases due to over-fitting. The top-ranked four engineered features (listed in Table 2) are kept for ML prediction with a classification error of 0.08. FE suppresses the error by 0.05 over No FE. The $HEA_{L21}$ database is included in the Section 1 of the Supplementary Materials.

The refractory Al-X-Y type B2 phase comprises at least three components: X is Ti, Zr, or Hf; and Y is Cr, Mo, Nb, or V[35]. The database consists of 52 HEAs with Al-X-Y type B2 phase ($HEA_{AlXY\text{-}B2}$) and 35 without Al-X-Y type B2 phase ($HEA_{Non\text{-}AlXY\text{-}B2}$) but having Al, X, and Y elements. From thirty rounds of SL, the average classification errors for using FE and not using FE are presented in Fig. 4D. As more features are included, the FE error continuously drops while the No-FE error increases rapidly due to over-fitting. The best three engineered features (listed in Table 2) are kept with a classification error of 0.2. The $HEA_{AlXY\text{-}B2}$ database can be found in the Section 2 of the Supplementary Materials.

**Part 3: Interpreting the Important Phase Determination Features**

Although ML is a powerful classification tool, it is a black box and does not show the input and output relationships. Therefore, appropriate techniques are needed to evaluate the features' importance in determining phase formation[56]. The single accuracy method, which uses only one feature for ML at a time and takes the classification accuracy as the feature importance, is utilized in this work. Although FE is found to reduce the prediction error, the feature variants generated are not amenable to direct physical meaning interpretation. Therefore, FE is not applied in this part. Moreover, PD features as phenomenological parameters [3,21] do not directly infer the physical mechanism of phase formation. On the other hand, Thermo and HR features can reflect the physics and are deemed to play an important role in the classification of specific IM. Therefore, we will identify the three most important IM formation determining Thermo and HR features from the feature importance values shown in Fig. 5, and plot the HEA distribution probability density function based on the values of these features in Fig. 6 to interpret their influence on specific IM formation. In the single accuracy method, the SVM algorithm is once again deployed to assess the classification accuracy of individual features.

From Fig. 5A, the Heusler phase formation is mainly controlled by VEC, $\Phi$, and $\frac{E_2}{E_0}$. $HEA_{L21}$ generally have lower VEC values than $HEA_{Non\text{-}L21}$ (Fig. 6-H1). The low VEC implies that a BCC-prone environment[48–50] is favored for the Heusler phase formation, potentially due to the structural similarity between the Heusler and BCC lattices. $\Phi$ is a Thermo feature controlling IM/SS formation tendency. IM formation is favored when $\Phi$ is small[43]. Fig. 6-H2 shows that $HEA_{L21}$ are generally low in $\Phi$ and energetically favored to form IM. Finally, $HEA_{L21}$ generally have larger $\frac{E_2}{E_0}$ values (Fig. 6-H3), which represent larger atomic size difference[46]. This makes specific elements, such as Al, whose atomic size is different from the transition-metal elements, confined to certain sites on a crystal lattice, forming the ordered Heusler phase.

Al-X-Y type B2 formation is predominantly controlled by η, $\Delta S_{mix}$, and Ω (Fig. 5B). $HEA_{AlXY\text{-}B2}$ generally have more negative η values (Fig. 6-B1), which indicates the IM formation tendency, consistent with DFT results[44]. $HEA_{AlXY\text{-}B2}$ also have a wide $\Delta S_{mix}$ distribution spectrum (Fig. 6-B2) while $HEA_{Non\text{-}AlXY\text{-}B2}$ are clustered at the high $\Delta S_{mix}$ value region. Higher $\Delta S_{mix}$ prompts the disordering and suppresses the ordered $HEA_{AlXY\text{-}B2}$ formation. Ω is another Thermo feature showing the SS and IM formation tendencies[42]. $HEA_{AlXY\text{-}B2}$ generally have low Ω values (Fig. 6-B3), which favors the ordered IM phase such as the B2 formation. More importantly, all three dominant features are thermodynamic, and HR features show limited influence. Electron environment-related HR features, VEC and Δχ, are found to be correlated to FCC, BCC[57], and topological close-packed Sigma and Laves[50,58] but not B2 formation. Lattice distortion-related HR features, $\frac{E_2}{E_0}$ and δ, are relatively more important for predicting the IM with non-cubic structures (e.g., Laves) which can accommodate the severe atomic size mismatch. The B2 phase retains the BCC structure, where small lattice distortion should be expected for both disordered BCC and B2 phases. Despite the low effectiveness of the HR features, the key to ML predicting Al-X-Y type B2 is to distinguish it from the disordered BCC, where enthalpy and thermodynamic consideration are proven to be crucial in determining BCC/B2 ordering by a Monte Carlo and DFT combined study[59]. Our ML model draws a similar conclusion. In future, first-principles methods such as *ab-initio* simulations and DFT are promising to give an accurate, in-depth analysis of the order-disorder transition of such alloy systems.

For the Laves phase formation, $\frac{E_2}{E_0}$, η, δ, and $\Delta H_{mix}$ are the four most important features (Fig. 5C). $\frac{E_2}{E_0}$ and δ both indicate the atomic size difference and the internal strain. As shown in Fig. 6-L1 and L3, $HEA_{Laves}$ have higher atomic size mismatches than the $HEA_{Non\text{-}Laves}$. The severe lattice distortion favors the ordered IM formation. From the thermodynamic aspects, the inset box plot in Fig. 6-L2 shows that $HEA_{Laves}$ all cluster at a region with low η absolute values while $HEA_{Non\text{-}Laves}$ has wide η distribution. Besides, $HEA_{Laves}$ also show more negative $\Delta H_{mix}$ values than $HEA_{Non\text{-}Laves}$. The η and $\Delta H_{mix}$ distribution trends of $HEA_{Laves}$ favors the IM formation.

Fig. 5D shows that multiple features have weak impacts on Sigma formation. However, when these features are combined using FE, a low classification error of 0.05 is attained, illustrating the efficacy of the FE methodology used herein. The important roles of these features can be seen primarily in η, VEC, Δχ, and $\Delta H_{mix}$ as examples. The inset of Fig. 6-S1 shows that $HEA_{Sigma}$ cluster at a region with low η absolute values, indicating a higher IM formation tendency. Similarly, $HEA_{Sigma}$ shows more negative $\Delta H_{mix}$ values that favors IM formation. The influence of VEC, and Δχ, both electron-related features, is shown in Fig. 6-S2 and S3. It is previously reported that the formation of the topological close-packed Sigma phase formation is favored when $\Delta\chi > 0.133$[58] and $6.88 < VEC < 7.84$[50]. The current work obtains similar results based on a larger database. The first, second (i.e., median), and third quartiles of VEC distribution are 7.36, 7.61, and 7.86 (7.36 < VEC < 7.86 is the region for the middle 50 % of the distribution). This new Sigma-prone VEC region overlaps the FCC-prone VEC region[57]. A further review of the database also shows that ~80 % of $HEA_{Sigma}$ contain FCC phase. Finally, the larger $\Delta\chi$ values of $HEA_{Sigma}$ provide clear separation from the $HEA_{Non\text{-}Sigma}$. Therefore, one should consider decreasing the electronegativity discrepancy of the constituent elements to avoid Sigma formation during HEA design. The current work identifies the electron configuration as the most important HR factor in controlling Sigma formation.

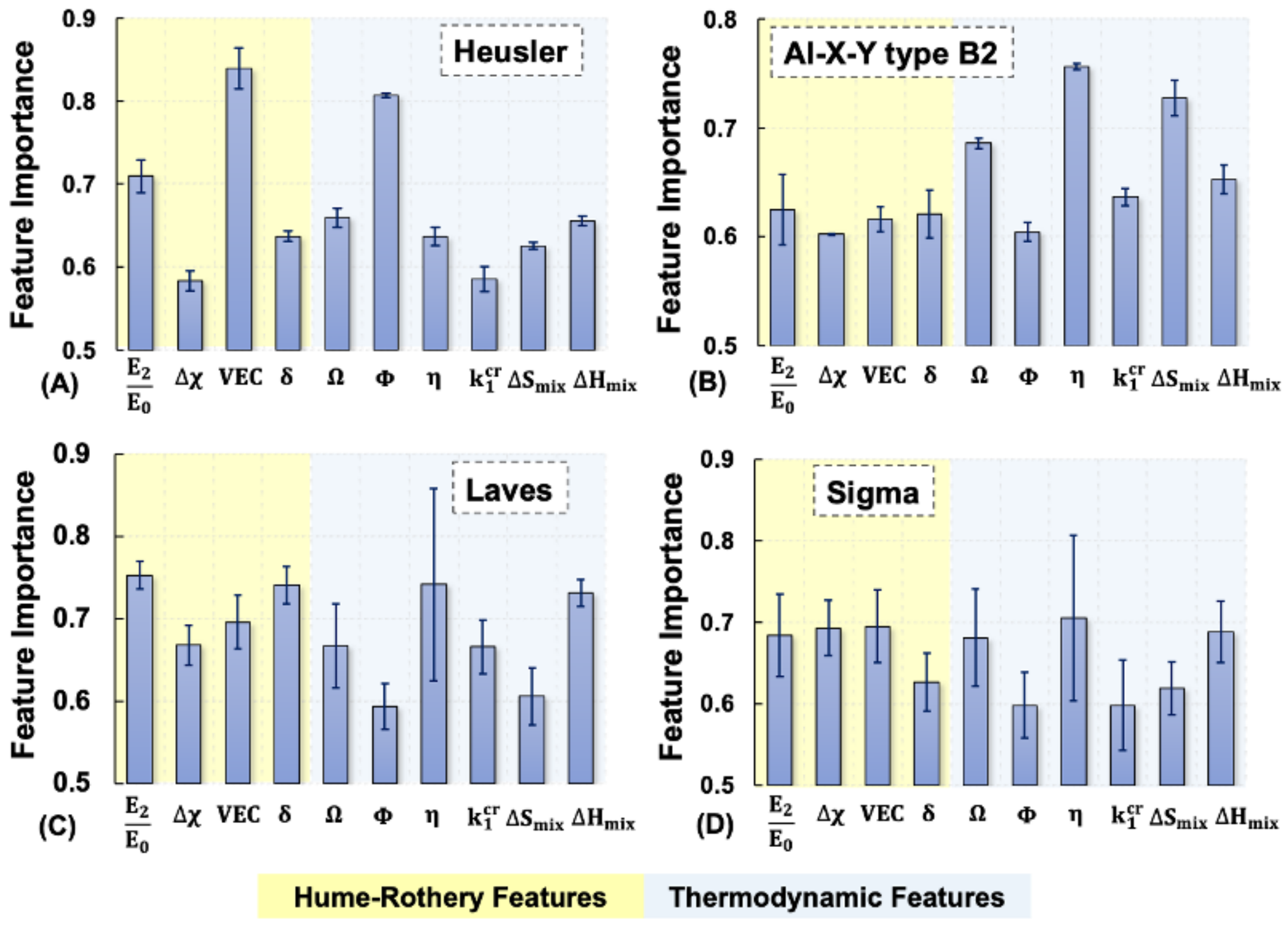

Figure 5. Feature importance in determining different phases' formation. Figures A-D are plotted for Heusler, Al-X-Y type B2, Laves, and Sigma phases. Yellow and blue backgrounds correspond to HR and Thermo features. Error bars (standard deviation) are shown.

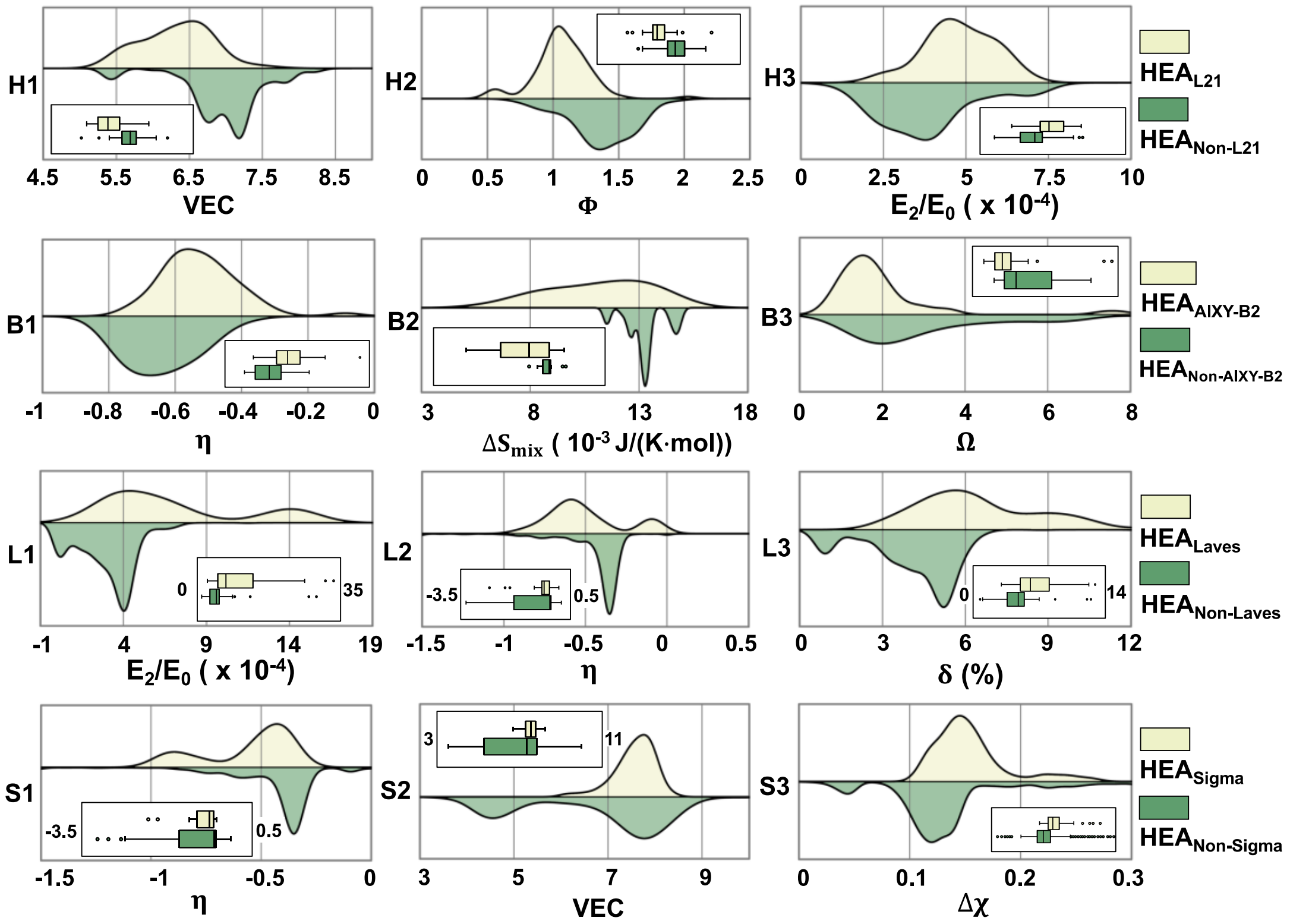

Figure 6: HEA distribution probability density functions based on the values of the three most important top-ranked features. The horizontal axis represents feature values. The vertical axis represents distribution probability density. Insets are box plots showing the relative positions of the two categories' distribution. The upper and lower bounds of box plots are labeled if different from the main plots. Figures H1-H3 show the HEA distribution based on VEC, Φ, and $\frac{E_2}{E_0}$ values in the Heusler+ prediction model. Figures B1-3 show the HEA distribution based on η, $\Delta S_{mix}$, and Ω values in the Al-X-Y type B2+ prediction model. Figures L1-L3 show the HEA distribution based on $\frac{E_2}{E_0}$, η, and δ values in the Laves+ prediction model. Figures S1-3 show the HEA distribution based on η, VEC, and Δχ values in the Sigma+ prediction model.

**Part 4: Experimental Validation**

Experimental validation is important to provide an unbiased evaluation of a ML model trained on available databases. As such, the palette of elements for the validation alloys should be an unbiased representation of the compositional space where the model is trained. Accordingly, of the 86 validation alloys, the multi-phase prediction model will have 60 alloys (Table 3A) with randomly chosen compositions based on the common element in the training database, located both inside and outside the feature space covered by the current database. The distributions of validation HEAs in each predicted phase category are proportional to the database phase distribution. 50 alloys are predicted correctly, giving a validation accuracy of 83 %. Since the Laves+, Sigma+, and multi-phase prediction models are trained on the same database, the same 60 HEAs also validate the Laves+ and Sigma+ models, with validation accuracies of 92 % and 95 %, respectively (Table 3A). To validate the Al-X-Y B2+ prediction model, another 14 new HEAs containing the Al-X-Y type B2 phase essential elements are randomly chosen that involves two or more refractory elements (Table 3B). For Al-X-Y type B2 formation, 12 out of the 14 HEAs are predicted correctly with an accuracy of 86 %. For a similar consideration, another 12 HEAs (Table 3C) containing the Heusler phase essential elements were synthesized to validate the Heusler+ prediction model. For the Heusler phase formation, 11 out of the 12 HEAs are predicted correctly, with an accuracy of 92 %. Overall, the validation accuracies essentially match the classification accuracies. All the XRD patterns can be found in the Section 5 of Supplementary Materials.

Table 3: Validation HEAs compositions, phase prediction results, and experimental phase characterization results obtained from XRD are listed. The number subscripts in compositions are elemental atomic percentages. Detailed phase contents are listed in the experimental results column. True or False in Laves+, Sigma+, Al-X-Y B2+, and Heusler+ prediction columns represent forming or not forming the corresponding IM phases, respectively. Abbreviations AlNi B2+, A1, A2, Mix A1-A2, Al-X-Y B2+, and $L2_1$ represent the AlNi type B2 forming with other solid solution phases, disordered FCC_A1 phase, BCC_A2 phase, mixed A1-A2 phase (coexistence of multiple A1 or A2, or mixture of A1 and A2), Al-X-Y type B2+ forming with other phases, and Heusler phase. The incorrect predictions are underlined and bolded.

A. Multi-phase, Laves+, and Sigma+ prediction models validation HEAs

| Composition | Multi-phase Prediction | Laves+ Prediction | Sigma+ Prediction | Experimental results |
|---|---|---|---|---|
| $Ag_{20}Al_{20}Cr_{20}Mn_{20}Ni_{20}$ | AlNi B2+ | False | False | B2+A1 |
| $Ag_5Al_{38}Cr_{19}Mn_{19}Ni_{19}$ | AlNi B2+ | False | False | B2+A1 |
| $Al_5Co_{20}Cr_{10}Fe_{40}Ni_{20}Ti_5$ | **<u>AlNi B2+</u>** | **<u>True</u>** | False | A1 |
| $Al_{10}Co_{20}Cu_{20}Fe_{20}Ni_{20}V_{10}$ | AlNi B2+ | False | False | B2+A1 |
| $Al_{11}Co_{22}Cr_{11}Cu_{11}Ni_{33}V_{12}$ | AlNi B2+ | False | **<u>True</u>** | B2+A1 |
| $Al_{15}Cr_{15}Mo_{15}Ni_{46}W_9$ | AlNi B2+ | False | False | B2+A1+A2 |
| $Al_{15}Cr_{31}Fe_{31}Mn_{15}Ni_8$ | AlNi B2+ | False | False | B2 |
| $Al_{16}Co_{20}Fe_{20}Mn_{18}Ni_{20}V_6$ | AlNi B2+ | False | False | B2 |
| $Al_{16}Co_{21}Cr_{21}Fe_{21}Ni_{21}$ | AlNi B2+ | False | False | B2+A1 |
| $Al_{16}Cr_{16}Fe_{16}Mn_{16}Ni_{31}V_5$ | AlNi B2+ | False | False | B2 |
| $Al_{19}Cr_{19}Cu_{19}Fe_{19}Ni_{19}Si_5$ | AlNi B2+ | False | False | B2+A1+A2 |
| $Al_{20}Co_{20}Cr_{20}Fe_{20}Mn_{20}$ | AlNi B2+ | False | False | B2 |
| $Al_{21}Co_{11}Cr_{21}Cu_5Fe_{21}Mn_{21}$ | AlNi B2+ | False | False | B2 |
| $Al_{22}Co_{26}Fe_{26}Ni_{26}$ | **<u>AlNi B2+</u>** | False | False | A2 |
| $Al_{23}Co_{23}Cu_{23}Fe_{23}V_8$ | AlNi B2+ | False | False | B2+A1 |
| $Al_{23}Cu_{23}Fe_{23}Ni_{23}V_8$ | AlNi B2+ | False | False | B2+A1 |
| $Al_{24}Co_{24}Cu_{23}Ni_{23}Ti_6$ | AlNi B2+ | False | False | B2+A1 |
| $Al_{25}Co_{25}Cr_{25}Fe_{25}$ | AlNi B2+ | False | False | B2 |
| $Al_{25}Cu_{25}Fe_{25}Ni_{25}$ | AlNi B2+ | False | False | B2+A1 |
| $Al_{29}Co_{29}Cu_{13}Fe_{29}$ | AlNi B2+ | False | False | B2+A1 |
| $Al_{33}Co_{17}Nb_{33}Ni_{17}$ | **AlNi B2+** | **False** | False | B2+Laves |
| $Co_7Ta_{31}Ti_{31}V_{31}$ | BCC | False | False | A2 |
| $Cr_6Ti_{56}V_{19}Zr_{19}$ | BCC | False | False | A2 |
| $Cr_{25}Mo_{25}Ti_{25}V_{25}$ | BCC | False | False | A2 |
| $Cr_{33}Mo_{22}Nb_{12}V_{33}$ | BCC | False | False | A2 |
| $Hf_{25}Nb_{25}Ta_{25}Zr_{25}$ | BCC | False | False | A2 |
| $Hf_{30}Nb_{30}Ti_{30}V_{10}$ | BCC | False | False | A2 |
| $Hf_{30}Ta_{30}Ti_{30}V_{10}$ | BCC | False | False | A2 |
| $Mo_{29}Nb_{13}Ti_{29}V_{29}$ | BCC | False | False | A2 |
| $Nb_{22}Ta_{22}Ti_{22}V_{22}Zr_{12}$ | BCC | False | False | A2 |
| $Nb_{29}Ta_{29}Ti_{29}Zr_{13}$ | BCC | False | False | A2 |
| $Co_{15}Cr_{15}Fe_{15}Mn_{15}Ni_{32}V_8$ | FCC | False | **True** | A1 |
| $Co_{18}Cu_{18}Fe_{18}Mn_{18}Ni_{18}V_{10}$ | **FCC** | False | False | A1+A1 |
| $Co_{19}Cr_{29}Fe_{29}Ni_{19}Si_4$ | FCC | False | False | A1 |
| $Co_{21}Cr_{11}Fe_{42}Ni_{21}Ti_5$ | FCC | **True** | False | A1 |
| $Co_{22}Fe_{22}Mn_{12}Ni_{44}$ | FCC | False | False | A1 |
| $Co_{24}Cr_{24}Fe_{24}Ni_{24}Si_4$ | FCC | False | False | A1 |
| $Co_{24}Fe_{24}Ni_{47}V_5$ | FCC | False | False | A1 |
| $Co_{25}Cr_8Cu_5Fe_{25}Ni_{25}V_{12}$ | FCC | False | False | A1 |
| $Cr_{19}Cu_{19}Fe_{19}Mn_{18}Ni_{19}Ti_6$ | **FCC** | **True** | False | A1+A2 |
| $Al_4Cr_{32}Cu_{32}Fe_{11}Mn_{21}$ | MIX A1-A2 | False | False | A1+A2 |
| $Al_8Cr_{56}Fe_{14}Mn_{22}$ | **MIX A1-A2** | False | False | A2 |
| $Al_{10}Co_{20}Cr_{10}Cu_{20}Mn_{20}Ni_{20}$ | MIX A1-A2 | False | False | A1+A2 |
| $Al_{24}Co_{24}Cr_{23}Fe_{23}Ti_6$ | **MIX A1-A2** | False | False | B2 |
| $Co_{16}Cr_{16}Cu_{16}Fe_{16}Mn_{14}Ni_{16}Ti_6$ | MIX A1-A2 | False | False | A1+A1 |
| $Co_{25}Cr_{25}Cu_{25}Fe_{25}$ | **MIX A1-A2** | False | False | A1+A1+Unknown |
| $Cr_{25}Cu_{25}Fe_{25}Mn_{25}$ | MIX A1-A2 | False | False | A1+A2 |
| $Cr_{40}Fe_{40}Mn_{10}Ni_{10}$ | **MIX A1-A2** | False | False | A2 |
| $Co_{20}Fe_{20}Mn_{20}Ni_{20}Ti_{10}V_{10}$ | Laves+ | True | False | Laves+A2 |
| $Co_{20}Fe_{20}Mo_{20}Ni_{20}Ti_{20}$ | Laves+ | True | False | Laves+A1+A2 |
| $Co_{21}Cr_{21}Cu_{21}Mn_{16}Ti_{21}$ | Laves+ | True | False | Laves+A1 |
| $Co_{25}Cr_{25}Fe_{25}Nb_{13}Ti_{12}$ | Laves+ | True | False | Laves+A1+A2 |
| $Cr_{20}Nb_{20}Ni_{20}Ti_{20}Zr_{20}$ | Laves+ | True | False | Laves+A2 |
| $Cr_{40}Fe_{20}Ni_{20}Ti_{20}$ | Laves+ | True | False | Laves+A1+A2 |
| $Cu_{17}Fe_{17}Mn_{17}Ni_{17}Ti_{32}$ | Laves+ | True | False | Laves+A1+A2 |
| $Co_{15}Cr_{15}Cu_8Fe_{15}Ni_{31}Ti_8V_8$ | **Sigma+** | **True** | **True** | A1 |
| $Co_{18}Cr_{18}Fe_{18}Mo_{18}Ni_{18}V_{10}$ | Sigma+ | False | True | Sigma+A1 |
| $Co_{20}Cr_{20}Fe_{20}Mo_{20}V_{20}$ | Sigma+ | False | True | Sigma+A2 |
| $Co_{26}Cr_{26}Fe_{26}Mo_{22}$ | Sigma+ | False | True | Sigma+A2 |
| $Cu_{20}Fe_{20}Mn_{20}Ni_{20}V_{20}$ | Sigma+ | False | True | Sigma+A1 |

B. Al-X-Y type B2+ prediction model validation HEAs

| Composition | Al-X-Y B2 + prediction | Experimental results | | Composition | Al-X-Y B2 + prediction | Experimental results |
|---|---|---|---|---|---|---|
| $Al_{10}Hf_{20}Nb_{22}Ti_{33}V_{15}$ | True | B2 | | $Al_{30}Nb_{20}Ta_{15}Ti_{20}V_{10}Zr_5$ | True | B2 |
| $Al_{15}Hf_{25}Nb_{32}Ti_{28}$ | True | B2 | | $Al_{30}Nb_{20}Ta_{20}Ti_{20}Zr_{10}$ | True | B2+Unknown |
| $Al_{20}Hf_{24}Nb_{29}Ti_{27}$ | True | B2 | | $Al_4Hf_6Nb_{42}Ti_{18}V_{24}W_6$ | False | A2 |
| $Al_{23}Hf_{23}Nb_{23}Ti_{23}V_8$ | True | B2 | | $Al_8Cr_{15}Mo_{15}Nb_{15}Ti_{15}V_{32}$ | False | A2 |
| $Al_{23}Hf_{23}Ta_{23}Ti_{23}V_8$ | True | B2 | | $Al_{10}Hf_{18}Nb_{18}Ta_{18}Ti_{18}Zr_{18}$ | False | A2 |
| $Al_{26}Mo_{21}Nb_{11}Ti_{21}V_{21}$ | **True** | A2 | | $Al_{32}Nb_{17}Ta_{17}Ti_{17}V_{17}$ | **False** | B2 |
| $Al_{30}Mo_{20}Nb_{20}Ti_{30}$ | True | B2 | | $Al_6Nb_{21}Ta_{21}Ti_{21}V_{21}Zr_{10}$ | False | A2 |

C. Heusler+ prediction model validation HEAs

| Composition | Heusler+ prediction | Experimental results | | Composition | Heusler+ prediction | Experimental results |
|---|---|---|---|---|---|---|
| $Al_{10}Co_{25}Fe_{25}Mn_{25}Ti_{15}$ | **True** | A2+Unknown | | $Al_{25}Cr_{10}Fe_{20}Mn_{10}Ni_{20}Ti_{15}$ | True | $L2_1$ |
| $Al_{10}Cr_5Fe_{45}Mn_{12}Ni_{20}Ti_8$ | True | $L2_1$+A1 | | $Al_{10}Co_{20}Mn_{20}Ni_{30}Ti_{10}$ | False | A1+A2 |
| $Al_{12}Co_{28}Fe_{19}Ni_{29}Ti_{12}$ | True | $L2_1$+A1 | | $Al_{10}Co_{30}Fe_{20}Ni_{32}Ti_8$ | False | A1 |
| $Al_{14}Cr_4Fe_{17}Mn_4Mo_1Ni_{44}Ti_{16}$ | True | $L2_1$+A1 | | $Al_{15}Co_{30}Fe_{30}Ni_{10}Ti_{15}$ | False | B2 |
| $Al_{15}Cr_{10}Fe_{30}Ni_{30}Ti_{15}$ | True | $L2_1$+A2 | | $Al_{20}Fe_{10}Mn_{30}Ni_{40}$ | False | B2+A1 |
| $Al_{15}Fe_{40}Mn_{20}Ni_{10}Ti_{10}$ | True | $L2_1$ | | $Al_7Co_{30}Fe_{30}Mn_{25}Ti_8$ | False | B2+A1 |

## 5. Summary

This work demonstrates a machine learning methodology assisted by feature engineering (FE) in predicting the common high entropy alloy (HEA) phases.

1) The multiphase prediction model, utilizing six engineered features in conjunction with the Random Forest algorithm, which is determined to exhibit the lowest prediction error amongst various algorithms, currently stands out as one of the top-performing methods and precisely predicts seven distinct phase categories.
2) Mixed-phase compositions are further evaluated by four other models trained to predict the formation of four commonly occurring intermetallic phases that include Sigma, Laves, Heusler, and Al-X-Y type B2 phases with high accuracies. These models, with high degree of accuracies, incorporate the use of four engineered features and the Support Vector Machine algorithm which is the optimal performing algorithm for these scenarios.
3) The models are experimentally validated with 86 new compositions. The experimental accuracy aligns with the model accuracy, further attesting to their reliability.
4) We identify the most relevant thermodynamic (Thermo) and Hume-Rothery rule (HR) features that control the formation of the four intermetallic phases. Thermo feature $\Phi$, and the valence electron and atomic size discrepancies in HR features have an impact on the Heusler phase formation. Al-X-Y type B2 phase formation is mainly determined by the Thermo features, implying that the ordering transformation from BCC to B2 is a thermodynamic process with limited influence from HR features. Laves phase is determined by the Thermo feature $\eta$ and the atomic size discrepancy in HR features, while the Sigma phase is mainly influenced by Thermo feature η and the electronic effect encoded in the HR features.

We have developed feature variants-based models that can enhance the phase classification and prediction accuracies while also providing insight into the physics behind these predictions. The creation of the machine learning toolset is the practical value of the present study. The scientific significance is the discovery of links between scientific parameters and phase formation inside the ML black box. Thus, the machine learning method in this work can be further developed to explore other material phases. Currently, the ML Heusler and Al-X-Y type B2 phase prediction models are trained to predict the HEAs with corresponding IM formation elements. Active learning can be employed to explore novel elemental combinations. Additionally, a comprehensive HEA design model can be constructed with the help of the properties prediction models to automatically search for compositions that fulfill specific phase and property requirements.

## 6. Acknowledgment

The authors thank Dr. Andrew M. Cheung for his critical reading of the manuscript. This work is supported by the Office of Naval Research grant N00014-18-1-2621.

## 7. Author contributions

SJP supervised the project. JQ performed the simulation and materials synthesis. DIH assisted with the materials synthesis. JQ and SJP co-wrote the manuscript. All authors discussed the research and edited the manuscript.

## 8. Data availability

The data supporting this work are available from the authors upon reasonable request. Part of the data used in this study is available within the Supplementary Materials.

## 9. Competing Interests

The authors declare no competing interests.

## 10. Reference

[1] O.N. Senkov, J.D. Miller, D.B. Miracle, C. Woodward, Accelerated exploration of multi-principal element alloys for structural applications, Calphad Comput. Coupling Phase Diagrams Thermochem. 50 (2015) 32–48. https://doi.org/10.1016/j.calphad.2015.04.009.

[2] K. Guruvidyathri, K.C.H. Kumar, J.W. Yeh, B.S. Murty, Topologically Close-packed Phase Formation in High Entropy Alloys: A Review of Calphad and Experimental Results, (n.d.). https://doi.org/10.1007/s11837-017-2566-5.

[3] S.J. Poon, J. Qi, A.M. Cheung, Harnessing the Complex Compositional Space of High-Entropy Alloys, in: Jamieson Brechtl, Peter K. Liaw (Eds.), High-Entropy Mater. Theory, Exp. Appl., Springer, Cham, Cham, 2021: pp. 63–113. https://doi.org/10.1007/978-3-030-77641-1_3.

[4] C. Wen, Y. Zhang, C. Wang, D. Xue, Y. Bai, S. Antonov, L. Dai, T. Lookman, Y. Su, Machine learning assisted design of high entropy alloys with desired property, Acta Mater. 170 (2019) 109–117. https://doi.org/10.1016/j.actamat.2019.03.010.

[5] Y.-J.J. Chang, C.-Y.Y. Jui, W.-J.J. Lee, A.-C.C. Yeh, Prediction of the Composition and Hardness of High-Entropy Alloys by Machine Learning, Jom. 71 (2019) 3433–3442. https://doi.org/10.1007/s11837-019-03704-4.

[6] J. Xiong, S.Q. Shi, T.Y. Zhang, Machine learning of phases and mechanical properties in complex concentrated alloys, J. Mater. Sci. Technol. 87 (2021) 133–142. https://doi.org/10.1016/j.jmst.2021.01.054.

[7] D. Dai, T. Xu, X. Wei, G. Ding, Y. Xu, J. Zhang, H. Zhang, Using machine learning and feature engineering to characterize limited material datasets of high-entropy alloys, Comput. Mater. Sci. 175 (2020). https://doi.org/10.1016/j.commatsci.2020.109618.

[8] Y.V. Krishna, U.K. Jaiswal, R.M. R, Machine learning approach to predict new multiphase high entropy alloys, Scr. Mater. 197 (2021) 113804. https://doi.org/10.1016/J.SCRIPTAMAT.2021.113804.

[9] S. Risal, W. Zhu, P. Guillen, L. Sun, Improving phase prediction accuracy for high entropy alloys with Machine learning, Comput. Mater. Sci. 192 (2021) 110389. https://doi.org/10.1016/J.COMMATSCI.2021.110389.

[10] A. Roy, T. Babuska, B. Krick, G. Balasubramanian, Machine learned feature identification for predicting phase and Young's modulus of low-, medium- and high-entropy alloys, Scr. Mater. 185 (2020) 152–158. https://doi.org/10.1016/J.SCRIPTAMAT.2020.04.016.

[11] U.K. Jaiswal, Y. Vamsi Krishna, M.R. Rahul, G. Phanikumar, Machine learning-enabled identification of new medium to high entropy alloys with solid solution phases, Comput. Mater. Sci. 197 (2021) 110623. https://doi.org/10.1016/J.COMMATSCI.2021.110623.

[12] K. Lee, M. V. Ayyasamy, P. Delsa, T.Q. Hartnett, P. V. Balachandran, Phase classification of multi-principal element alloys via interpretable machine learning, Npj Comput. Mater. 8 (2022) 1–12. https://doi.org/10.1038/s41524-022-00704-y.

[13] F. Tancret, I. Toda-Caraballo, E. Menou, P.E.J. Rivera Díaz-Del-Castillo, Designing high entropy alloys employing thermodynamics and Gaussian process statistical analysis, Mater. Des. 115 (2017) 486–497. https://doi.org/10.1016/j.matdes.2016.11.049.

[14] D. Beniwal, P.K. Ray, FCC vs. BCC phase selection in high-entropy alloys via simplified and interpretable reduction of machine learning models, Materialia. 26 (2022) 101632. https://doi.org/10.1016/J.MTLA.2022.101632.

[15] T. Jin, I. Park, T. Park, J. Park, J.H. Shim, Accelerated crystal structure prediction of multi-elements random alloy using expandable features, Sci. Reports |. 11 (2021) 5194. https://doi.org/10.1038/s41598-021-84544-8.

[16] D. Beniwal, P.K. Ray, Learning phase selection and assemblages in High-Entropy Alloys through a stochastic ensemble-averaging model, Comput. Mater. Sci. 197 (2021) 110647. https://doi.org/10.1016/J.COMMATSCI.2021.110647.

[17] N. Islam, W. Huang, H.L. Zhuang, Machine learning for phase selection in multi-

principal element alloys, Comput. Mater. Sci. 150 (2018) 230–235. https://doi.org/10.1016/J.COMMATSCI.2018.04.003.
[18] Z. Pei, J. Yin, J.A. Hawk, D.E. Alman, M.C. Gao, Machine-learning informed prediction of high-entropy solid solution formation: Beyond the Hume-Rothery rules, Npj Comput. Mater. 6 (2020) 1–8. https://doi.org/10.1038/s41524-020-0308-7.
[19] Z. Zhou, Y. Zhou, Q. He, Z. Ding, F. Li, Y. Yang, Machine learning guided appraisal and exploration of phase design for high entropy alloys, Npj Comput. Mater. 5 (2019) 1–9. https://doi.org/10.1038/s41524-019-0265-1.
[20] A. Agarwal, A.K. Prasada Rao, Artificial Intelligence Predicts Body-Centered-Cubic and Face-Centered-Cubic Phases in High-Entropy Alloys, Jom. 71 (2019) 3424–3432. https://doi.org/10.1007/s11837-019-03712-4.
[21] J. Qi, A.M. Cheung, S.J. Poon, High Entropy Alloys Mined From Binary Phase Diagrams, Sci. Rep. 9 (2019) 1–10. https://doi.org/10.1038/s41598-019-50015-4.
[22] Y. Li, W. Guo, Machine-learning model for predicting phase formations of high-entropy alloys, Phys. Rev. Mater. 3 (2019) 95005. https://doi.org/10.1103/PhysRevMaterials.3.095005.
[23] W. Huang, P. Martin, H.L. Zhuang, Machine-learning phase prediction of high-entropy alloys, Acta Mater. 169 (2019) 225–236. https://doi.org/10.1016/j.actamat.2019.03.012.
[24] M.H. Tsai, R.C. Tsai, T. Chang, W.F. Huang, Intermetallic Phases in High-Entropy Alloys: Statistical Analysis of their Prevalence and Structural Inheritance, Met. 2019, Vol. 9, Page 247. 9 (2019) 247. https://doi.org/10.3390/MET9020247.
[25] T. Graf, C. Felser, S.S.P. Parkin, Simple rules for the understanding of Heusler compounds, Prog. Solid State Chem. 39 (2011) 1–50. https://doi.org/10.1016/j.progsolidstchem.2011.02.001.
[26] S. Wolff-Goodrich, A. Marshal, K.G. Pradeep, G. Dehm, J.M. Schneider, C.H. Liebscher, Combinatorial exploration of B2/L21 precipitation strengthened AlCrFeNiTi compositionally complex alloys, J. Alloys Compd. 853 (2021) 156111. https://doi.org/10.1016/J.JALLCOM.2020.156111.
[27] S. Inman, J. HAN, A. Gerard, J. Qi, M. Wischhusen, S. Agnew, S. Poon, K. Ogle, J. Scully, Effect of Mn Content on the Passivation and Corrosion of Al0.3Cr0.5Fe2MnxMo0.15Ni1.5Ti0.3 FCC Compositionally Complex Alloys, Corrosion. (2021). https://doi.org/10.5006/3906.
[28] R. Feng, C. Zhang, M.C. Gao, Z. Pei, F. Zhang, Y. Chen, D. Ma, K. An, J.D. Poplawsky, L. Ouyang, Y. Ren, J.A. Hawk, M. Widom, P.K. Liaw, High-throughput design of high-performance lightweight high-entropy alloys, Nat. Commun. 2021 121. 12 (2021) 1–10. https://doi.org/10.1038/s41467-021-24523-9.
[29] P. Shi, W. Ren, T. Zheng, Z. Ren, X. Hou, J. Peng, P. Hu, Y. Gao, Y. Zhong, P.K. Liaw, Enhanced strength–ductility synergy in ultrafine-grained eutectic high-entropy alloys by inheriting microstructural lamellae, Nat. Commun. 10 (2019) 1–8. https://doi.org/10.1038/s41467-019-08460-2.
[30] R. Feng, Y. Rao, C. Liu, X. Xie, D. Yu, Y. Chen, M. Ghazisaeidi, T. Ungar, H. Wang, K. An, P.K. Liaw, Enhancing fatigue life by ductile-transformable multicomponent B2 precipitates in a high-entropy alloy, Nat. Commun. 2021 121. 12 (2021) 1–10.

https://doi.org/10.1038/s41467-021-23689-6.
[31] R.S. Polvani, W.S. Tzeng, P.R. Strutt, High temperature creep in a semi-coherent NiAl-Ni2AlTi alloy, Metall. Trans. A. 7 (1976) 33–40. https://doi.org/10.1007/BF02644036.
[32] P.R. Strutt, R.S. Polvani, J.C. Ingram, Creep behavior of the heusler type structure alloy Ni2AlTi, Metall. Trans. A. 7 (1976) 23–31. https://doi.org/10.1007/BF02644035.
[33] C. Li, Y. Ma, J. Hao, Y. Yan, Q. Wang, C. Dong, P.K. Liaw, Microstructures and mechanical properties of body-centered-cubic (Al,Ti)0.7(Ni,Co,Fe,Cr)5 high entropy alloys with coherent B2/L21 nanoprecipitation, Mater. Sci. Eng. A. 737 (2018) 286–296. https://doi.org/10.1016/j.msea.2018.09.060.
[34] Y. Qi, Y. Wu, T. Cao, L. He, F. Jiang, J. sun, L21-strengthened face-centered cubic high-entropy alloy with high strength and ductility, Mater. Sci. Eng. A. 797 (2020) 140056. https://doi.org/10.1016/J.MSEA.2020.140056.
[35] D.B. Miracle, M.H. Tsai, O.N. Senkov, V. Soni, R. Banerjee, Refractory high entropy superalloys (RSAs), Scr. Mater. 187 (2020) 445–452. https://doi.org/10.1016/J.SCRIPTAMAT.2020.06.048.
[36] G. Qin, Z. Li, R. Chen, H. Zheng, C. Fan, L. Wang, Y. Su, H. Ding, J. Guo, H. Fu, CoCrFeMnNi high-entropy alloys reinforced with Laves phase by adding Nb and Ti elements, J. Mater. Res. 34 (2019) 1011–1020. https://doi.org/10.1557/JMR.2018.468.
[37] V.K. Soni, S. Sanyal, K.R. Rao, S.K. Sinha, A review on phase prediction in high entropy alloys, Proc. Inst. Mech. Eng. Part C J. Mech. Eng. Sci. 235 (2021) 6268–6286. https://doi.org/10.1177/09544062211008935.
[38] S.R. Xie, Y. Quan, A.C. Hire, B. Deng, J.M. DeStefano, I. Salinas, U.S. Shah, L. Fanfarillo, J. Lim, J. Kim, G.R. Stewart, J.J. Hamlin, P.J. Hirschfeld, R.G. Hennig, Machine learning of superconducting critical temperature from Eliashberg theory, Npj Comput. Mater. 2022 81. 8 (2022) 1–8. https://doi.org/10.1038/s41524-021-00666-7.
[39] R. Ouyang, S. Curtarolo, E. Ahmetcik, M. Scheffler, L.M. Ghiringhelli, SISSO: A compressed-sensing method for identifying the best low-dimensional descriptor in an immensity of offered candidates, Phys. Rev. Mater. 2 (2018) 083802. https://doi.org/10.1103/PHYSREVMATERIALS.2.083802/FIGURES/5/MEDIUM.
[40] J.W. Yeh, S.K. Chen, S.J. Lin, J.Y. Gan, T.S. Chin, T.T. Shun, C.H. Tsau, S.Y. Chang, Nanostructured high-entropy alloys with multiple principal elements: Novel alloy design concepts and outcomes, Adv. Eng. Mater. 6 (2004) 299-303+274. https://doi.org/10.1002/adem.200300567.
[41] Y. Zhang, Y.J. Zhou, J.P. Lin, G.L. Chen, P.K. Liaw, Solid-solution phase formation rules for multi-component alloys, Adv. Eng. Mater. 10 (2008) 534–538. https://doi.org/10.1002/adem.200700240.
[42] Y. Zhang, Z.P. Lu, S.G. Ma, P.K. Liaw, Z. Tang, Y.Q. Cheng, M.C. Gao, Guidelines in predicting phase formation of high-entropy alloys, MRS Commun. 4 (2014) 57–62. https://doi.org/10.1557/mrc.2014.11.
[43] D.J.M. King, S.C. Middleburgh, A.G. McGregor, M.B. Cortie, Predicting the formation and stability of single phase high-entropy alloys, Acta Mater. 104 (2016) 172–179. https://doi.org/10.1016/j.actamat.2015.11.040.
[44] M.C. Troparevsky, J.R. Morris, P.R.C. Kent, A.R. Lupini, G.M. Stocks, Criteria for

predicting the formation of single-phase high-entropy alloys, Phys. Rev. X. 5 (2015) 1–6. https://doi.org/10.1103/PhysRevX.5.011041.
[45] O.N. Senkov, D.B. Miracle, A new thermodynamic parameter to predict formation of solid solution or intermetallic phases in high entropy alloys, J. Alloys Compd. 658 (2016) 603–607. https://doi.org/10.1016/j.jallcom.2015.10.279.
[46] Z. Wang, W. Qiu, Y. Yang, C.T. Liu, Atomic-size and lattice-distortion effects in newly developed high-entropy alloys with multiple principal elements, Intermetallics. 64 (2015) 63–69. https://doi.org/10.1016/j.intermet.2015.04.014.
[47] S. Fang, X. Xiao, L. Xia, W. Li, Y. Dong, Relationship between the widths of supercooled liquid regions and bond parameters of Mg-based bulk metallic glasses, J. Non. Cryst. Solids. 321 (2003) 120–125. https://doi.org/10.1016/S0022-3093(03)00155-8.
[48] S. Guo, C. Ng, J. Lu, C.T. Liu, Effect of valence electron concentration on stability of fcc or bcc phase in high entropy alloys, J. Appl. Phys. 109 (2011). https://doi.org/10.1063/1.3587228.
[49] S. Yang, J. Lu, F. Xing, L. Zhang, Y. Zhong, Revisit the VEC rule in high entropy alloys (HEAs) with high-throughput CALPHAD approach and its applications for material design-A case study with Al-Co-Cr-Fe-Ni system, Acta Mater. 192 (2020) 11–19. https://doi.org/10.1016/j.actamat.2020.03.039.
[50] M.-H.H. Tsai, K.-Y.Y. Tsai, C.-W.W. Tsai, C. Lee, C.-C.C. Juan, J.-W.W. Yeh, Criterion for sigma phase formation in Cr- and V-Containing high-entropy alloys, Mater. Res. Lett. 1 (2013) 207–212. https://doi.org/10.1080/21663831.2013.831382.
[51] P. McCullagh, J.A. Nelder, Generalized Linear Models, Chapman & Hall, New York, 1990.
[52] S.S.Y. Lee, S. Byeon, H.S. Kim, H. Jin, S.S.Y. Lee, Deep learning-based phase prediction of high-entropy alloys: Optimization, generation, and explanation, Mater. Des. 197 (2021) 109260. https://doi.org/10.1016/J.MATDES.2020.109260.
[53] H. He, Y. Bai, E.A. Garcia, S. Li, ADASYN: Adaptive synthetic sampling approach for imbalanced learning, Proc. Int. Jt. Conf. Neural Networks. (2008) 1322–1328. https://doi.org/10.1109/IJCNN.2008.4633969.
[54] N. V. Chawla, K.W. Bowyer, L.O. Hall, W.P. Kegelmeyer, SMOTE: Synthetic Minority Over-sampling Technique, J. Artif. Intell. Res. 16 (2002) 321–357. https://doi.org/10.1613/JAIR.953.
[55] C. Felser, G.H. Fecher, B. Balke, Spintronics: A Challenge for Materials Science and Solid-State Chemistry, Angew. Chemie Int. Ed. 46 (2007) 668–699. https://doi.org/10.1002/ANIE.200601815.
[56] K. Lee, M. V. Ayyasamy, Y. Ji, P. V. Balachandran, A comparison of explainable artificial intelligence methods in the phase classification of multi-principal element alloys, Sci. Reports 2022 121. 12 (2022) 1–15. https://doi.org/10.1038/s41598-022-15618-4.
[57] M.C. Gao, C. Zhang, P. Gao, F. Zhang, L.Z. Ouyang, M. Widom, J.A. Hawk, Thermodynamics of concentrated solid solution alloys, Curr. Opin. Solid State Mater. Sci. 21 (2017) 238–251. https://doi.org/10.1016/j.cossms.2017.08.001.
[58] Y. Dong, Y. Lu, L. Jiang, T. Wang, T. Li, Effects of electro-negativity on the stability

of topologically close-packed phase in high entropy alloys, Intermetallics. 52 (2014) 105–109. https://doi.org/10.1016/j.intermet.2014.04.001.

[59] L.J. Santodonato, P.K. Liaw, R.R. Unocic, H. Bei, J.R. Morris, Predictive multiphase evolution in Al-containing high-entropy alloys, Nat. Commun. 9 (2018) 1–10. https://doi.org/10.1038/s41467-018-06757-2.

## Supplementary Materials

Jie Qi, Diego Ibarra Hoyos, and S. Joseph Poon, Machine Learning-Based Classification, Interpretation, and Prediction of High-Entropy-Alloy Intermetallic Phases, 2023

**Section 1: Database for HEA containing Heusler phase**

Table.S1. Database for 77 HEA containing Heusler ($L2_1$) phase. HEA element systems, compositions, preparation method, phases and reference are listed in columns. Preparation method abbreviations: Cold roll (CR, thickness reduction is parentheses.), As-cast (AC), Water-quenched (WQ), and Furnace-cooled (FC). Annealing temperatures are included with unit Celsius (C). Annealing time is included with unit minute (min), hour (h), or day (D). Phase structure notation: the Strukturbericht designations are used except the phases FCC, BCC, HCP, Sigma, χ, and η, which correspond to Strukturbericht designations A1, A2, A3, $D8_b$, A12, and $D0_{24}$. Laves phase corresponds to C14, C15, or C36. Phase content with unknown IM structure is labeled as IM.

| System | Composition | Preparation method | Phase | Ref |
|---|---|---|---|---|
| AlBCoCrFeNiTi | $Al_6B_{0.1}Co_{30}Cr_{15}Fe_{13}Ni_{29.9}Ti_6$ | AC + 1165 C / 2 h + CR (-65 %) + 1165 / 2 min + 800 C / 720 h | FCC + $L1_2$ + $L2_1$ | 1 |
| | | AC + 1165 C / 2 h + CR (-65 %) + 1165 / 2 min + 900 C / 720 h | FCC + $L1_2$ + $L2_1$ | 1 |
| AlBCrFeMoNiTiZr | $Al_{12.602}B_{0.024}Cr_{10.059}Fe_{59.85}Mo_{1.853}Ni_{8.911}Ti_{6.557}Zr_{0.14}$ | AC / AC + 1200C / 4h | BCC + $L2_1$ | 2 |
| | $Al_{12.651}B_{0.024}Cr_{10.099}Fe_{62.436}Mo_{1.860}$ $Ni_{8.946}Ti_{3.84}Zr_{0.14}$ | AC / AC + 1200C / 4h | BCC + B2 + $L2_1$ | 2 |
| | $Al_{12.661}B_{0.024}Cr_{10.106}Fe_{62.956}Mo_{1.861}Ni_{8.953}Ti_{3.294}Zr_{0.14}$ | AC / AC + 1200C / 4h | BCC + B2 + $L2_1$ | 2 |
| | $Al_{12.7}B_{0.024}Cr_{10.1}Fe_{64}Mo_{1.9}Ni_{9.0}Ti_{2.2}Zr_{0.14}$ | AC / AC + 1200C / 4h | BCC + B2 + $L2_1$ | 2 |
| | $Al_{12.7}B_{0.024}Cr_{10.1}Fe_{63.5}Mo_{1.863}Ni_{8.96}Ti_{2.75}Zr_{0.14}$ | AC / AC + 1200C / 4h | BCC + B2 + $L2_1$ | 2 |
| AlCoCrCuFeNiTi | $AlCo_{0.5}CrCu_{0.5}FeNi_{1.5}Ti_{0.4}$ | AC | BCC + $L2_1$ | 3 |
| | $Al_{0.3}CoCrCu_{0.3}FeNiTi_{0.2}$ | AC + 1150 C / 1 h + CR (-70 %) + 1150 C / 5 min | FCC + B2 + $L2_1$ | 4 |
| | | AC + 1150 C / 1 h + CR (-70 %) + 1150 C / 5 min + 600 C / 150 h | FCC + B2 + Sigma + $L1_2$ + BCC + $L2_1$ | 4 |
| | | AC + 1150 C / 1 h + CR (-70 %) + 1150 C / 5 min + 800 C / 0.5 h | FCC + B2 + Sigma + $L1_2$ + $L2_1$ | 4 |
| AlCoCrCuMnTi | AlCoCrCuMnTi | AC | BCC + BCC + FCC + $L2_1$ | 5 |
| AlCoCrCuNiTi | $AlCoCrCuNiTiY_{0.5}$ | AC | BCC + FCC + C15 + $L2_1$ | 6 |
| | $AlCoCrCuNiTiY_{0.8}$ | AC | BCC + FCC + $L2_1$ | 6 |
| | AlCoCrCuNiTiY | AC | BCC + FCC + IM + $L2_1$ | 6 |
| AlCoCrFeHfNiTi | $Al_{9.5}Co_{25}Cr_8Fe_{15}Hf_{0.5}Ni_{36}Ti_6$ | AC + 1220 C / 20 h | FCC + $L1_2$ + $L2_1$ | 7 |
| | | AC + 1140 C / 20 h | FCC + $L1_2$ + $L2_1$ | 7 |
| | | AC + 1220 C / 20 h + 900 C / 50 h | FCC + $L1_2$ + $L2_1$ | 7 |
| | | AC + 1220 C / 20 h + 950 C / 100 h | FCC + $L1_2$ + $L2_1$ | 7 |
| AlCoCrFeMnNiTi | $Al_{10}Co_{20}Cr_{10}Fe_{26}Mn_{14}Ni_{10}Ti_{10}$ | AC + 700 C / 24 h / WQ | FCC + χ + $L2_1$ | 8 |
| | | AC + 1000 C / 504 h + WQ | BCC + C14 + $L2_1$ | 23 |
| | $Al_2CrFeMnTi$ | AC | BCC + C14 + $L2_1$ | 22 |
| | $Al_2CrFeMnTi_{0.25}$ | AC | BCC + $L2_1$ | 22 |
| | $Al_3CrFeMnTi_{0.25}$ | AC | BCC + A12 + A7 + $L2_1$ | 22 |
| | $Al_4CrFeMnTi_{0.25}$ | AC | BCC + A12 + A7 + $L2_1$ | 22 |
| | $Al_{15}Cr_5Fe_{50}Mn_{25}Ti_5$ | AC / AC + 1200 C / 30 min | BCC + $L2_1$ | 24 |
| | $Al_{20}Cr_5Fe_{50}Mn_{20}Ti_5$ | AC / AC + 1200 C / 30 min | BCC + $L2_1$ | 24 |
| | $Al_{25}Cr_5Fe_{50}Mn_{15}Ti_5$ | AC / AC + 1200 C / 30 min | BCC + $L2_1$ | 24 |
| | $Al_{30}Cr_5Fe_{40}Mn_{15}Ti_{10}$ | AC / AC + 1200 C / 30 min | BCC + $L2_1$ | 24 |
| | $Al_{30}Cr_5Fe_{45}Mn_{10}Ti_{10}$ | AC / AC + 1200 C / 30 min | BCC + $L2_1$ | 24 |
| | $Al_{30}Cr_5Fe_{50}Mn_{10}Ti_5$ | AC / AC + 1200 C / 30 min | BCC + $L2_1$ | 24 |
| | $Al_{30}Cr_{10}Fe_{35}Mn_{15}Ti_{10}$ | AC / AC + 1200 C / 30 min | BCC + $L2_1$ | 24 |
| | $Al_{35}Cr_5Fe_{40}Mn_{10}Ti_{10}$ | AC / AC + 1200 C / 30 min | BCC + $L2_1$ | 24 |
| AlCrFeNiTi | $AlCrFeNiTi_{0.5}$ | AC + 650 / 850 / 1200 C / 4 h | BCC + B2 + FCC + $L2_1$ | 25 |
| | $Al_5Cr_{32}Fe_{35}Ni_{22}Ti_6$ | AC | BCC + FCC + $L2_1$ | 26 |
| | | AC + 1100 C / 6 h + WQ | BCC + FCC + $L2_1$ | 26 |
| | | AC + 1100 C / 6 h + WQ + 700 C / 100 h | BCC + FCC + Sigma + $L2_1$ | 26 |
| | | AC + 1100 C / 6 h + WQ + 800 C / 100 h | BCC + FCC + Sigma + η + $L2_1$ | 26 |
| | | AC + 1100 C / 6 h + WQ + 900 C / 100 h | BCC + FCC + Sigma + η + $L2_1$ | 26 |
| | $Al_{10}Cr_{15}Fe_{35}Ni_{20}Ti_{20}$ | AC | BCC + FCC + C14 + $L2_1$ | 27 |
| | $Al_{10}Cr_{20}Fe_{35}Ni_{25}Ti_{10}$ | AC | BCC + FCC + $L2_1$ | 27 |
| | $Al_{10}Cr_{30}Fe_{30}Ni_{20}Ti_{10}$ | AC | BCC + FCC + $L2_1$ | 27 |
| | $Al_{15}Cr_5Fe_{30}Ni_{30}Ti_{20}$ | AC + 700 C / 24 h / WQ | C14 + χ + $L2_1$ | |
| | $Al_{15}Cr_{10}Fe_{35}Ni_{25}Ti_{15}$ | AC | BCC + FCC + C14 + $L2_1$ | 27 |
| | $Al_{15}Cr_{20}Fe_{35}Ni_{15}Ti_{15}$ | AC | BCC + C14 + $L2_1$ | 27 |
| | $Al_{16}Cr_{25}Fe_{35}Ni_{20}Ti_4$ | AC / AC+900 C / 100 h / FC | BCC + $L2_1$ | 28 |
| | $Al_{20}Cr_{10}Fe_{35}Ni_{10}Ti_{25}$ | AC | C14 + $L2_1$ | 27 |
| | $Al_{20}Cr_{10}Fe_{35}Ni_{20}Ti_{15}$ | AC | BCC + C14 + $L2_1$ | 27 |
| | $Al_{20}Cr_{20}Fe_{20}Ni_{20}Ti_{20}$ | AC | BCC + C14 + $L2_1$ | 27 |
| | $Al_{20}Cr_{20}Fe_{35}Ni_{20}Ti_5$ | AC | BCC + $L2_1$ | 27 |

**Section 2: Database for HEA containing Al-X-Y type B2 phase**

Table.S2. Database for 52 HEA containing Al-X-Y type B2 phase. *The database is build based on expansion of a previous database*[34]. HEA element systems, compositions, preparation method, phases and reference are listed in columns. Preparation method abbreviations: Cold roll (CR, thickness reduction is parentheses.), As-cast (AC), Water-quenched (WQ), and Furnace-cooled (FC). Unknown preparation method is labeled as "Unknown". Annealing temperatures are included with unit Celsius (C). Hot pressing pressure is included with unit Megapascal (MPa). Annealing time is included with unit minute (min), hour (h), or day (D). Phase structure notation: the Strukturbericht designations are used except the phases FCC, BCC, HCP, and Sigma, which correspond to Strukturbericht designations A1, A2, A3, $D8_b$, A12, and $D0_{24}$. Laves phase corresponds to C14, C15, or C36. Unknown phase content is labeled as "?".

| System | Composition | Preparation method | Phase | Ref |
|---|---|---|---|---|
| AlCrMoNbTi | Al20Cr20Mo20Nb20Ti20 | AC + 1300 C / 20 h | B2 + Laves | 35 |
| AlCrMoTaTi | Al20Cr20Mo20Ta20Ti20 | AC + 1500 C / 20 h | BCC + B2 + C14 + C15 + C36 | 36 |
| AlCrMoTi | Al10Cr30Mo30Ti30 | AC + 1200 C / 20 h | B2 + Laves | 37 |
| | Al15Cr28.3Mo28.3Ti28.3 | AC + 1200 C / 20 h | B2 + Laves | 37 |
| | Al25Cr25Mo25Ti25 | AC + 1200 C / 20 h | B2 + Laves | 35 |
| AlCrNbTiV | Al20Cr20Nb20Ti20V20 | AC + 1200 C / 24 h + 800 C / 1000 C / 100 h | B2 + C14 + Sigma | 38 |
| | Al22.2Cr11.1Nb22.2Ti22.2V22.2 | AC + 1200 C / 24 h + 800 C / 1000 C / 100 h | B2 + Sigma | 38 |
| AlCrNbTiVZr | Al20Cr10Nb15Ti20V25Zr10 | AC + 1200 C / 24 h | B2 + C14 + Al3Zr5 | 39 |
| AlCrTiV | AlCrTiV | AC | B2 | 40 |
| AlFeNbTi | Al10Fe10Nb60Ti20 | Unknown | B2 + ? | 34 |
| AlHfNbTi | Al25Hf25Nb25Ti25 | AC | B2 | 41 |
| AlHfNbW | Al10Hf20Nb60W10 | Unknown | B2 + ? | 34 |
| AlHfTaTi | Al25Hf25Ta25Ti25 | AC | B2 | 41 |
| AlMnNbTiV | AlMnNbTiV | Unknown | B2 + Laves | 42 |
| AlMoNbTaTi | Al22.5Mo5Nb15Ta5Ti52.5 | Unknown | B2 + ? | 34 |
| AlMoNbTaTiZr | Al11.1Mo11.1Nb22.2Ta11.1Ti22.2Zr22.2 | AC + 1400 C / 2 h / 207 MPa + 1400 C / 6 h + FC | BCC + B2 + Zr | 43 |
| | Al20Mo10Nb20Ta10Ti20Zr20 | AC + 1400 C / 2 h / 207 MPa + 1400 C / 24 h + FC | BCC + B2 + AlxZry | 44 |
| | | AC + 1400 C / 2 h / 207 MPa + 1400 C / 24 h + FC | BCC + B2 | 45 |
| | Al22.2Mo11.1Nb22.2Ta11.1Ti22.2Zr11.1 | AC + 1400 C / 2 h / 207 MPa + 1400 C / 6 h + FC | B2+AlxZry | 43 |
| AlMoNbTi | Al15Mo15Nb40Ti30 | Unknown | B2 + ? | 34 |
| | Al18.5Mo5Nb24.5Ti52 | Unknown | B2 + ? | 34 |
| | Al24.5Mo10Nb16Ti49.5 | Unknown | B2 + ? | 34 |
| | Al25Mo25Nb25Ti25 | AC + 1500 C / 20 h | B2 | 35 |
| AlMoNbTiV | Al1.5MoNbTiV | AC | B2 | 46 |
| | Al15.8Mo21.2Nb20.9Ti20.9V21.2 | AC + 1400 C / 24 h | B2 | 47 |
| | Al25Mo4Nb14Ti53V4 | Unknown | B2 + ? | 34 |
| AlMoTaTi | Al17Mo17Ta32Ti34 | Unknown | B2 + ? | 34 |
| | Al20Mo10Ta15Ti55 | Unknown | B2 + ? | 34 |
| | Al20Mo20Ta20Ti40 | Unknown | B2 + ? | 34 |
| AlMoTi | Al23.5Mo6Ti70.5 | Unknown | B2 + ? | 34 |
| | Al25Mo20Ti55 | Unknown | B2 + ? | 34 |
| | Al25Mo25Ti50 | Unknown | B2 + ? | 34 |
| AlNbTaTi | Al22Nb20Ta7Ti51 | Unknown | B2+ Orthorhombic phase | 48 |
| AlNbTaTiVZr | Al10Nb20Ta16Ti30V4Zr20 | AC + 1200 C / 2 h / 207 MPa + 1200 C / 24 h | BCC + B2 | 49 |
| | | AC + 1200 C / 2 h / 207 MPa + 1200 C / 24 h | BCC + B2 | 50 |
| | | AC + 1200 C / 24 h | BCC + B2 | 51 |
| AlNbTaTiZr | Al5.7Nb23.5Ta17.6Ti27.2Zr26 | AC + 1200 C / 2 h / 207 MPa + 1200 C / 24 h | BCC + B2 | 49 |
| | Al5.9Nb23.5Ta23.5Ti23.5Zr23.5 | AC + 1400 C / 2 h / 207 MPa + 1400 C / 6 h + FC | BCC + B2 | 43 |
| | Al10Nb15Ta5Ti30Zr40 | AC + 1100 C / 24 h | BCC + B2 | 52 |
| | Al25Nb25Ta12.5Ti25Zr12.5 | AC + 1400 C / 2 h / 207 MPa + 1400 C / 6 h + FC | B2 | 43 |
| AlNbTi | Al15Nb55Ti30 | Unknown | B2 + ? | 34 |
| | Al15Nb75Ti10 | Unknown | B2 + ? | 34 |
| | Al20.5Nb18Ti61.5 | Unknown | B2 + ? | 34 |
| | Al24Nb20Ti54 | Unknown | B2 + Orthorhombic phase,<br>B2 + Alpha 2,<br>B2 + Orthorhombic phase + Alpha 2 | 34 |
| | Al25Nb25Ti50 | Unknown | B2 + ? | 34 |
| AlNbTiV | Al25Nb25Ti25V25 | AC + 1200 C / 24 h + 800 C / 1000 C / 100 h | B2 + Sigma | 38 |
| | | AC + 1200 C / 24 h | B2 | 53 |
| AlNbTiVZr | Al18.2Nb18.2Ti18.2V18.2Zr27.3 | AC + 1200 C / 24 h | B2 + Al3Zr5 | 53 |
| | Al20Nb20Ti20V20Zr20 | AC + 1200 C / 24 h | B2 + Al3Zr5 | 53 |
| | | AC + 1200 C / 24 h + 800 C / 100 h | B2 + Al3Zr5 + C14 | 38 |
| | Al22.2Nb22.2Ti22.2V22.2Zr11.1 | AC + 1200 C / 24 h | B2 + Al3Zr5 | 53 |

**Section 3: Machine Learning algorithm details**

The following figure illustrates a comparison of F1 errors when utilizing the Random Forest (RF), Neural Network (NN), and Support Vector Machine (SVM) algorithms for both Layer 1 and Layer 2 models. These computations were executed through MATLAB, leveraging built-in machine learning (ML) classification algorithms from toolboxes such as the Statistics and Machine Learning Toolbox. For each ML algorithm, a 5-fold cross-validation method was employed. The recorded average F1 errors are complemented by standard deviation as error bars, which result from repeating the sequential learning process at least 10 times.

For RF model training, we employ the bagging (bootstrap aggregating) approach. Gini Impurity is used as the criterion for the quality of the split. Hyperparameters such as *NumLearningCycles*, *MaxNumSplits*, *MinLeafSize*, and *NumVariablesToSample* are automatically fine-tuned and optimized with the *fitcensemble()* function. The training of the NN model was accomplished using the Stochastic Gradient Descent with Momentum (SGDM) optimization method, with an initial learning rate set at 0.01. The mini-batch size employed was 64. The network architecture incorporated 30 fully connected hidden layers, with the Leaky ReLU function serving as the activation mechanism for these layers. The final output layer utilized the softmax activation function for its operations. For the SVM model training, various hyperparameters, including *BoxConstraint*, *KernelScale*, *KernelFunction*, and *PolynomialOrder*, were optimized automatically by the *fitcsvm()* or *fitcecoc()* function (for layer 2 and 1 models, respectively).

RF demonstrated superior performance for the Layer 1 model, while SVM proved to be more effective for Layer 2 models.

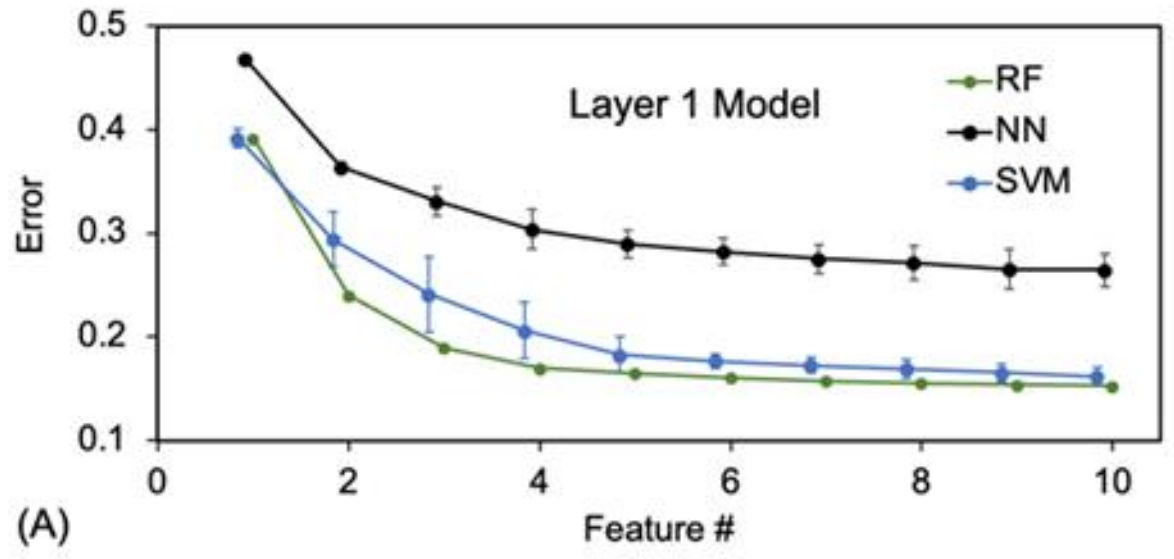

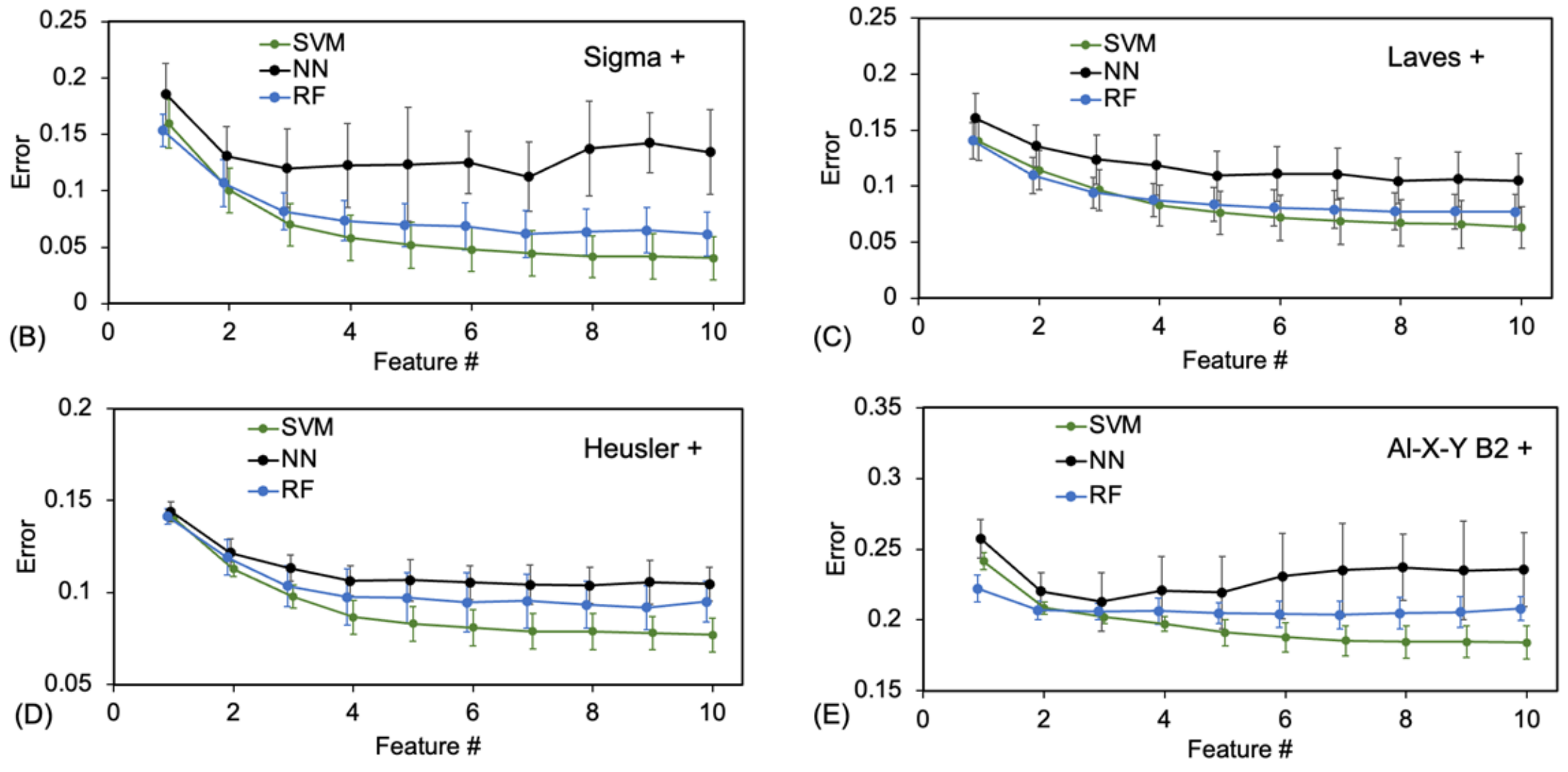

Figure S1. ML classification F1 errors vary as the number of engineered features increases. The comparisons of the errors are presented for: (A) Layer 1 model, (B) Sigma+, (C) Laves+, (D) Heusler+, and (E) Al-X-Y B2+ models, among using Random Forest (RF), Neural Network (NN), and Support Vector Machine (SVM) algorithms.

**Section 4: Under/Over-sampling methods implementation and comparison for Laves+ and Sigma+ prediction model**

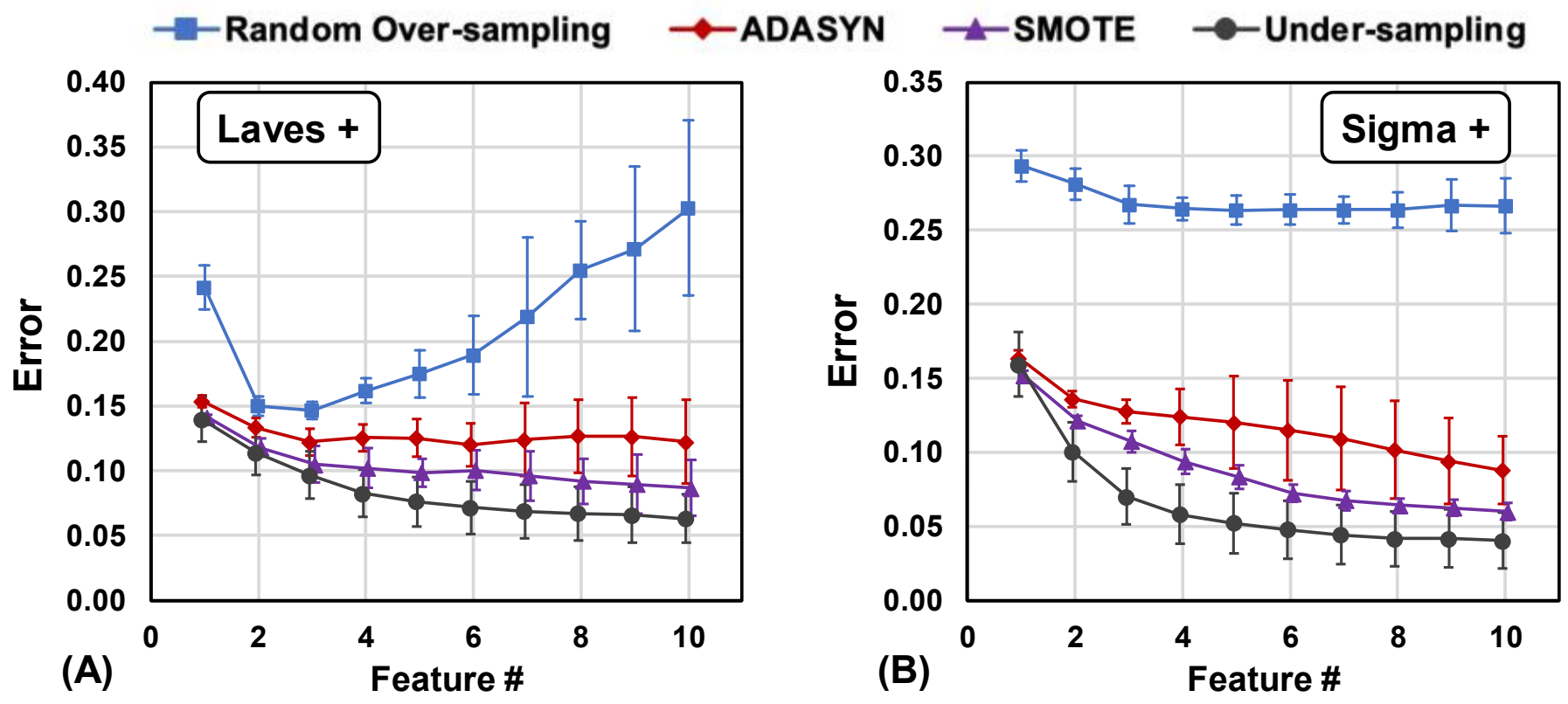

Figure S2. ML classification F1 errors vary as the number of engineered features increases. The comparisons of the errors are presented for: (A) Laves+, and (B) Sigma+ models, among using Random Over-sampling, ADASYN, SMOTE, and Under-sampling methods. Error bars are presented in both plots.

Four methods of handling imbalanced databases are compared using errors from 5-fold cross-validations. Random Over-sampling method randomly generates new samples by repeating the samples in the minor dataset. *Adaptive Synthetic Sampling Approach for Imbalanced Learning* (ADASYN)[55] and *Synthetic Minority Over-sampling Technique* (SMOTE)[56] are synthetic over-sampling methods, which create virtual samples based on samples in the minor database. Under-sampling method randomly draws samples from the major dataset, to form a balanced training database with the minor dataset.

The authors have designed and executed the random oversampling implementation, which is straightforward to utilize. The corresponding code can be procured from the authors upon a reasonable request. For the SMOTE and ADASYN algorithms, the code can be accessed from the given reference: Michio (2023). Oversampling Imbalanced Data: SMOTE related algorithms (https://github.com/minoue-xx/Oversampling-Imbalanced-Data/releases/tag/1.0.1), GitHub.

As shown in Fig.S1, the under-sampling method shows the lowest error compared to other over-sampling methods. More importantly, although under-sampling method may have info loss due to data removal in the majority class in each round of simulation, this information loss can be overcome by bootstrapping the database and training multiple ML models based on the bootstrapped sub-database. Random Over-sampling method that creates repeated data for minority class may cause overfitting. ADASYN and SMOTE would expand the minority class by creating virtual data that are not physically existed. These new data do not have any physical meaning. Based on these reasons, we choose under-sampling as our unbalanced database handling method.

**Section 5: XRD patterns for validation HEAs**

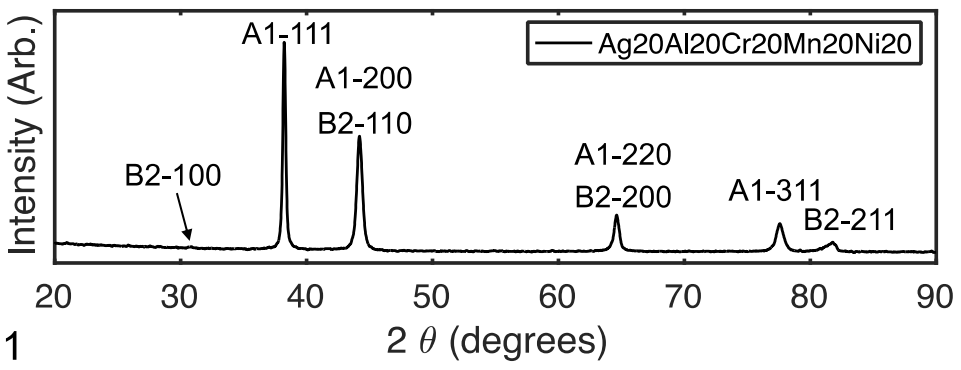

1

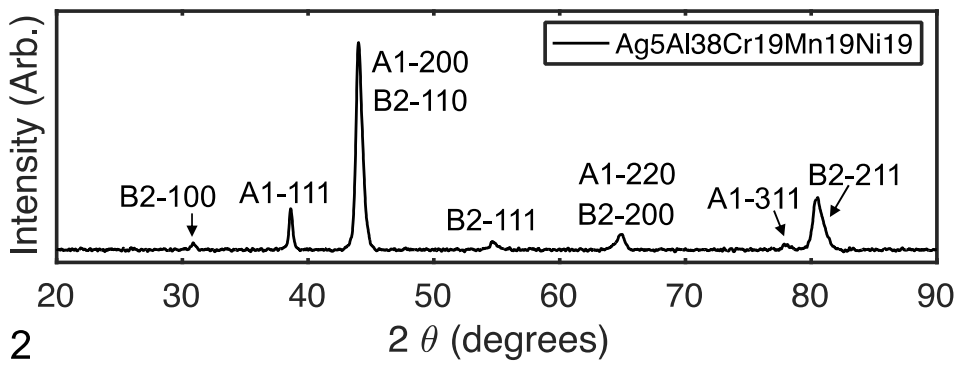

2

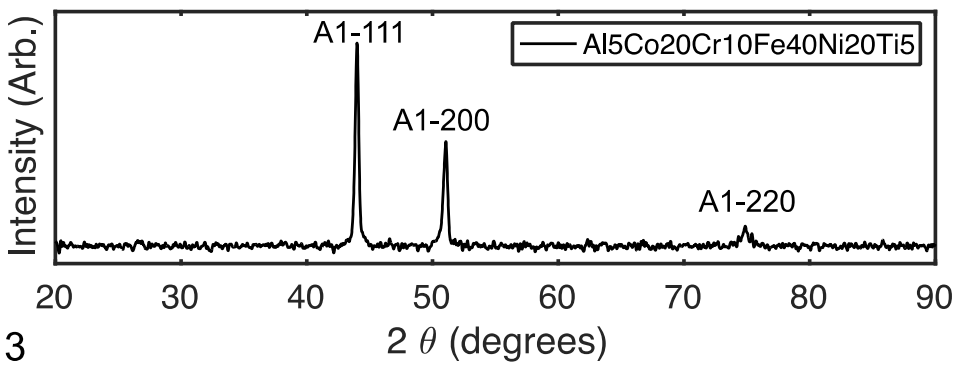

3

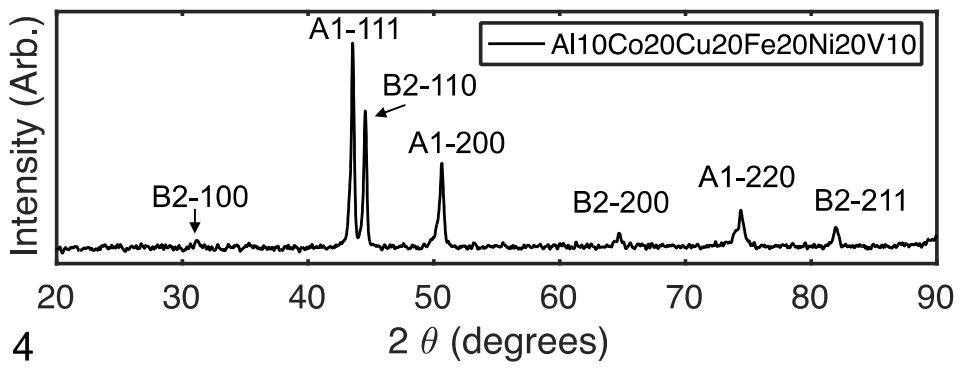

4

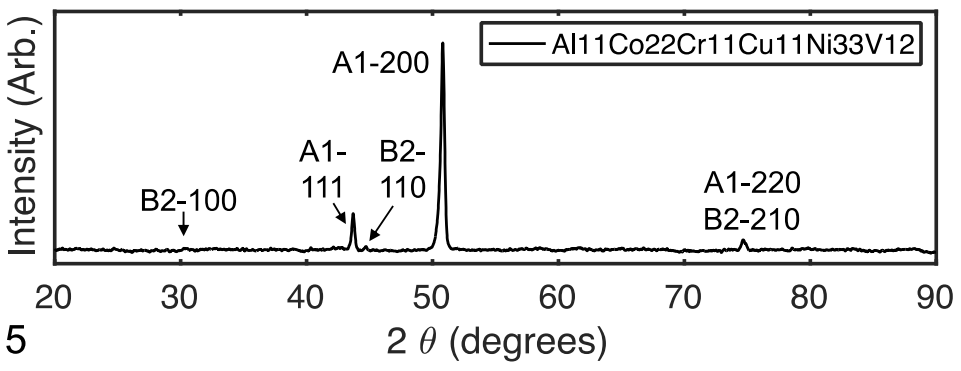

5

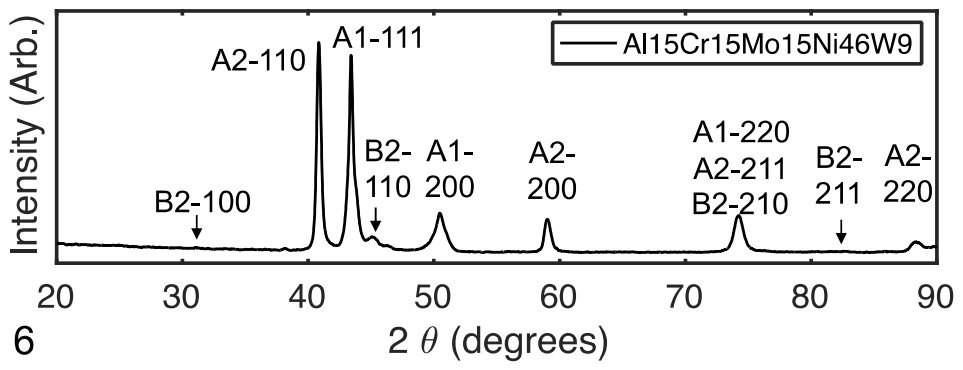

6

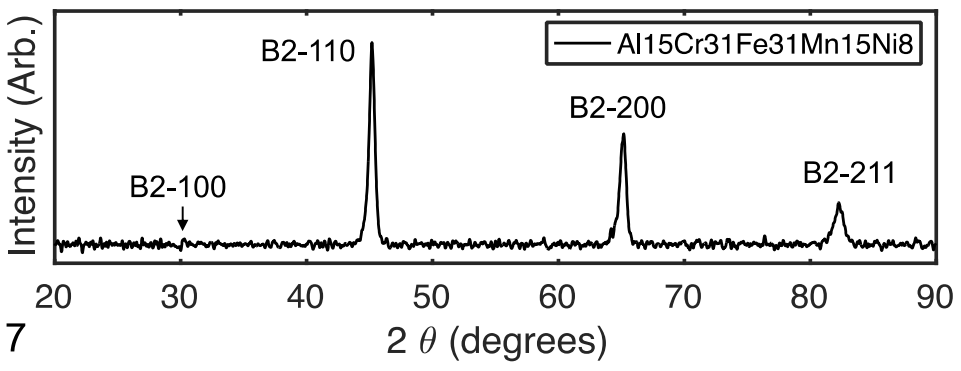

7

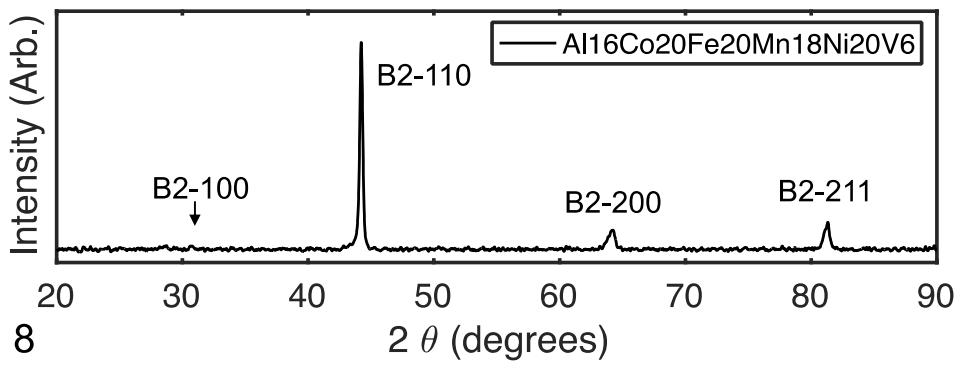

8

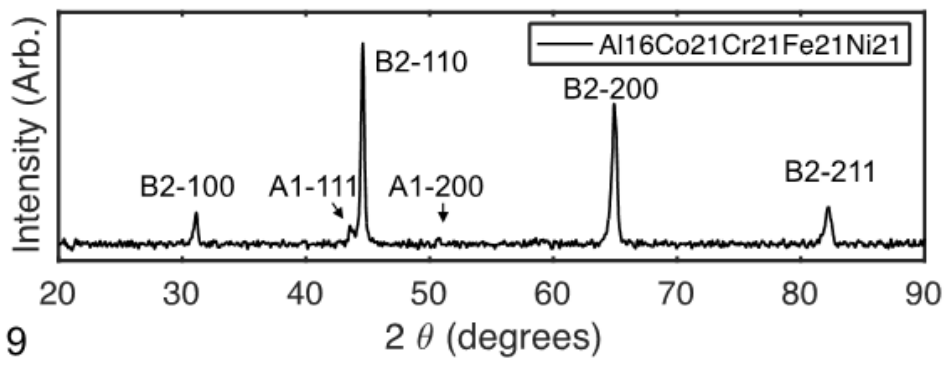

9

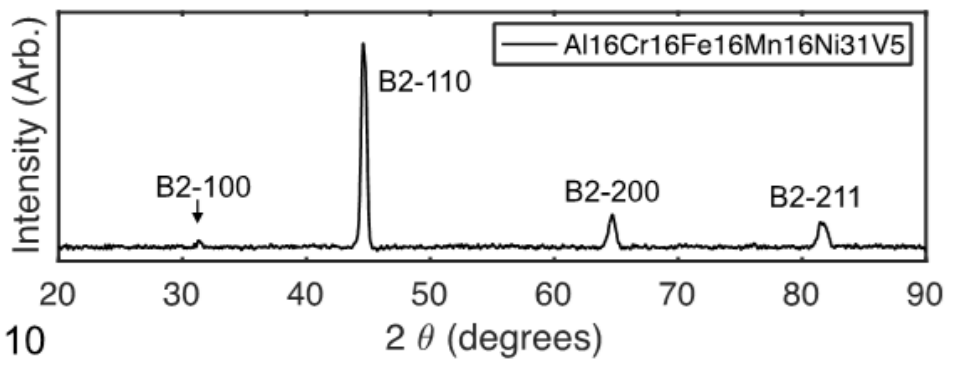

10

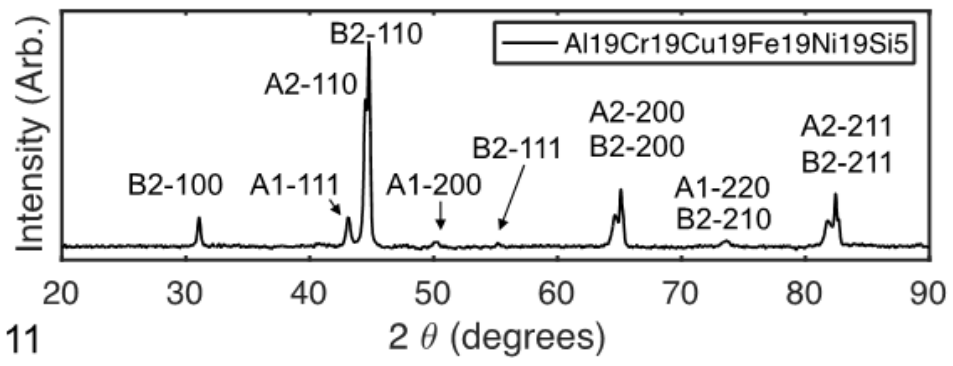

11

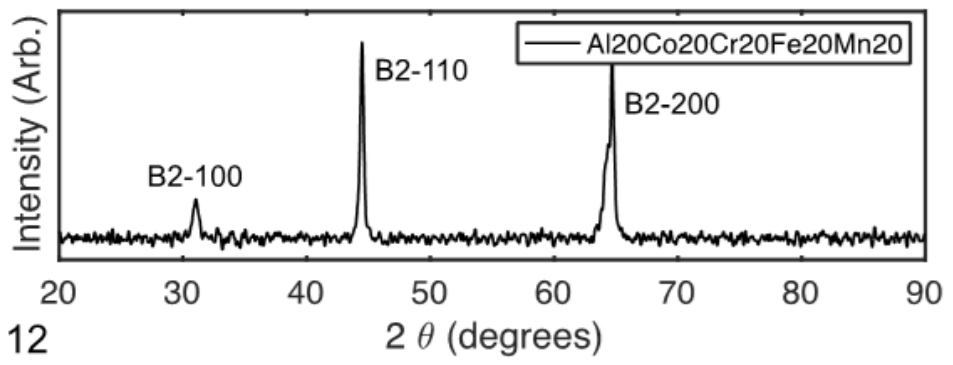

12

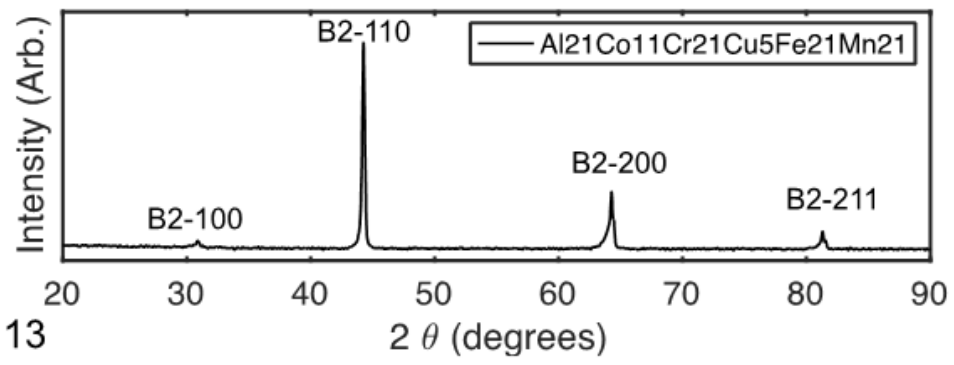

13

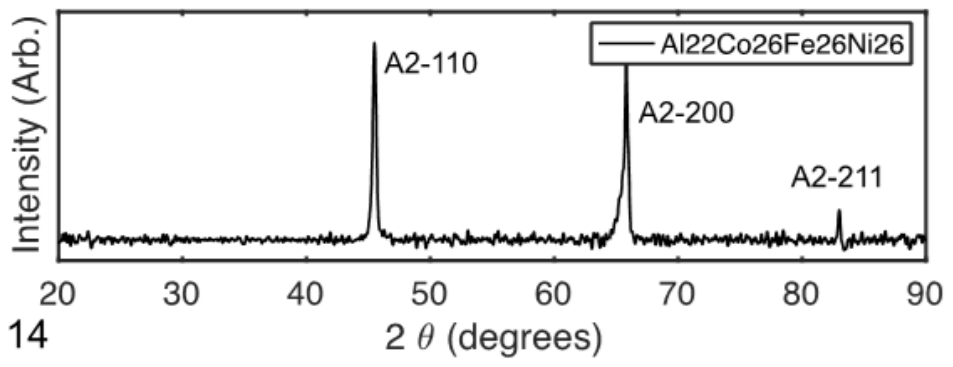

14

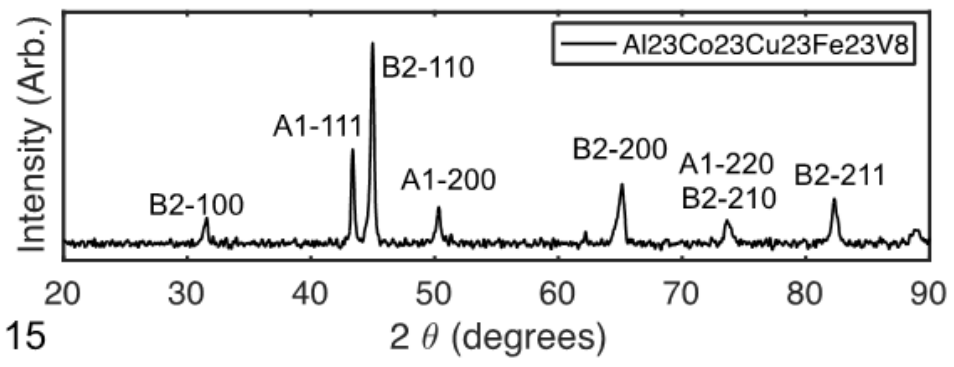

15

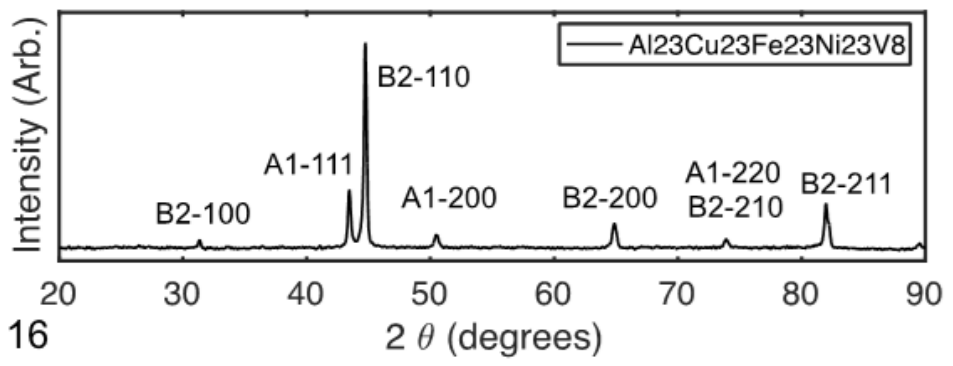

16

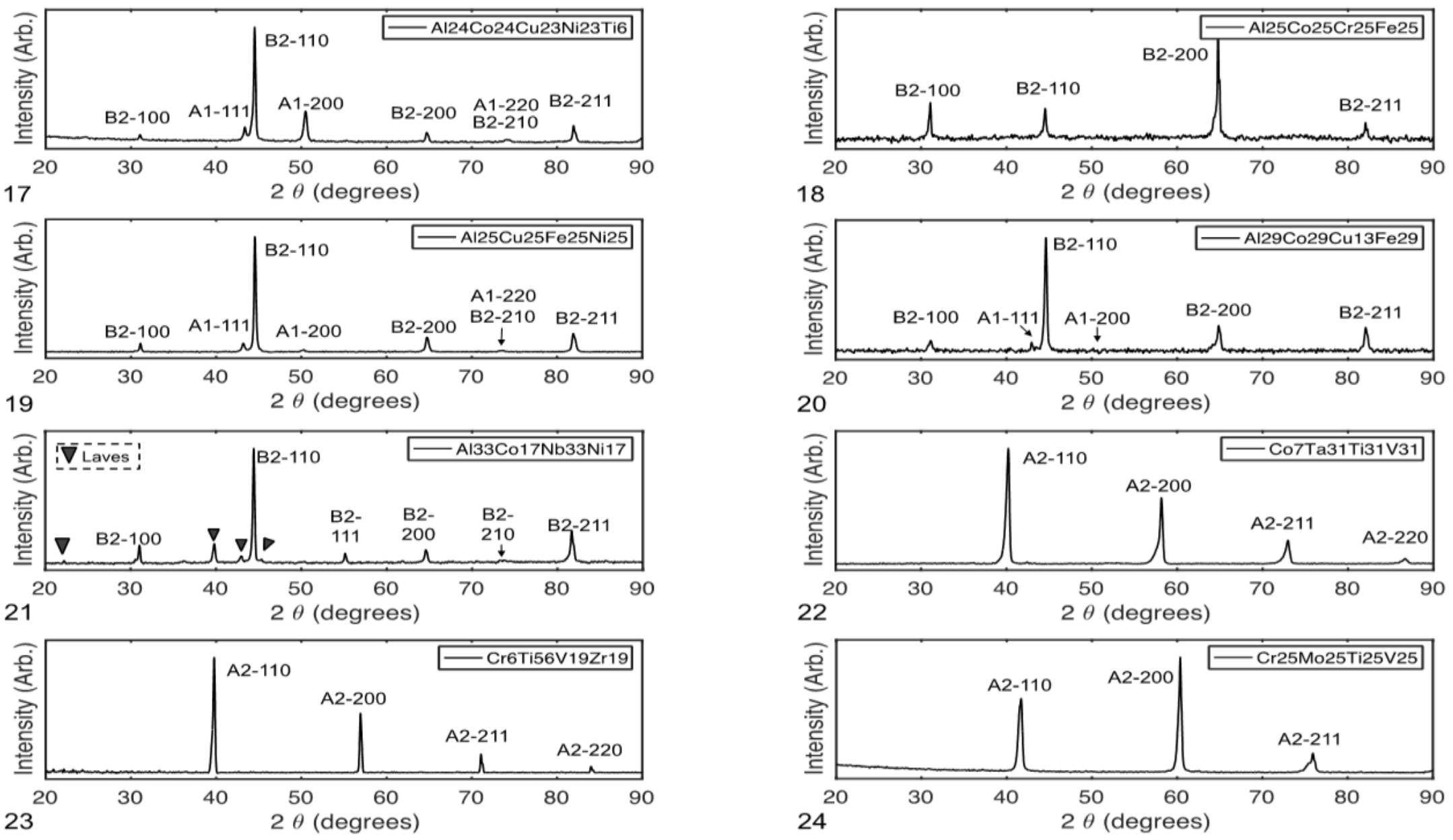

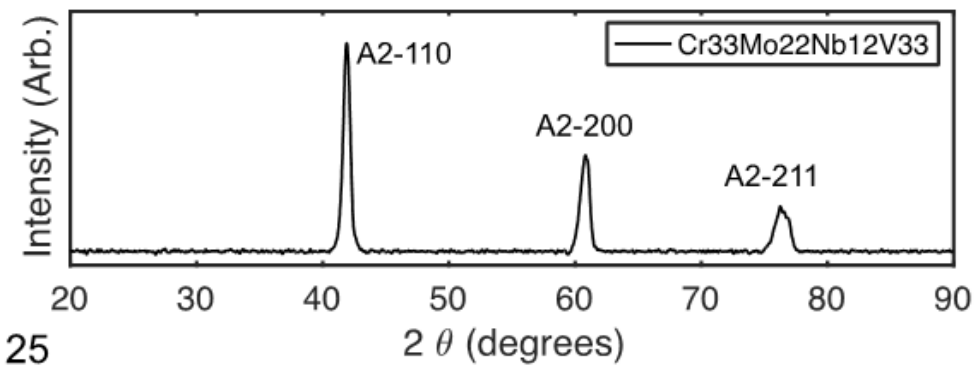

25

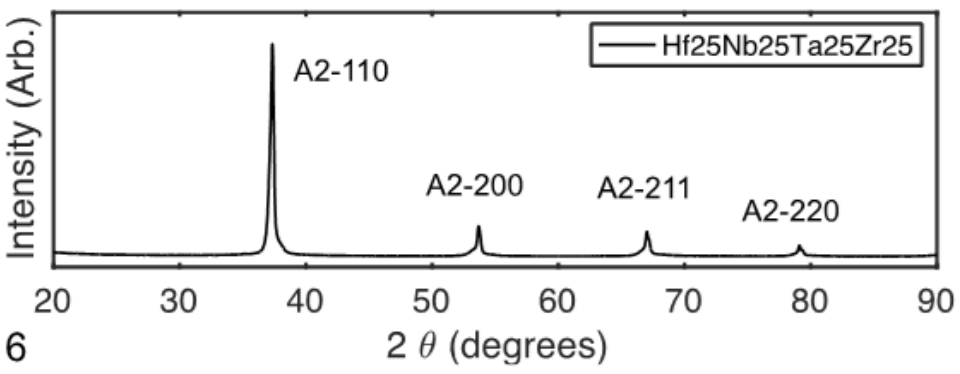

6

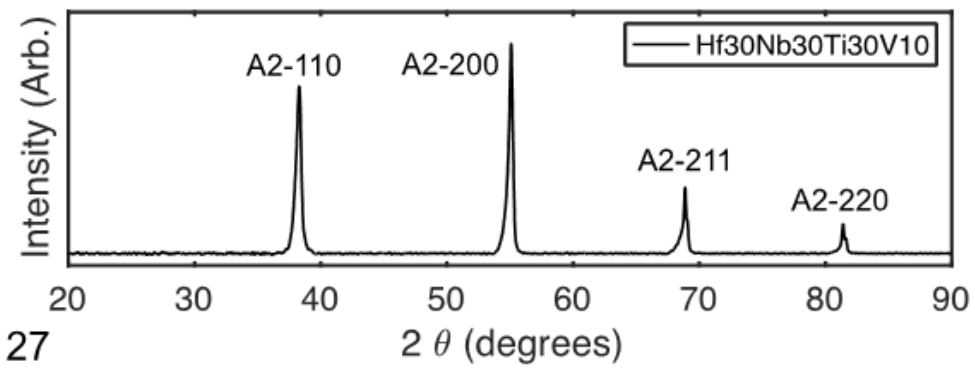

27

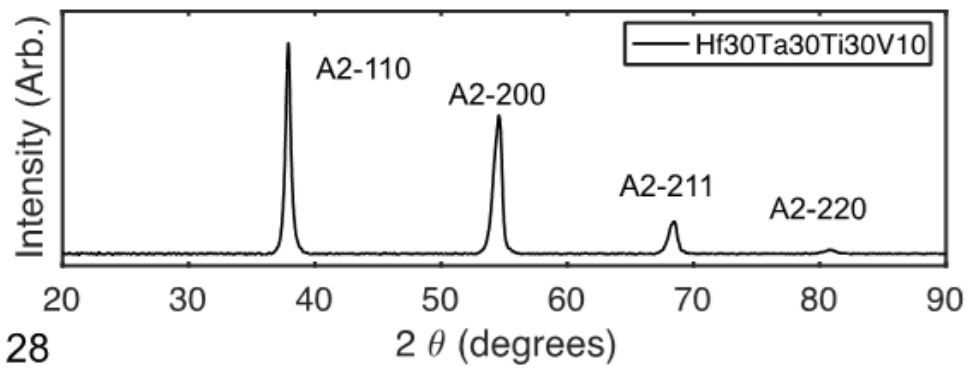

28

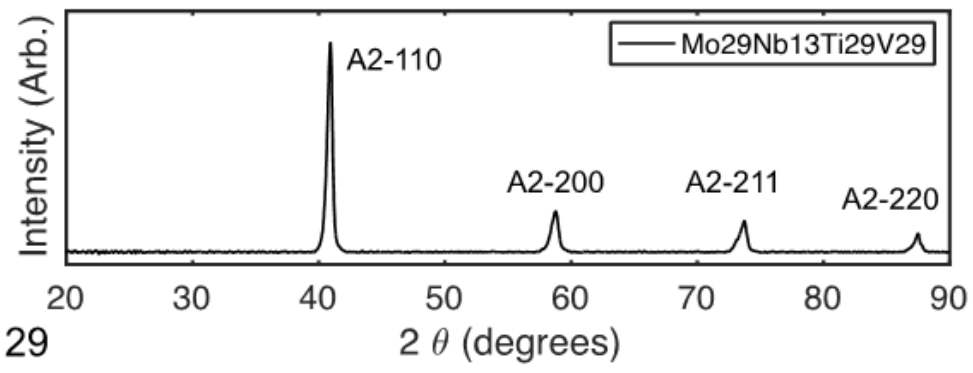

29

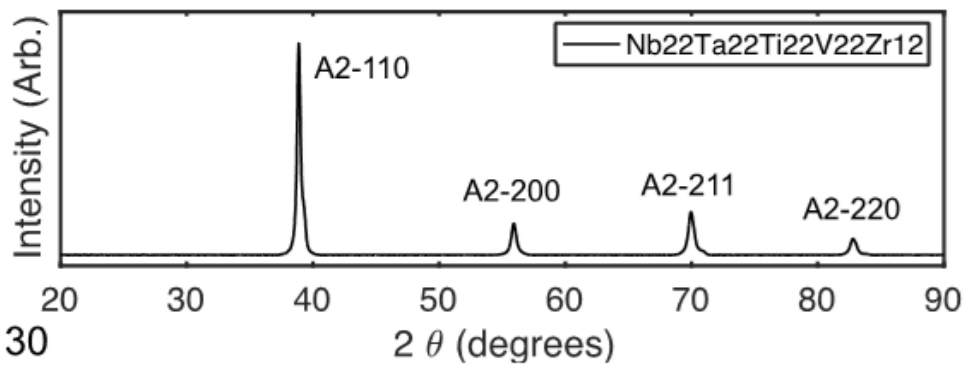

30

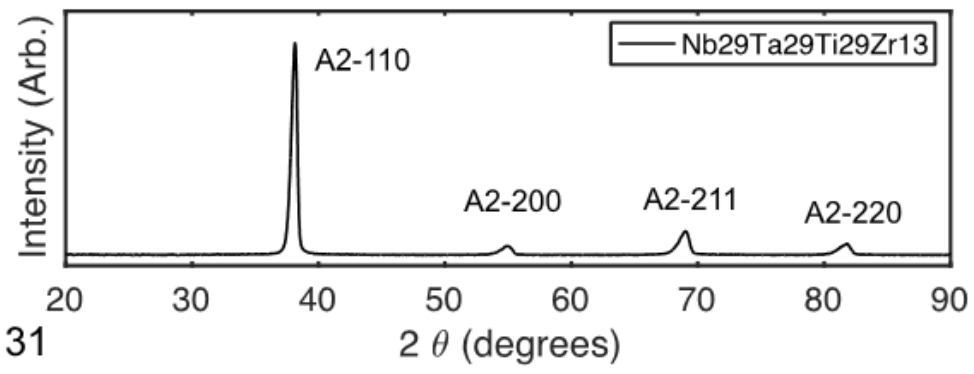

31

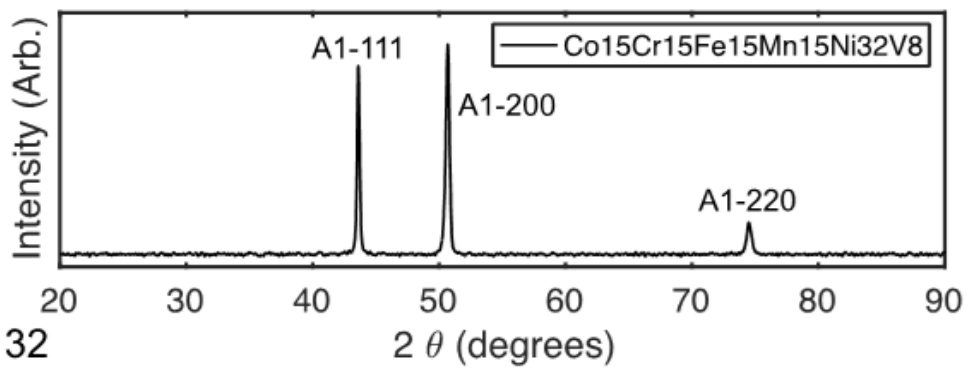

32

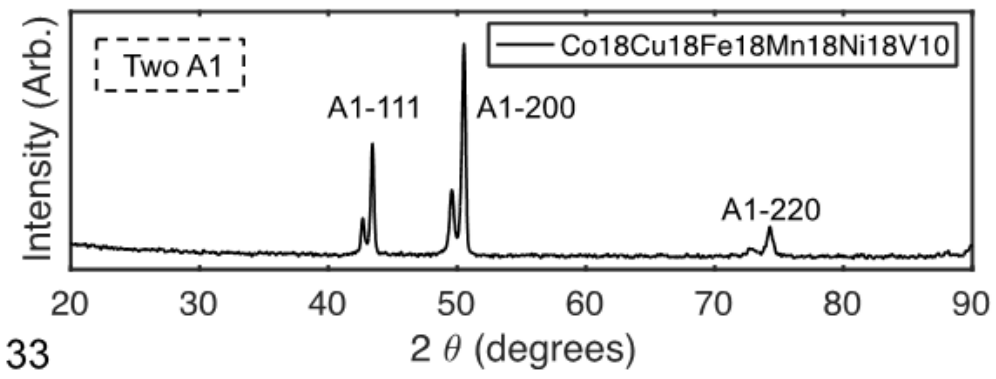

33

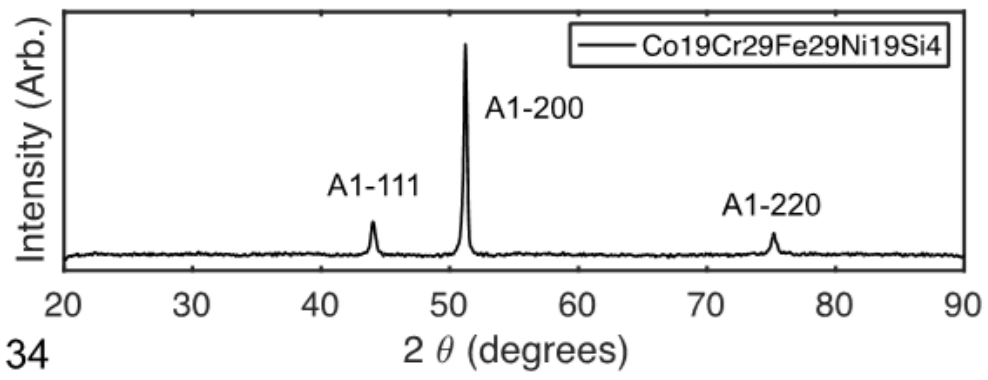

34

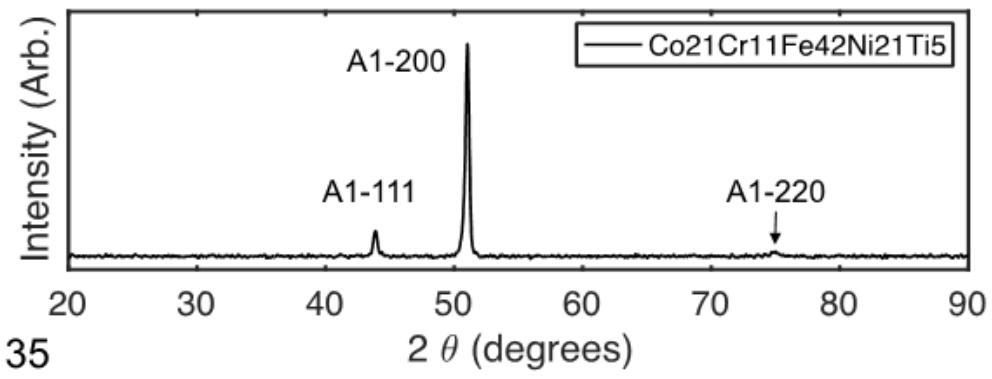

35

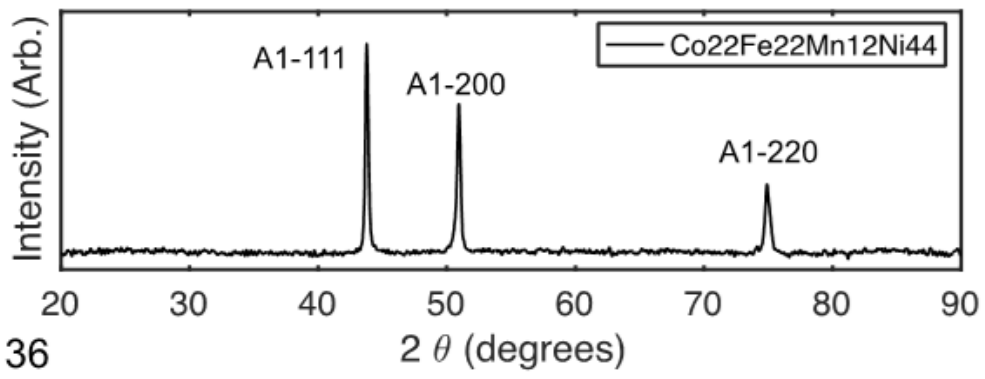

36

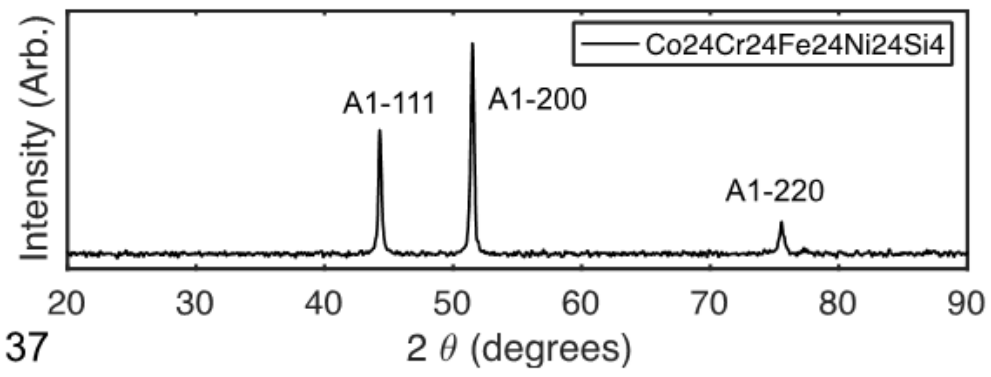

37

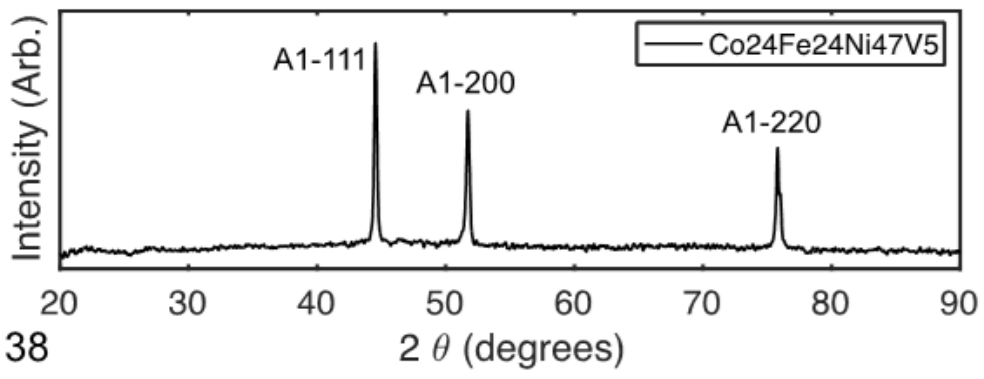

38

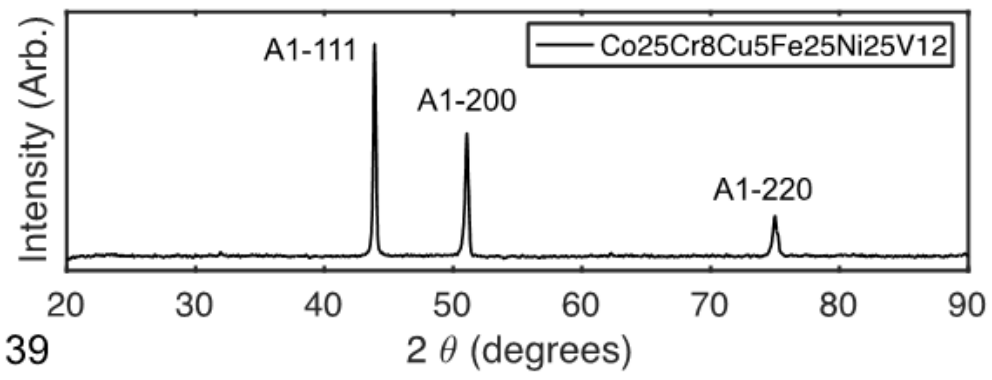

39

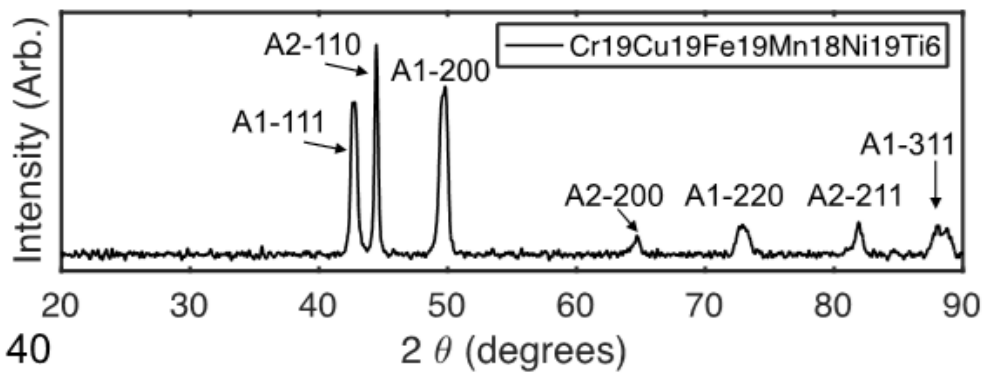

40

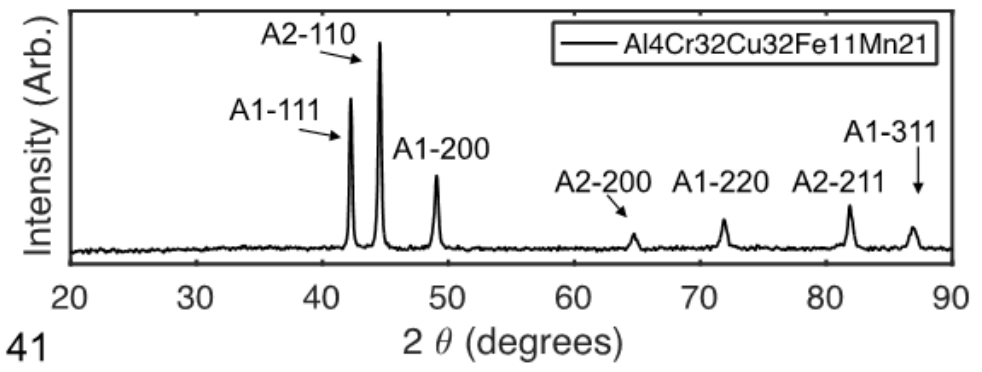

41

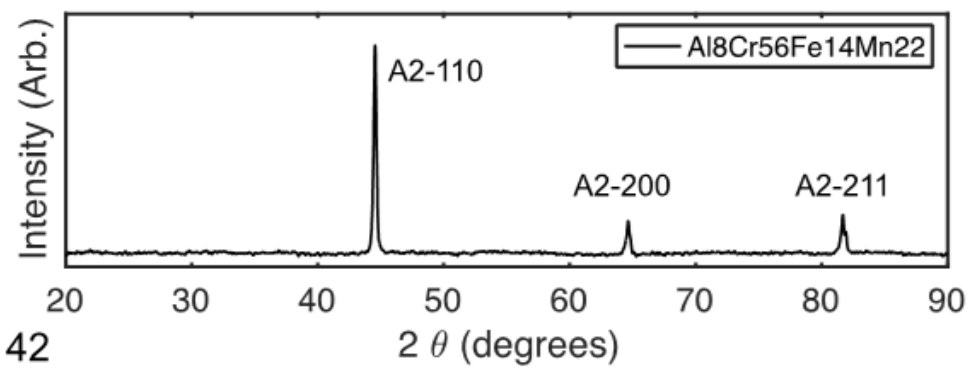

42

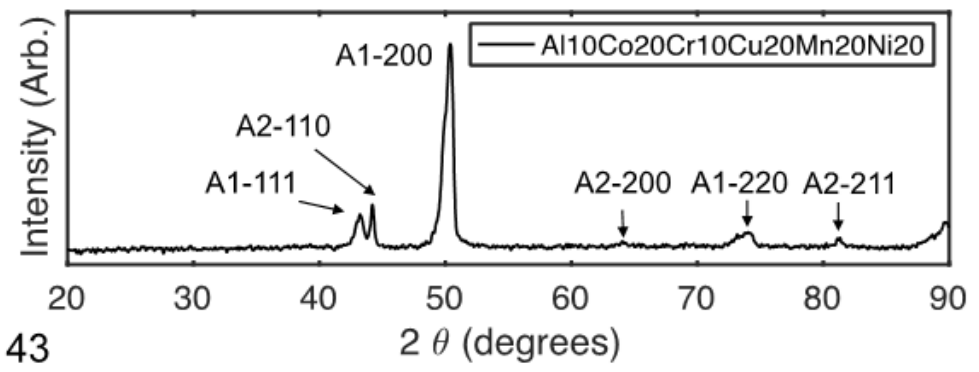

43

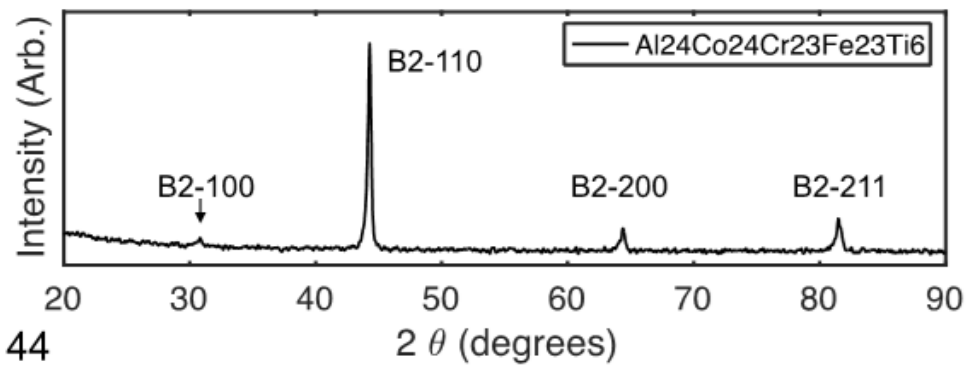

44

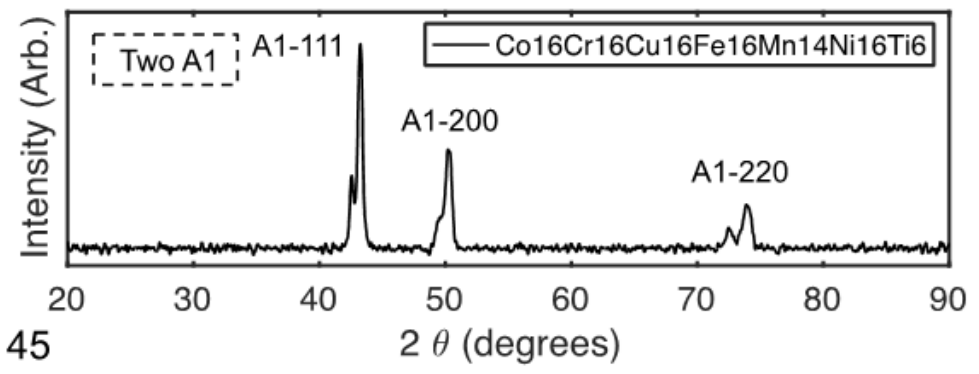

45

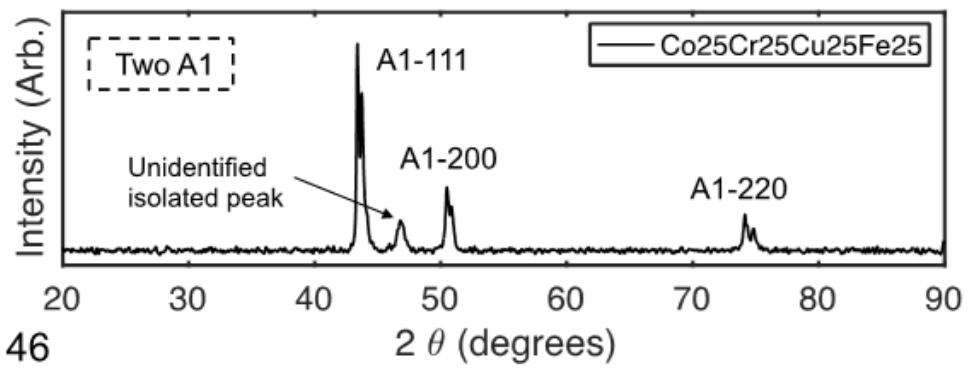

46

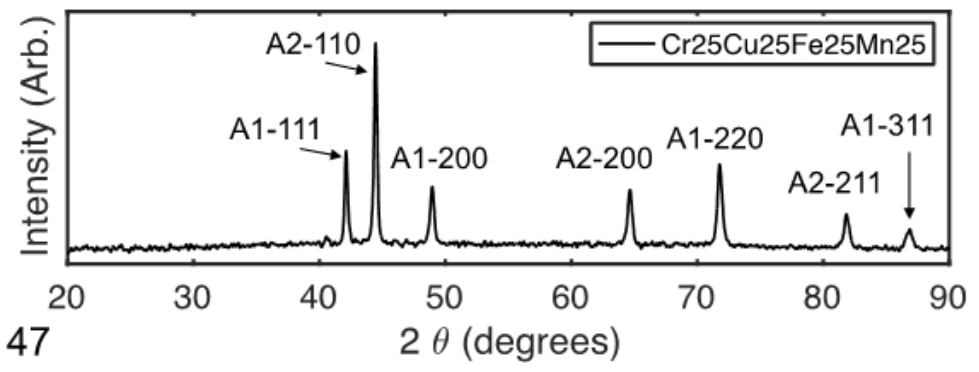

47

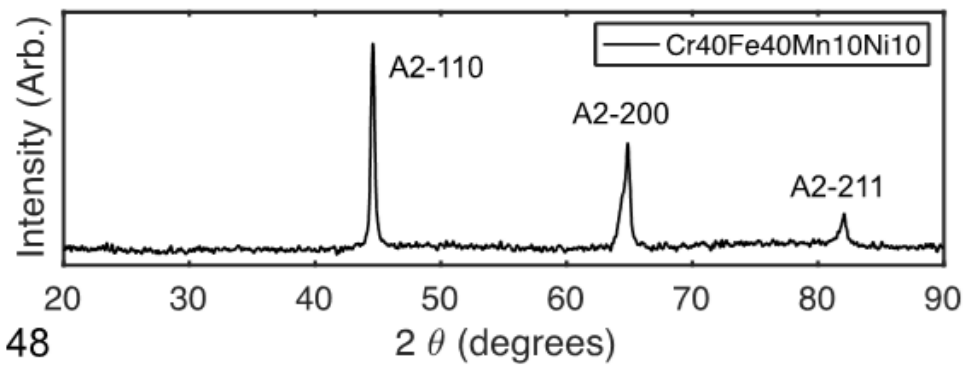

48

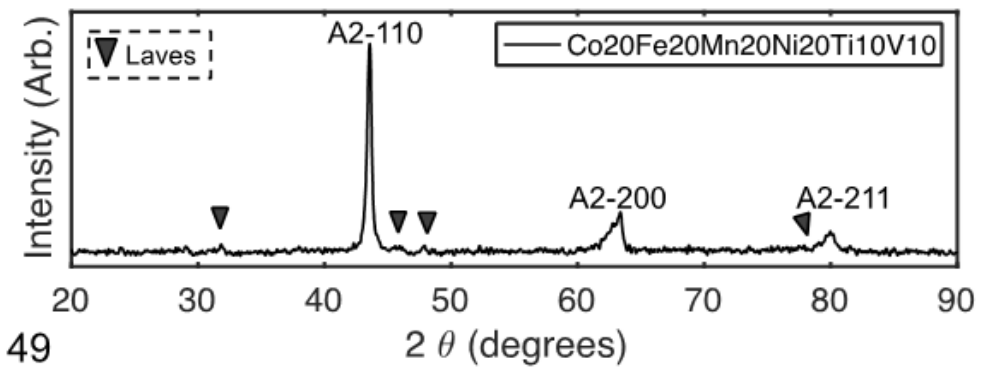

49

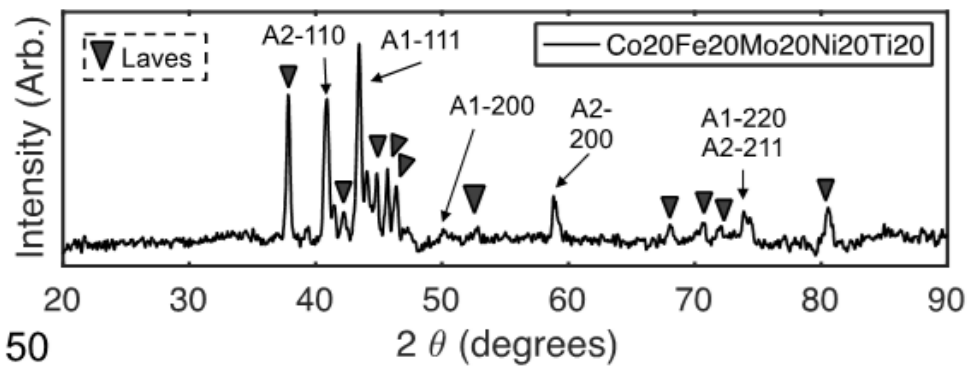

50

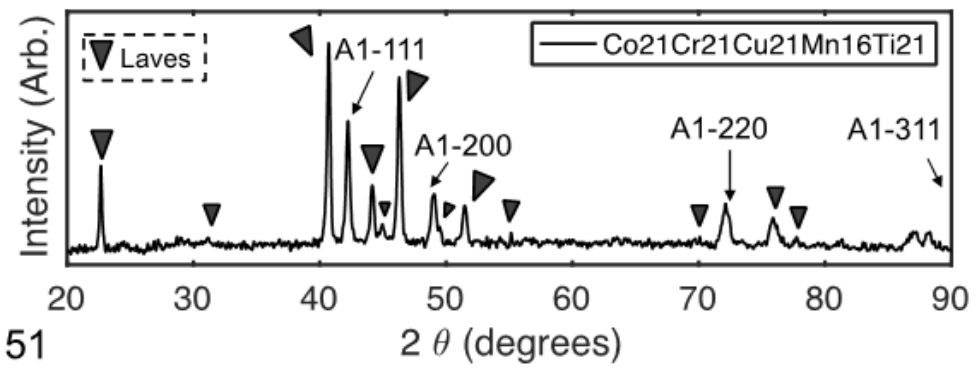

51

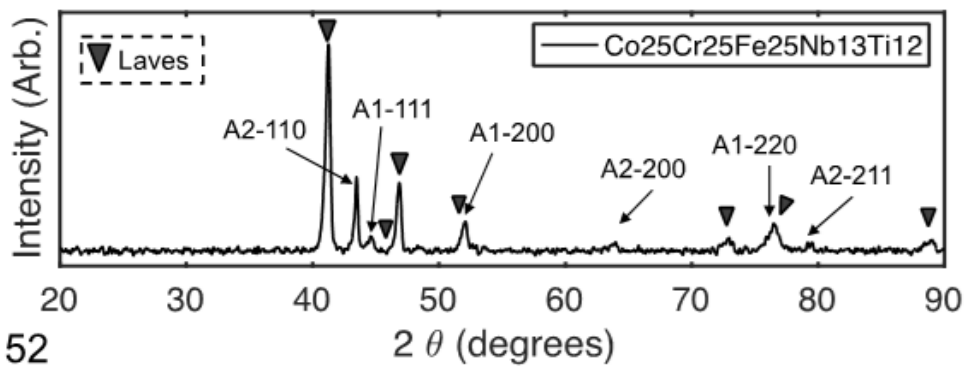

52

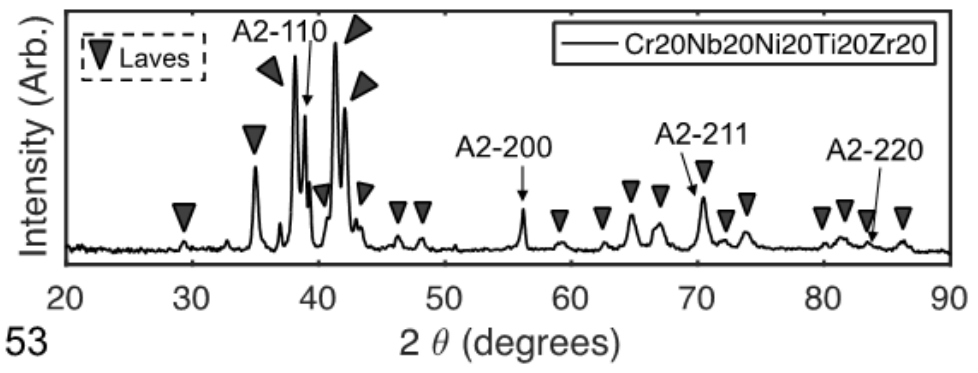

53

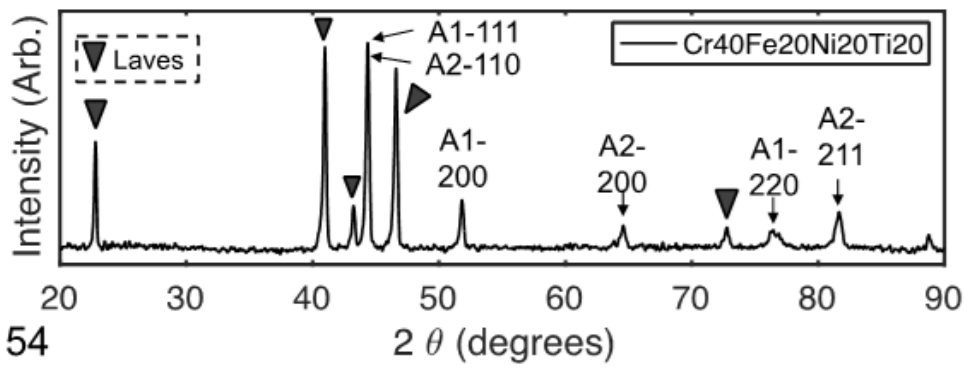

54

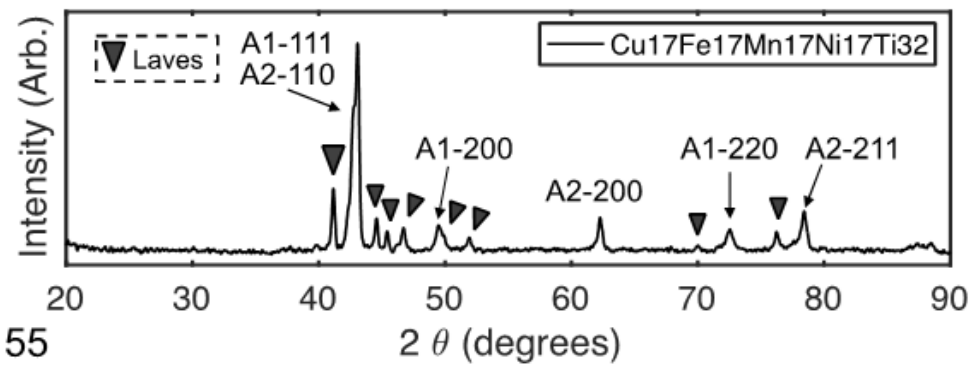

55

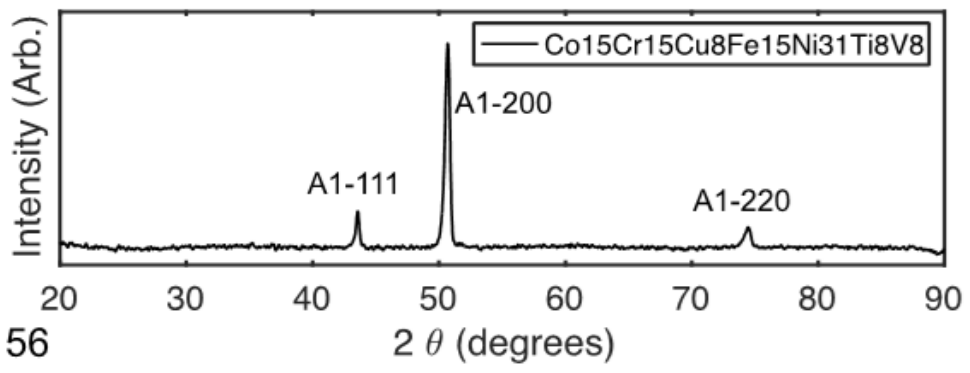

56

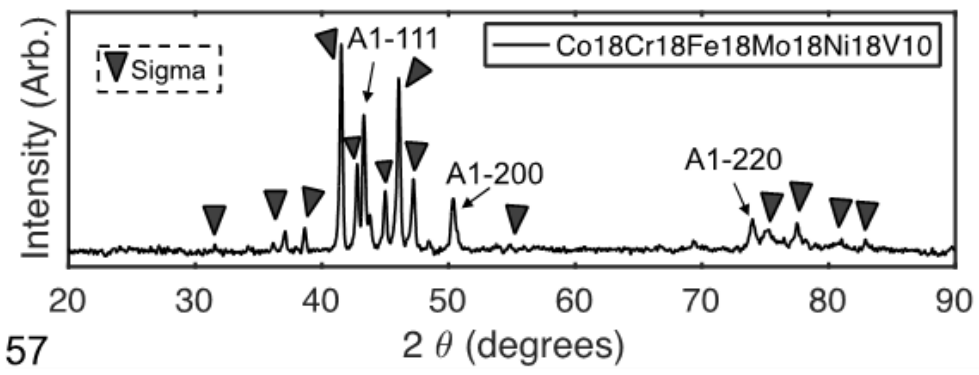

57

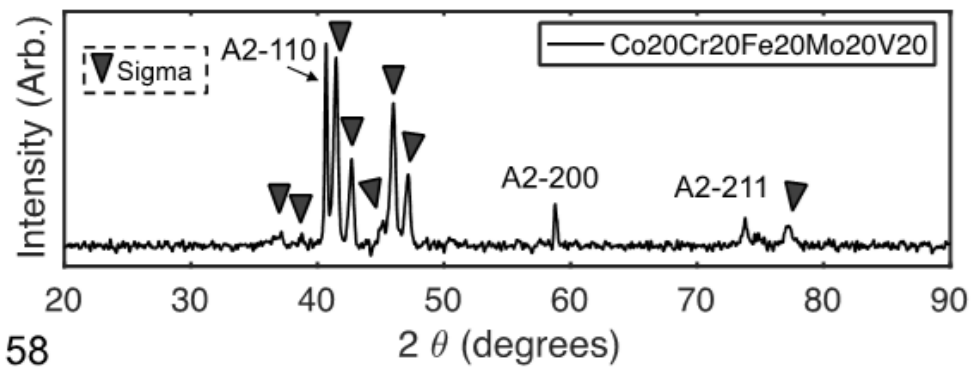

58

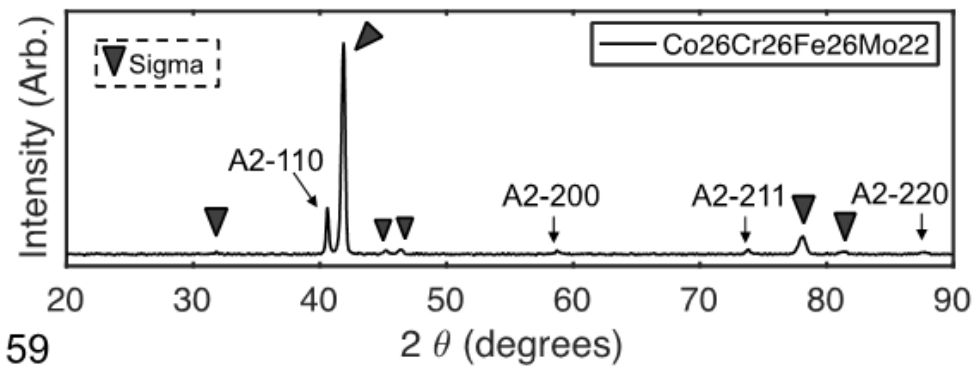

59

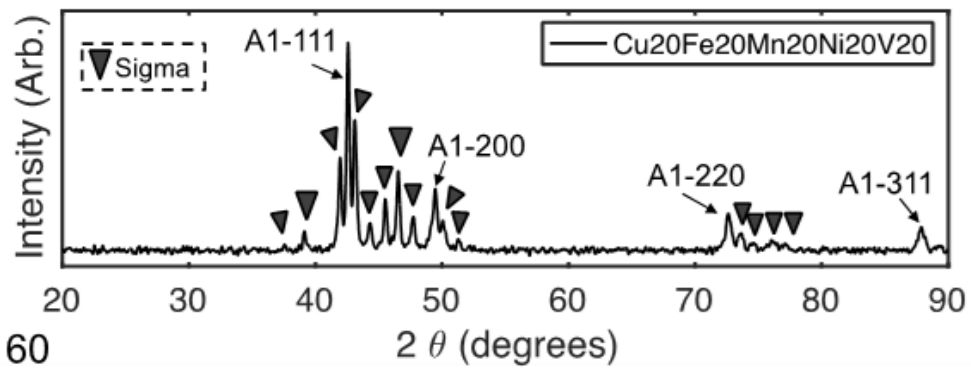

60

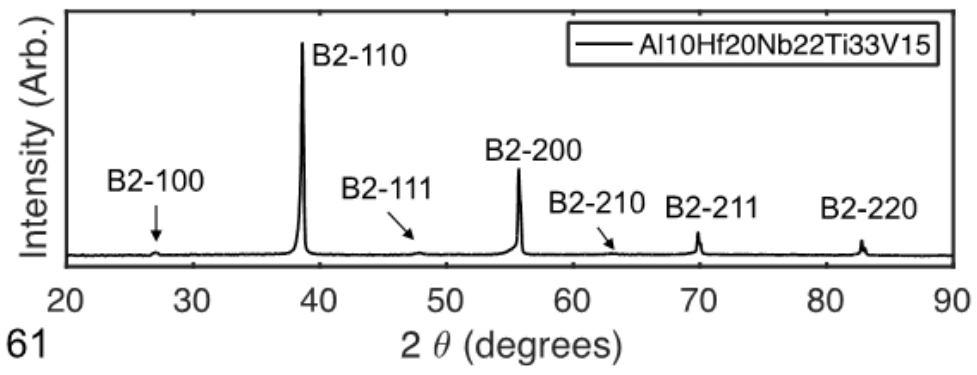

61

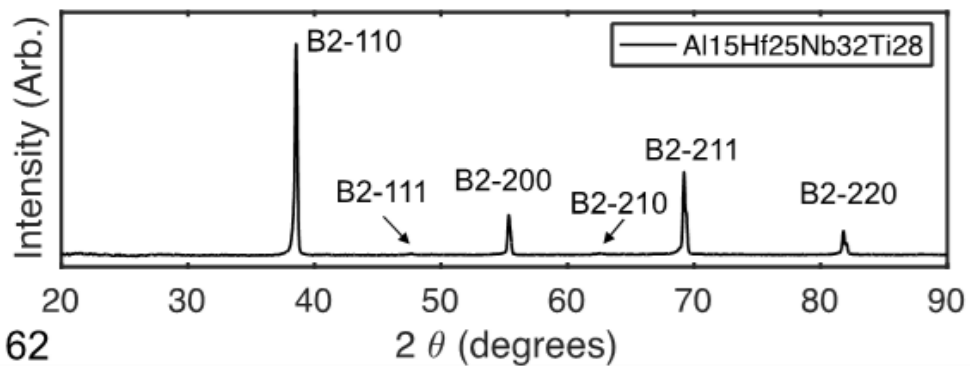

62

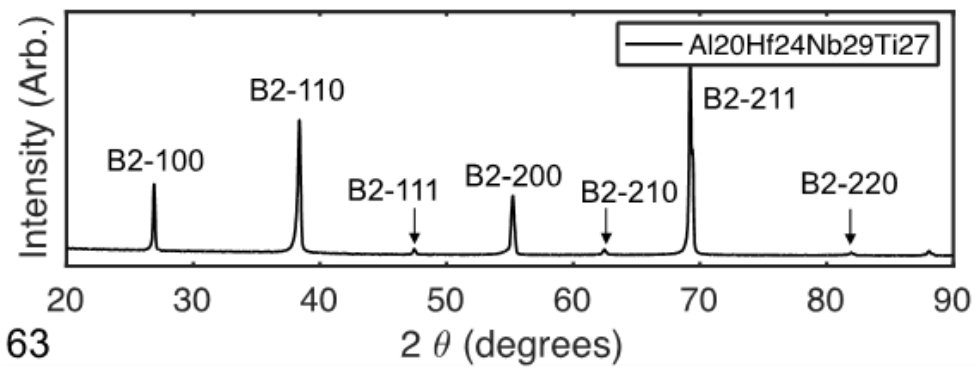

63

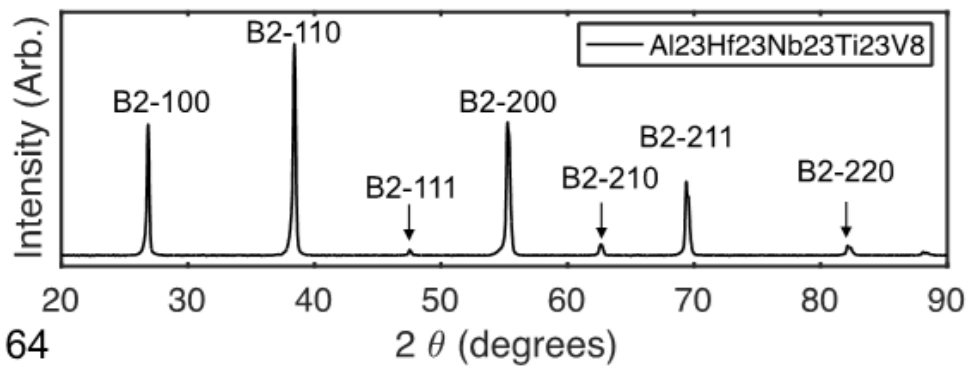

64

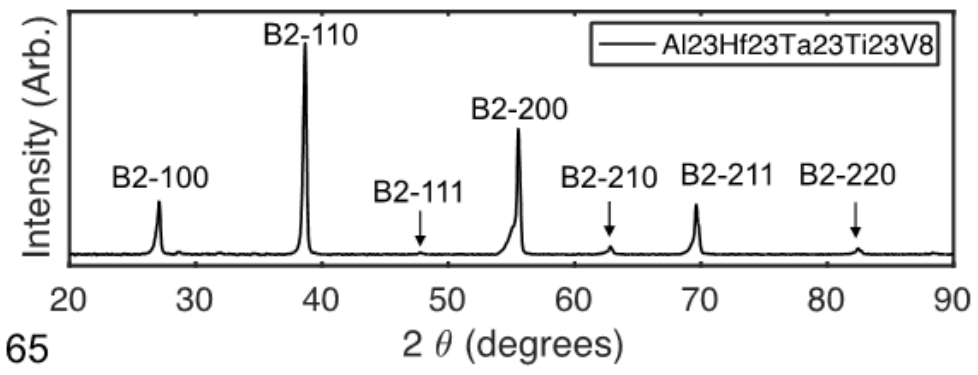

65

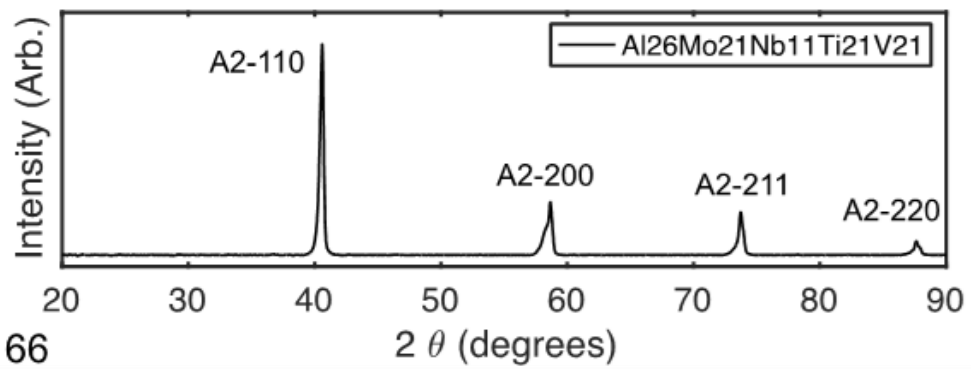

66

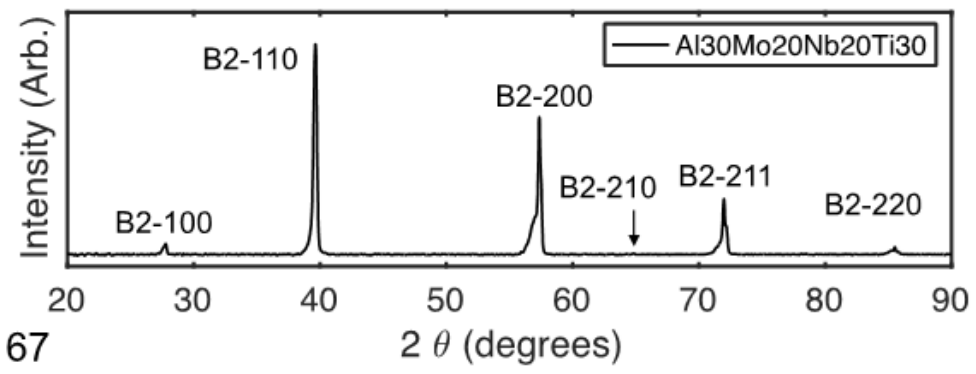

67

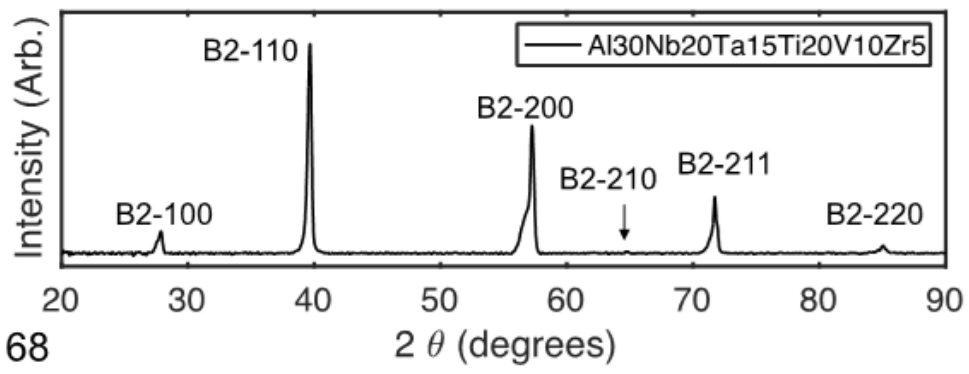

68

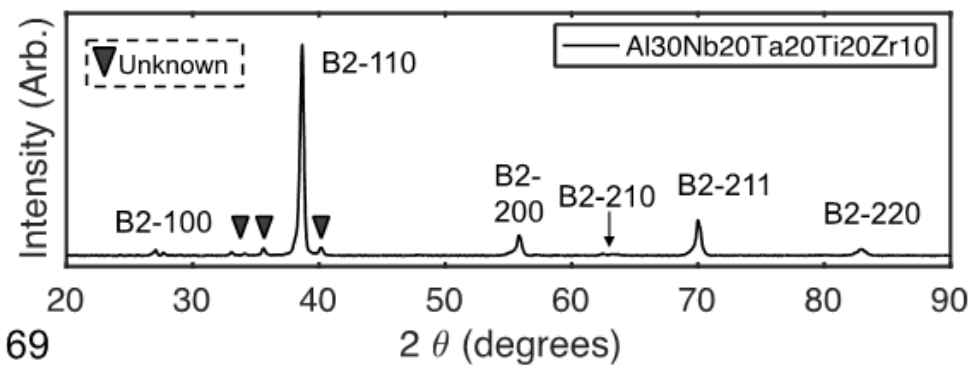

69

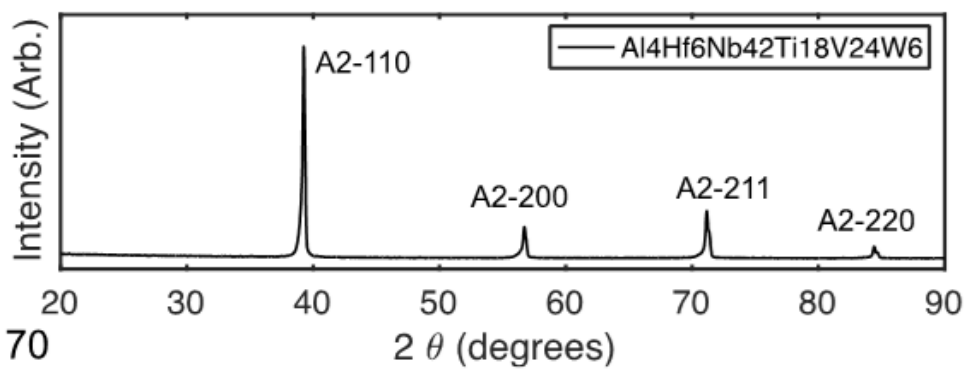

70

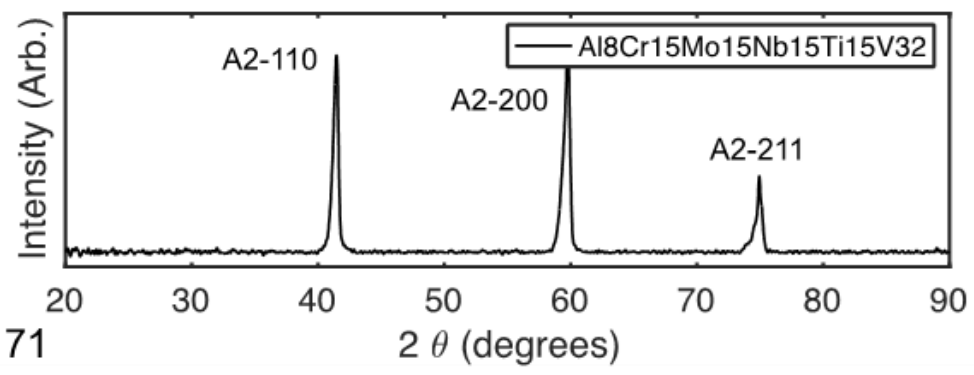

71

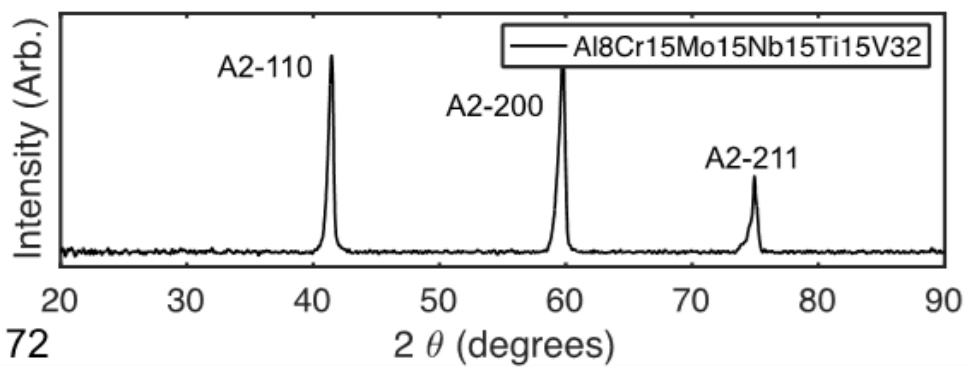

72

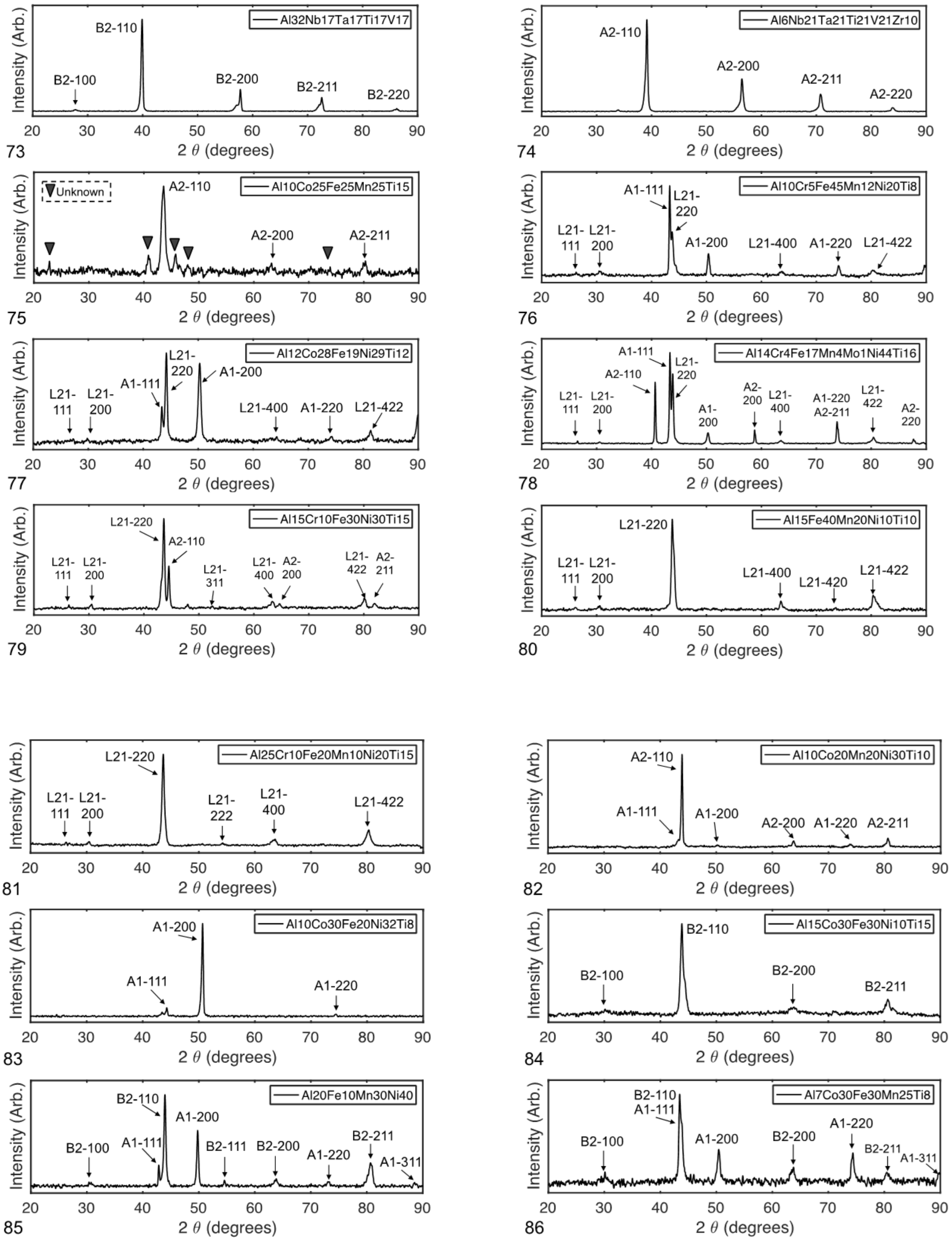

Figure S3. XRD patterns for newly synthesized validation HEAs.

## References:

1. Yang, T. *et al.* Control of nanoscale precipitation and elimination of intermediate-temperature embrittlement in multicomponent high-entropy alloys. *Acta Mater.* **189**, 47–59 (2020).
2. Wischhusen, M., Glover, C., Scully, J., Liaw, P. K. & Agnew, S. An Investigation into the Link Between Microstructure and Pitting Corrosion of Novel Alloy FBB8+Ti. *Miner. Met. Mater. Ser.* 1561–1571 (2020) doi:10.1007/978-3-030-36296-6_144.
3. Choudhuri, D. *et al.* Formation of a Huesler-like L21 phase in a CoCrCuFeNiAlTi high-entropy alloy. *Scr. Mater.* **100**, 36–39 (2015).
4. Gwalani, B. *et al.* Role of copper on L1 2 precipitation strengthened fcc based high entropy alloy. *Materialia* **6**, 100282 (2019).
5. Wang, Z., Wang, X., Yue, H., Shi, G. & Wang, S. Microstructure, thermodynamics and compressive properties of AlCoCrCuMn-x (x=Fe, Ti) high-entropy alloys. *Mater. Sci. Eng. A* **627**, 391–398 (2015).
6. Hu, Z., Zhan, Y., Zhang, G., She, J. & Li, C. Effect of rare earth Y addition on the microstructure and mechanical properties of high entropy AlCoCrCuNiTi alloys. *Mater. Des.* **31**, 1599–1602 (2010).
7. Haas, S., Manzoni, A. M., Krieg, F. & Glatzel, U. Microstructure and mechanical properties of precipitate strengthened high entropy alloy Al10Co25Cr8Fe15Ni36Ti6 with additions of Hafnium and molybdenum. *Entropy* **21**, 169 (2019).
8. Tsai, M. H., Chang, K. C., Li, J. H., Tsai, R. C. & Cheng, A. H. A second criterion for sigma phase formation in high-entropy alloys. *Mater. Res. Lett.* **4**, 90–95 (2016).
9. Dasari, S. *et al.* Discontinuous precipitation leading to nano-rod intermetallic precipitates in an Al0.2Ti0.3Co1.5CrFeNi1.5 high entropy alloy results in an excellent strength-ductility combination. *Mater. Sci. Eng. A* **805**, 140551 (2021).
10. Kuo, C. M. & Tsai, C. W. Effect of cellular structure on the mechanical property of Al0.2Co1.5CrFeNi1.5Ti0.3 high-entropy alloy. *Mater. Chem. Phys.* **210**, 103–110 (2018).
11. Gwalani, B. *et al.* Microstructure and wear resistance of an intermetallic-based Al0.25Ti0.75CoCrFeNi high entropy alloy. *Mater. Chem. Phys.* **210**, 197–206 (2018).
12. Bai, X. *et al.* Effects of Al and Ti additions on precipitation behavior and mechanical properties of Co35Cr25Fe40-xNix TRIP high entropy alloys. *Mater. Sci. Eng. A* **767**, 138403 (2019).
13. He, J. Y. *et al.* A precipitation-hardened high-entropy alloy with outstanding tensile properties. *Acta Mater.* **102**, 187–196 (2016).
14. He, J. Y. *et al.* Precipitation behavior and its effects on tensile properties of FeCoNiCr high-entropy alloys. *Intermetallics* **79**, 41–52 (2016).
15. He, J. Y. *et al.* High-temperature plastic flow of a precipitation-hardened FeCoNiCr high entropy alloy. *Mater. Sci. Eng. A* **686**, 34–40 (2017).
16. Qi, Y. *et al.* L21-strengthened face-centered cubic high-entropy alloy with high strength and ductility. *Mater. Sci. Eng. A* **797**, 140056 (2020).
17. Li, C. *et al.* Microstructures and mechanical properties of body-centered-cubic (Al,Ti)0.7(Ni,Co,Fe,Cr)5 high entropy alloys with coherent B2/L21 nanoprecipitation. *Mater. Sci. Eng. A* **737**, 286–296 (2018).
18. Stepanov, N. D., Shaysultanov, D. G., Tikhonovsky, M. A. & Zherebtsov, S. V. Structure and high temperature mechanical properties of novel non-equiatomic Fe-(Co, Mn)-Cr-Ni-Al-(Ti) high entropy alloys. *Intermetallics* **102**, 140–151 (2018).
19. Zhang, H., Pan, Y. & He, Y. Effects of annealing on the microstructure and properties of 6FeNiCoCrAlTiSi high-entropy alloy coating prepared by laser cladding. *J. Therm. Spray Technol.* **20**, 1049–1055 (2011).
20. Baker, P. I. I. Understanding the Deformation Mechanisms of FeNiMnAlCr High Entropy Alloys. 1–17 (2018) doi:10.2172/1458757.
21. Zhao, Y. L. *et al.* Development of high-strength Co-free high-entropy alloys hardened by nanosized precipitates. *Scr. Mater.* **148**, 51–55 (2018).
22. Feng, R. *et al.* Design of light-weight high-entropy alloys. *Entropy* **18**, 16–29 (2016).
23. Feng, R. *et al.* Phase stability and transformation in a light-weight high-entropy alloy. *Acta Mater.* **146**, 280–293 (2018).
24. Feng, R. *et al.* High-throughput design of high-performance lightweight high-entropy alloys. *Nat. Commun. 2021 121* **12**, 1–10 (2021).
25. Munitz, A., Salhov, S., Guttmann, G., Derimow, N. & Nahmany, M. Heat treatment influence on the microstructure and mechanical properties of AlCrFeNiTi0.5 high entropy alloys. *Mater. Sci. Eng. A* **742**, 1–14 (2019).
26. Lin, C. W., Tsai, M. H., Tsai, C. W., Yeh, J. W. & Chen, S. K. Microstructure and aging behaviour of Al5Cr32Fe35Ni22Ti6 high entropy alloy. *Mater. Sci. Technol. (United Kingdom)* **31**, 1165–1170 (2015).
27. Wolff-Goodrich, S. *et al.* Combinatorial exploration of B2/L21 precipitation strengthened AlCrFeNiTi compositionally complex alloys. *J. Alloys Compd.* **853**, 156111 (2021).
28. Wolff-Goodrich, S., Haas, S., Glatzel, U. & Liebscher, C. H. Towards superior high temperature properties in low density ferritic AlCrFeNiTi compositionally complex alloys. *Acta Mater.* **216**, 117113 (2021).
29. Tsai, M. H. *et al.* Criterion for sigma phase formation in Cr- and V-Containing high-entropy alloys. *Mater. Res. Lett.* **1**, 207–212 (2013).
30. Wang, M. *et al.* A novel bulk eutectic high-entropy alloy with outstanding as-cast specific yield strengths at elevated temperatures. *Scr. Mater.* **204**, 114132 (2021).
31. Wu, X. *et al.* Microstructure and mechanical behavior of directionally solidified Fe 35Ni 15Mn 25Al 25. *Intermetallics* **32**, 413–422 (2013).
32. Wu, X. *et al.* Microstructure and mechanical properties of two-phase Fe 30Ni20Mn20Al30. Part I: Microstructure. *J. Mater. Sci.* **48**, 7435–7445 (2013).
33. Zuo, T. *et al.* Tailoring magnetic behavior of CoFeMnNiX (X = Al, Cr, Ga, and Sn) high entropy alloys by metal doping. *Acta Mater.* **130**, 10–18 (2017).
34. Miracle, D. B., Tsai, M. H., Senkov, O. N., Soni, V. & Banerjee, R. Refractory high entropy superalloys (RSAs). *Scr. Mater.* **187**, 445–452 (2020).
35. Chen, H. *et al.* Crystallographic ordering in a series of Al-containing refractory high entropy alloys Ta–Nb–Mo–Cr–Ti–Al. *Acta Mater.* **176**, 123–133 (2019).
36. Gorr, B. *et al.* Development of Oxidation Resistant Refractory High Entropy Alloys for High Temperature Applications: Recent Results and Development Strategy. *Miner. Met. Mater. Ser.* **Part F12**, 647–659 (2018).
37. Laube, S. *et al.* Controlling crystallographic ordering in Mo–Cr–Ti–Al high entropy alloys to enhance ductility. *J. Alloys Compd.* **823**, 153805 (2020).
38. Yurchenko, N. Y. *et al.* Effect of Cr and Zr on phase stability of refractory Al-Cr-Nb-Ti-V-Zr

high-entropy alloys. *J. Alloys Compd.* **757**, 403–414 (2018).

39. Yurchenko, N. *et al.* Structure and mechanical properties of an in situ refractory Al20Cr10Nb15Ti20V25Zr10 high entropy alloy composite. *Mater. Lett.* **264**, 127372 (2020).
40. Huang, X., Miao, J. & Luo, A. A. Lightweight AlCrTiV high-entropy alloys with dual-phase microstructure via microalloying. *J. Mater. Sci.* **54**, 2271–2277 (2019).
41. Coury, F. G. *et al.* Phase equilibria, mechanical properties and design of quaternary refractory high entropy alloys. *Mater. Des.* **155**, 244–256 (2018).
42. Ye, Y. F., Wang, Q., Lu, J., Liu, C. T. & Yang, Y. High-entropy alloy: challenges and prospects. *Mater. Today* **19**, 349–362 (2016).
43. Senkov, O. N., Jensen, J. K., Pilchak, A. L., Miracle, D. B. & Fraser, H. L. Compositional variation effects on the microstructure and properties of a refractory high-entropy superalloy AlMo0.5NbTa0.5TiZr. *Mater. Des.* **139**, 498–511 (2018).
44. Jensen, J. K. *et al.* Characterization of the microstructure of the compositionally complex alloy Al1Mo0.5Nb1Ta0.5Ti1Zr1. *Scr. Mater.* **121**, 1–4 (2016).
45. Senkov, O. N., Isheim, D., Seidman, D. N. & Pilchak, A. L. Development of a Refractory High Entropy Superalloy. *Entropy 2016, Vol. 18, Page 102* **18**, 102 (2016).
46. Chen, S. Y., Yang, X., Dahmen, K. A., Liaw, P. K. & Zhang, Y. Microstructures and crackling noise of AlxNbTiMoV high entropy alloys. *Entropy* **16**, 870–884 (2014).
47. Lacour-Gogny-Goubert, A., Zhao-Huvelin, Z., Bachelier-Locq, A., Guillot, I. & Denquin, A. Effect of Al Content on Microstructure and Properties of AlxMoNbTiV RCCA's Alloys. *Mater. Sci. Forum* **941**, 1111–1116 (2018).
48. Peng, J., Li, S., Mao, Y. & Sun, X. Phase transformation and microstructures in Ti–Al–Nb–Ta system. *Mater. Lett.* **53**, 57–62 (2002).
49. Senkov, O. N., Woodward, C. & Miracle, D. B. Microstructure and Properties of Aluminum-Containing Refractory High-Entropy Alloys. *JOM* **66**, 2030–2042 (2014).
50. Soni, V. *et al.* Phase inversion in a two-phase, BCC+B2, refractory high entropy alloy. *Acta Mater.* **185**, 89–97 (2020).
51. Soni, V. *et al.* Phase stability as a function of temperature in a refractory high-entropy alloy. *J. Mater. Res.* **33**, 3235–3246 (2018).
52. Soni, V. *et al.* Phase stability and microstructure evolution in a ductile refractory high entropy alloy Al10Nb15Ta5Ti30Zr40. *Materialia* **9**, 100569 (2020).
53. Yurchenko, N. Y., Stepanov, N. D., Zherebtsov, S. V., Tikhonovsky, M. A. & Salishchev, G. A. Structure and mechanical properties of B2 ordered refractory AlNbTiVZrx (x = 0–1.5) high-entropy alloys. *Mater. Sci. Eng. A* **704**, 82–90 (2017).
54. Kral, P. *et al.* Creep behavior of an AlTiVNbZr0.25 high entropy alloy at 1073 K. *Mater. Sci. Eng. A* **783**, 139291 (2020).
55. He, H., Bai, Y., Garcia, E. A. & Li, S. ADASYN: Adaptive synthetic sampling approach for imbalanced learning. *Proc. Int. Jt. Conf. Neural Networks* 1322–1328 (2008) doi:10.1109/IJCNN.2008.4633969.
56. Chawla, N. V., Bowyer, K. W., Hall, L. O. & Kegelmeyer, W. P. SMOTE: Synthetic Minority Over-sampling Technique. *J. Artif. Intell. Res.* **16**, 321–357 (2002).